\documentclass[prd,superscriptaddress,nofootinbib,secnumarabic,preprintnumbers]{revtex4-2}[10pt]
\pdfoutput=1
\usepackage{latexsym}
\usepackage{mathrsfs, amsmath, amssymb}
\usepackage{makecell, multirow, diagbox, slashed}
\usepackage{hyperref, url}
\usepackage{color}
\usepackage{pstricks}
\usepackage{graphicx}
\usepackage{appendix}
\usepackage{verbatim}
\usepackage{float}
\usepackage{placeins}
\usepackage{ucs}
\usepackage{indentfirst, layout, geometry}
\usepackage{enumitem}

\newcommand{\mr}{\mathrm}
\newcommand{\mc}{\mathcal}
\newcommand{\ds}{\displaystyle}

\graphicspath{{figures/}}

\begin{document}
\title{Charting the new physics through one-loop effects in the muon magnetic dipole moment}
\author{Shi-Ping He}
\email{heshiping@tyut.edu.cn}
\affiliation{College of Physics and Optoelectronic Engineering, Taiyuan University of Technology, Taiyuan 030024, China}
\affiliation{Asia Pacific Center for Theoretical Physics, Pohang 37673, Korea}

\date{\today}

\begin{abstract}

Since the announcement of the muon $g-2$ measurements, numerous studies have been devoted to exploring the new physics through this precision probe. In specific models, the new physics contributions depend on the couplings and mass scales. Approximate formulae are widely adopted when determining the new physics contributions. This manuscript is dedicated to comprehensive analytical results and approximations for the canonical interactions at one-loop level, which is universal for the spin-half fermions and can serve as a useful reference for the magnetic dipole moment calculations. We present the analytic and approximate expressions for the scalar and vector mediator cases, supplemented by explicit examples in specific models to illustrate their applications. For given muon interactions, we can estimate both the sign and magnitude of the contributions conveniently. Furthermore, we investigate the broader physical implications of these results, offering insights into potential new physics.

\end{abstract}
\maketitle
\tableofcontents
\clearpage


\section{Introduction}
The muon anomalous magnetic dipole moment (aMDM), denoted as $a_\mu\equiv(g-2)_{\mu}/2$, is one of the crucial low energy observables sensitive to new physics, where $g$ is the Land\'e $g$-factor. The experimental introduction and theoretical determination are reviewed in the articles \cite{Miller:2007kk, Jegerlehner:2009ry, Jegerlehner:2017gek}. Within the framework of effective field theory (EFT), this moment corresponds to the dimension-5 dipole interaction $\pm ea_\mu/(4m_\mu)\bar{\mu}\sigma^{\mu\nu}\mu F_{\mu\nu}$. Here, $F_{\mu\nu}$ is the electromagnetic field tensor and the sign depends on the convention of covariant derivative. In the standard model (SM), this quantity can be predicted precisely by summing the quantum electrodynamics (QED), electroweak, and hadronic contributions. It is calculated as $a_{\mu}^{\mr{SM}}=116591810(43)\times10^{-11}$ from the 2020 white paper \cite{Aoyama:2020ynm}. In the 2025 white paper \cite{Aliberti:2025beg}, the most precise prediction is updated as $a_{\mu}^{\mr{SM}}=116592033(62)\times10^{-11}$. It means that the central value of theoretical prediction is shifted sizably, which impacts the observed discrepancy with experiment data.

The $(g-2)_{\mu}$ anomaly was first reported by the E821 experiment at Brookhaven National Laboratory (BNL) \cite{Muong-2:2006rrc}. In 2021, the Fermilab (FNAL) muon $g-2$ experiment increased the tension between the SM prediction and experiment data \cite{Muong-2:2021ojo}. Combining the BNL and FNAL Run-1 data, the averaged experimental result is $a_{\mu}^{\mr{Exp}}=116592061(41)\times10^{-11}$, corresponding to a $4.2\sigma$ discrepancy from the 2020 white paper prediction, namely $\Delta a_{\mu}\equiv a_{\mu}^{\mr{Exp}}-a_{\mu}^{\mr{SM}}=(251\pm59)\times10^{-11}$. In 2023, the FNAL muon $g-2$ experiment released the Run-2 and Run-3 data, which is consistent with the BNL and FNAL Run-1 data. The refined world average is $a_{\mu}^{\mr{Exp}}=116592059(22)\times10^{-11}$ at that time, which shows a $5.1\sigma$ tension with the 2020 white paper prediction \cite{Muong-2:2023cdq}. The persistent deviation from the SM prediction seems to indicate the indirect signal of new physics. The new physics scale can be light to produce considerable contributions. When the new physics scale is heavy, there should be a chiral enhancement to achieve the observable contributions. There are numerous new physics models motivated to address this anomaly, for example, the two-Higgs-doublet model \cite{Cherchiglia:2017uwv, Iguro:2023tbk}, dark photon model \cite{Pospelov:2008zw}, supersymmetric models \cite{Moroi:1995yh, Stockinger:2006zn, Stockinger:2006zn}, leptoquark extended models \cite{Dorsner:2016wpm, He:2021yck, He:2022zjz, He:2022glw}, and vector-like lepton extended models \cite{Crivellin:2018qmi, Crivellin:2021rbq, Cai:2024xvu}. Besides, more explanations are reviewed in the papers \cite{Czarnecki:2001pv, Jegerlehner:2009ry, Freitas:2014pua, Queiroz:2014zfa, Lindner:2016bgg, Calibbi:2018rzv, Athron:2021iuf}.

In 2025, the FNAL muon $g-2$ experiment released the latest data with the improvement of statistical precision by more than 2.5. Combined with the previous measurements, the new experimental world average is updated as $a_{\mu}^{\mr{Exp}}=116592071.5(14.5)\times10^{-11}$ \cite{Muong-2:2025xyk}. Adopting the most recent 2025 white paper prediction, the deviation from SM is obtained as $\Delta a_{\mu}\equiv a_{\mu}^{\mr{Exp}}-a_{\mu}^{\mr{SM}}=(38\pm63)\times10^{-11}$, which is consistent with the SM prediction \cite{Aliberti:2025beg}. Prior to release of the theoretical and experimental results in 2025, total new physics contributions should be positive and sizable such that $\Delta a_{\mu}=(251\pm59)\times10^{-11}$. At current stage, a key point is that both the positive and negative total new physics contributions are allowed to make $\Delta a_{\mu}=(38\pm63)\times10^{-11}$. Besides, the magnitude of new physics contributions is required to be small, which can still lead to strong constraints on some new physics models. In the light of new data of the $\Delta a_{\mu}$, the constraints on new physics models are significantly altered as shown in \cite{Athron:2025ets}. The simplest dark photon model revives under existing upper limit of kinetic mixing, where the dark photon mass is viable in certain parameter space. In supersymmetric models with heavy sparticle masses (such as TeV), the contributions can be small naturally, which is now in agreement with new results. For the leptoquark models, the representations without chiral enhancement are available currently under the perturbative couplings. For the representations with chirally enhanced contribution (see scalar leptoquark $R_2$ and $S_1$ \cite{Dorsner:2016wpm}), the new results can set upper bounds for the leptoquark Yukawa couplings. In the minimal vector-like lepton models, the mixing angle with SM lepton confronts the stringent constraints from electroweak precision observables. Hence, the new physics contributions are negligible, which means that they are permitted by the new results.

Obviously, the magnetic dipole operator $\bar{\mu}\sigma^{\mu\nu}\mu F_{\mu\nu}$ flips the chirality. Respecting to the SM gauge symmetry $SU_L(2)\otimes U_Y(1)$, such dipole interactions originate from the gauge invariant dimension-6 operators $\overline{L_L^i}\sigma^{\mu\nu}e_R^jHB_{\mu\nu}$ and $\overline{L_L^i}\sigma^{\mu\nu}\tau^ae_R^jHW_{\mu\nu}^a$ \cite{Grzadkowski:2010es}. To induce the dipole interactions, the SM gauge symmetry should be spontaneously broken. In ultraviolet complete models, there can be new particles or new interactions with modified SM couplings. To explore the new physics contributions, we need to consider all the muon-related interactions. Starting from these interactions, we can obtain the contributions emerging at loop level. Crucially, the chirality flip may occur in either the external muon line or the internal fermion line, which can cause distinct coupling patterns. When determining the new physics contributions, approximate formulae are usually adopted depending on the mass scale hierarchy. In Ref. \cite{Yu:2021suw}, the authors studied several approximations. Therefore, a comprehensive investigation of the relevant formulae is necessary, which is useful and convenient for the new physics study. Here, we provide the analytic and approximate expressions for the scalar and vector mediator contributions. Based on the analytic and approximate expressions, we also discuss the implications on potential new physics. With the decrease of theoretical and experimental uncertainties in the future, the $(g-2)_{\mu}$ anomaly might continue to be gone or reappear. However, the formulae remain universally applicable.

In Sec. \ref{sec:sim}, we introduce the most general renormalizable simplified interactions contributing to $(g-2)_{\mu}$. Then, we parameterize the contributions and perform the analytic forms for the scalar and vector mediators in Sec. \ref{sec:contributions}. The Sec. \ref{sec:expansionS} is dedicated to the expansion of the loop functions for scalar mediator case, namely, $I_{LL/LR}^f$ and $I_{LL/LR}^S$. The Sec. \ref{sec:expansionV} is dedicated to the expansion of the loop functions for vector mediator case, namely, $L_{LL/LR}^f$ and $L_{LL/LR}^V$. In Sec. \ref{sec:summary:examples}, we present examples in specific models to illustrate the applications of our formulae. Finally, we make the whole summary and conclusions in Sec. \ref{sec:summary}.

\section{The $(g-2)_\mu$ in simplified models}\label{sec:sim}
\subsection{Canonical interactions}\label{sec:sim:cano}
First, let us consider the muon interactions with a Dirac fermion $f$ mediated by a scalar $S$ or vector $V_{\mu}$. Then, the most general renormalizable Lagrangian can be written in two equivalent bases:
\begin{align}\label{eqn:int:mufM}
&\mathcal{L}\supset\bar{\mu}(y_S+i\gamma^5y_P)fS+\bar{\mu}\gamma^{\mu}(g_V+\gamma^5g_A)fV_{\mu}+\mathrm{h.c.}~\nonumber\\
&=\bar{\mu}(y_L\omega_-+y_R\omega_+)fS+\bar{\mu}\gamma^{\mu}(g_L\omega_-+g_R\omega_+)fV_{\mu}+\mathrm{h.c.}~,
\end{align}
where $\omega_\pm\equiv(1\pm\gamma^5)/2$ are the chirality projection operators. Generally, all the parameters above are complex. The charge conservation requires $Q_f+Q_{S(V)}=-1$. Especially, $f$ is neutral for $Q_{S(V)}=-1$. On the contrary, $S(V)$ is neutral for $Q_f=-1$. Hereafter, the first line in Eq. \eqref{eqn:int:mufM} is named as $CP$ basis, and the second line in Eq. \eqref{eqn:int:mufM} is called as chiral basis. The complex parameters in the two basis can be correlated through the relations
\begin{align}\label{eqn:sim:rep1}
y_S=\frac{y_L+y_R}{2},~y_P=\frac{y_R-y_L}{2i},~g_V=\frac{g_L+g_R}{2},~g_A=\frac{g_R-g_L}{2},
\end{align}
or equivalently,
\begin{align}\label{eqn:sim:rep2}
y_L=y_S-iy_P,~y_R=y_S+iy_P,~g_L=g_V-g_A,~g_R=g_V+g_A.
\end{align}
Each basis possesses its advantages and disadvantages. In the $CP$ basis, the contributions from scalar part $\bar{\mu}fS$ and pseudo-scalar part $\bar{\mu}(i\gamma^5)fS$ are separated, similar to the contributions from vector part $\bar{\mu}\gamma^{\mu}fV_{\mu}$ and axial vector part $\bar{\mu}\gamma^{\mu}\gamma^5fV_{\mu}$. In other words, there are only $|y_{S(P)}|^2$ and $|g_{V(A)}|^2$ terms; the terms like $y_Sy_P$ and $g_Vg_A$ do not show up, which is the consequence of $P$ invariance of the magnetic dipole interaction $\bar{\mu}\sigma^{\mu\nu}\mu F_{\mu\nu}$. In the chiral basis, the chiral flips from internal fermion and external muon are explicitly isolated. Because the SM gauge group is chiral,
the interactions are always given in terms of chiral fields. Moreover, it is convenient to capture the dominated contributions from the coupling of chiral symmetry breaking. Therefore, we prefer the contributions in the chiral basis in this manuscript.

For the special case of $f$ being muon, hermiticity necessitates that both $S$ and $V_{\mu}$ are self-conjugate. Consequently, the renormalizable Lagrangian in Eq. \eqref{eqn:int:mufM} should be parameterized as
\begin{align}\label{eqn:int:mumuM}
&\mathcal{L}\supset\bar{\mu}(y_S+i\gamma^5y_P)\mu S+\bar{\mu}\gamma^{\mu}(g_V+\gamma^5g_A)\mu V_{\mu}~\nonumber\\
&=\bar{\mu}(y_L\omega_-+y_L^{\ast}\omega_+)\mu S+\bar{\mu}\gamma^{\mu}(g_L\omega_-+g_R\omega_+)\mu V_{\mu}.
\end{align}
In this case, the $y_L$ is complex, while the other parameters $y_{S(P)},g_{V(A)},g_{L(R)}$ are all real.

Further, let us take the covariant derivative of electromagnetic interactions as $D_{\mu}\equiv\partial_{\mu}-iQeA_{\mu}$. Hence, we present the corresponding renormalizable QED interactions as\footnote{Here, we extract the QED interaction of charged vector based on the $W$ boson in the SM, and more general analysis is illustrated in Ref. \cite{Biggio:2016wyy}.}
\begin{align}\label{eqn:int:fMQED}
&eQ_f\bar{f}\gamma^{\mu}fA_{\mu}+ieQ_S[S^{\dag}(\partial^{\mu}S)-(\partial^{\mu}S)^{\dag}S]A_{\mu}+ieQ_V\big\{[V_{\mu\nu}(V^{\mu})^{\dag}-(V_{\mu\nu})^{\dag}V^{\mu}]A^{\nu}+V_{\mu}(V_{\nu})^{\dag}A^{\mu\nu}\big\},
\end{align}
where the $A_{\mu}$ labels the photon $\gamma$, and the field strength tensor $V_{\mu\nu}$ is defined as $(\partial_{\mu}V_{\nu}-\partial_{\nu}V_{\mu})$. Note that the standard interactions above are the minimal interactions respecting the electromagnetic gauge symmetry $U_{EM}(1)$, in which the QED interactions of fermions and bosons are chosen to be consistent.

Here, we consider the interactions in the unitary gauge. Thus, the $A^{\mu}V_{\mu}^{\dag}\phi$ interactions do not appear, in which $\phi$ is the Goldstone boson of $V_{\mu}$. For physical field $S$, the renormalizable $A^{\mu}V_{\mu}^{\dag}S$ interactions are forbidden because it violates the Ward identity.\footnote{We would like to thank Prof. Qing-Hong Cao for discussing about this problem.} For example, the singly charged scalar vertex $\gamma W^{\mp}H^{\pm}$ can only be induced by high-dimensional operators \cite{Diaz-Cruz:2001thx}. In renormalizable models, there are no tree level $\gamma W^{\mp}H^{\pm}$ interactions, which is verified in the charged Higgs singlet model \cite{Cao:2017ffm}, two Higgs doublet model \cite{Arhrib:2006wd, Aiko:2021can}, and Higgs triplet model \cite{Logan:2018wtm}.

Assuming Lorentz invariance and electromagnetic gauge invariance, we have written down the most general renormalizable interactions contributing to muon $g-2$ directly at one-loop level. For simplicity, the interactions in Eqs. \eqref{eqn:int:mufM} and \eqref{eqn:int:fMQED} are named as canonical interactions. These canonical interactions always appear in the renormalizable models, which are mainly investigated in this manuscript.

\subsection{Quasi-canonical interactions}\label{sec:sim:quasi-cano}
In previous part, we have given the canonical interactions in Eqs. \eqref{eqn:int:mufM} and \eqref{eqn:int:fMQED}. Some interactions may be different from the canonical interactions in form, while they can be transformed into canonical interactions through field redefinition, total derivative, equation of motion, and so on. These types of interactions are dubbed as quasi-canonical interactions, and this section is devoted to the brief statements of them.
\begin{itemize}[itemindent=0em, leftmargin=10pt, listparindent=1em]
\item \textit{Interactions involving Weyl and Majorana fermions}

For the Weyl fermion, the interactions can be written in left-handed or right-handed Dirac form. For the Majorana fermion, the interactions can also be written in Dirac form. Strict treatment of the spinors is discussed in Refs. \cite{Denner:1992vza, Dreiner:2008tw}.

\item \textit{Interactions involving other spin-half fermions}

The formulae are not only valid for the muon MDM, but also universal for any other spin-half fermions through proper substitutions of masses, couplings, and electric charge conservation conditions. For example, the $\bar{\chi}fS^{\ast}$ interactions will induce the MDM of $\chi$, which can be obtained by modifying the electric charge conservation condition as $Q_f+Q_{S^{\ast}}=Q_f-Q_S=Q_{\chi}$ and replacing $m_{\mu}$ by $m_{\chi}$.

\item \textit{Interactions involving charge conjugate fields}

For the field acted by the charge conjugate operator $\mc{C}$, there are fermion number violated ($F=2$) interactions such as $\bar{\mu}f^CS$ and $\bar{\mu}\gamma^{\mu}f^CV_{\mu}$. However, they can be mapped into the canonical interactions through field redefinition of $\psi\equiv f^C$. Because the current $\bar{f}\gamma^{\mu}f$ changes sign under the $\mc{C}$ transformation, we have the electric charge relation $Q_{\psi}=Q_{f^C}=-Q_f$. Analogously, the $\bar{t}f^CS$ interactions will induce the MDM of top quark, which can be obtained by modifying the electric charge conservation condition as $Q_{f^C}+Q_S=-Q_f+Q_S=Q_t=2/3$ and replacing $m_{\mu}$ by $m_t$.

\item \textit{Interactions involving equation of motion}

Some interactions can be transformed into the canonical interactions through total derivative and equation of motion, for example, $(\partial_{\mu}S)\bar{\mu}\gamma^{\mu}\gamma^5\mu=\partial_{\mu}(S\bar{\mu}\gamma^{\mu}\gamma^5\mu)-S\partial_{\mu}(\bar{\mu}\gamma^{\mu}\gamma^5\mu)$ and $(\partial_{\mu}S)\bar{\mu}\gamma^{\mu}\mu=\partial_{\mu}(S\bar{\mu}\gamma^{\mu}\mu)-S\partial_{\mu}(\bar{\mu}\gamma^{\mu}\mu)$. Total derivative vanishes in action. Applying the axial current equation $\partial_{\mu}(\bar{\mu}\gamma^{\mu}\gamma^5\mu)=2im_{\mu}\bar{\mu}\gamma^5\mu$ and vector current equation $\partial_{\mu}(\bar{\mu}\gamma^{\mu}\mu)=0$, the contributions are equivalent to those of canonical interactions. Similarly, the high-dimensional operators $(\partial_{\mu}S)\overline{\mu_L}\gamma^{\mu}f_L$ and $(\partial_{\mu}S)\overline{\mu_R}\gamma^{\mu}f_R$ can also be transformed into canonical interactions as illustrated in Ref. \cite{Galda:2023qjx}.
\end{itemize}

\subsection{Comments on non-canonical contributions}\label{sec:sim:non-cano}
In addition to the canonical and quasi-canonical interactions at one-loop level discussed in Eqs. \eqref{eqn:int:mufM} and \eqref{eqn:int:fMQED}, there can be more complex interactions in the high spin models, EFT framework or high-loop level. These contributions, dubbed as non-canonical, will not be studied in this manuscript. Here are some simple comments on the non-canonical contributions.
\begin{itemize}[itemindent=0em, leftmargin=10pt,listparindent=1em]

\item \textit{Interactions involving high spin particles}

The fermion $f$ in this work is spin half, while it can also be a spin 3/2 particle \cite{Criado:2021qpd}. In addition to the scalar and vector mediators, the mediator can also be a graviton \cite{Huang:2022zet}.

\item \textit{Interactions involving high-dimensional operators}

Besides the renormalizable interactions, there can be high-dimensional operators. The contributions from four fermion operators, two fermion operators, and bosonic operators are presented in  the SMEFT (standard model effective field theory) framework \cite{Jenkins:2013wua, Alonso:2013hga} and low energy EFT framework \cite{Jenkins:2017dyc, Aebischer:2021uvt}.

$\Longrightarrow$ Four-fermion interactions: In SMEFT, the dimension-6 operator $(\overline{L_{L,a}}\sigma^{\mu\nu}e_R)\epsilon_{ab}(\overline{Q_{L,b}}\sigma_{\mu\nu}u_R)$ can contribute to $(g-2)_\mu$ \cite{Jenkins:2013wua}. Here, we drop the generation indices and label the $SU_L(2)$ doublet indices as $a,b$. Similarly, the muon MDM may also receive contributions from the two muon and two fermion operators if $f$ is electrically charged.

$\Longrightarrow$ Dipole interactions: In SMEFT, the dimension-6 operators $\overline{L_L}\sigma^{\mu\nu}e_RHB_{\mu\nu}$ and $\overline{L_L}\sigma^{\mu\nu}\tau^ae_RHW_{\mu\nu}^a$ can contribute to $(g-2)_\mu$ \cite{Jenkins:2013wua}. Similarly, the muon MDM may also receive contributions from the dipole operators such as $\bar{\mu}\sigma^{\mu\nu}fV_{\mu\nu}$, $\bar{\mu}\sigma^{\mu\nu}\gamma^5fV_{\mu\nu}$, $\bar{f}\sigma^{\mu\nu}fA_{\mu\nu}$, and $\bar{f}\sigma^{\mu\nu}\gamma^5fA_{\mu\nu}$.

$\Longrightarrow$ Non-standard scalar QED interactions: In addition to the renormalizable QED interactions in Eq. \eqref{eqn:int:fMQED}, there can be $SS\gamma$ interactions of $i[(\partial^{\mu}S)^{\dag}(\partial^{\nu}S)-(\partial^{\nu}S)^{\dag}(\partial^{\mu}S)]A_{\mu\nu}$ type.

$\Longrightarrow$ Non-standard vector QED interactions: In general, the $WW\gamma$ interactions can be written as \cite{Hagiwara:1993ck}
\begin{align}
ieQ_{W^+}[g_{\gamma}(W_{\mu\nu}^+W^{-,\mu}-W_{\mu\nu}^-W^{+,\mu})A^{\nu}+\kappa_{\gamma}W_{\mu}^+W_{\nu}^-A^{\mu\nu}+\frac{\lambda_{\gamma}}{m_W^2}W_{\mu\nu}^+W^{-,\nu\rho}A_{\rho}^{~\mu}],
\end{align}
which returns to the SM for $g_{\gamma}=\kappa_{\gamma}=1$ and $\lambda_{\gamma}=0$. In Ref. \cite{Choudhury:2022iqz}, the authors studied the effects of non-minimal $WW\gamma$ interactions to $(g-2)_\mu$. Similarly, the standard $VV\gamma$ interactions can also be modified when $W^+(W^-)$ is replaced by $V(V^{\dag})$.

$\Longrightarrow$ Scalar and two field strength tensor interactions: In SMEFT, the dimension-6 operators $H^{\dag}HB^{\mu\nu}B_{\mu\nu}$, $H^{\dag}\tau^aHB^{\mu\nu}W_{\mu\nu}^a$, and $H^{\dag}HW^{a,\mu\nu}W_{a,\mu\nu}$ can contribute to $(g-2)_\mu$ \cite{Alonso:2013hga}. Similarly, the muon MDM may also receive contributions from the  bosonic operators such as $SA_{\mu\nu}A^{\mu\nu}$ and $SA_{\mu\nu}Z^{\mu\nu}$. In Ref. \cite{Bauer:2017ris}, the authors give the contributions from axion-like particle interactions $aA_{\mu\nu}\widetilde{A}^{\mu\nu}$ and $aA_{\mu\nu}\widetilde{Z}^{\mu\nu}$.

\item \textit{Interactions involving high-loop contributions}

Although we concentrate on the general one-loop contributions, there can also be high-loop contributions. For example, the new physics contributions are studied at two-loop level \cite{Barr:1990vd, Heinemeyer:2004yq, Iguro:2023tbk}, which can become important in some models. In fact, partial contributions from the high-dimensional operators are equivalent to those in the renormalizable models at two-loop level, because the inner loop can be contracted into a high-dimensional vertex effectively.

\item \textit{Beyond the local quantum field theory}

Actually, most of the explanations are based on the interactions in traditional local quantum field theory. However, there are also some quite different ideas. In Refs. \cite{Capolupo:2022awe, Abu-Ajamieh:2023txh}, the non-local QED interactions are investigated. In Ref. \cite{Morishima:2018bqz}, the authors even studied the effects of gravity.

\end{itemize}

\section{Contributions to $(g-2)_\mu$ for the canonical interactions}\label{sec:contributions}
The lepton MDM arises from loop corrections to the $\gamma\ell^+\ell^-$ vertex. For the $\ell^-\rightarrow\ell^-\gamma$ process, the amplitude can be parameterized as
\begin{align}\label{eqn:contri:MDMdef}
i\mc{M}=(ieQ_{\ell})\bar{u}(p_1)[\gamma^{\mu}F_1(q^2)+\frac{i\sigma^{\mu\nu}(p_1-p_2)_{\nu}}{2m_{\ell}}F_2(q^2)]u(p_2)\epsilon_{\mu}^{\ast}(q)~(q\equiv p_2-p_1).
\end{align}
Then, the lepton MDM is given as $a_{\ell}=F_2(0)$. Typically, radiative corrections generate the tensor structures of $\bar{u}(p_1)p_1^{\mu}u(p_2)$ and $\bar{u}(p_1)p_2^{\mu}u(p_2)$, which can be decomposed as the $(p_1-p_2)^{\mu}$ and $(p_1+p_2)^{\mu}$ terms. The $(p_1-p_2)^{\mu}$ term is eliminated by the photon polarization vector, while the $(p_1+p_2)^{\mu}$ term can be converted into the dipole form through the following Gordon identity:
\begin{align}
&(2m_{\ell})\bar{u}(p_1)\gamma^{\mu}u(p_2)=\bar{u}(p_1)[(p_1+p_2)^{\mu}+i\sigma^{\mu\nu}(p_1-p_2)_{\nu}]u(p_2).
\end{align} 
 Here are features of the form factor $F_2$:
 
\begin{itemize}[itemindent=0em, leftmargin=10pt,listparindent=1em]

\item \textit{Universal definition}: The definition of MDM is not only for the lepton in QED but also for any fermions in other models.

\item \textit{Convention independence}: The factor $(ieQ_{\ell})$ is extracted, which means that there is a ratio between the one-loop form factor and tree level QED coupling. Hence, MDM is free of the convention of covariant derivative.

\item \textit{Cross-channel equivalence}: The MDM can also be extracted in the other processes, for example, the $\gamma\rightarrow\ell^-\ell^+$ process, for which the amplitude can be written as
\begin{align}
i\mc{M}=(ieQ_{\ell})\bar{u}(p_1)[\gamma^{\mu}F_1(q^2)+\frac{i\sigma^{\mu\nu}(p_1+p_2)_{\nu}}{2m_{\ell}}F_2(q^2)]v(p_2)\epsilon_{\mu}(q)~(q\equiv p_1+p_2).
\end{align}
Then, the $(p_1+p_2)^{\mu}$ term is eliminated by the photon polarization vector. For the $(p_1-p_2)^{\mu}$ term, we can transform it into the dipole form through the following Gordon identity:
\begin{align}
&(2m_{\ell})\bar{u}(p_1)\gamma^{\mu}v(p_2)=\bar{u}(p_1)[(p_1-p_2)^{\mu}+i\sigma^{\mu\nu}(p_1+p_2)_{\nu}]v(p_2).
\end{align}

\item \textit{Complex structure}: The form factor $F_2(q^2)$ generally is complex; thus, there can be an absorptive or imaginary part for certain parameter space and $q^2$.\footnote{The imaginary part detection of top quark form factor $F_2(q^2)$ is studied in \cite{Atwood:1991ka} In Refs \cite{Bernreuther:2013aga, Montano-Dominguez:2021zmy}, the authors investigate chromo form factor of the top quark. There are also some studies on the $\tau$ weak MDM in Refs. \cite{Bernabeu:1994wh, ALEPH:2002kbp}.}.

\end{itemize}

\subsection{General structure of the one-loop contributions}
\begin{figure}
\begin{center}
\includegraphics[scale=0.5]{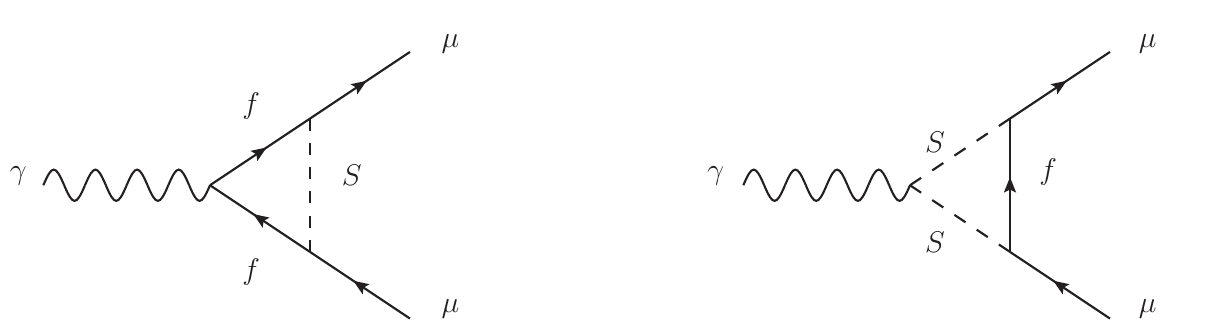}\\
\includegraphics[scale=0.5]{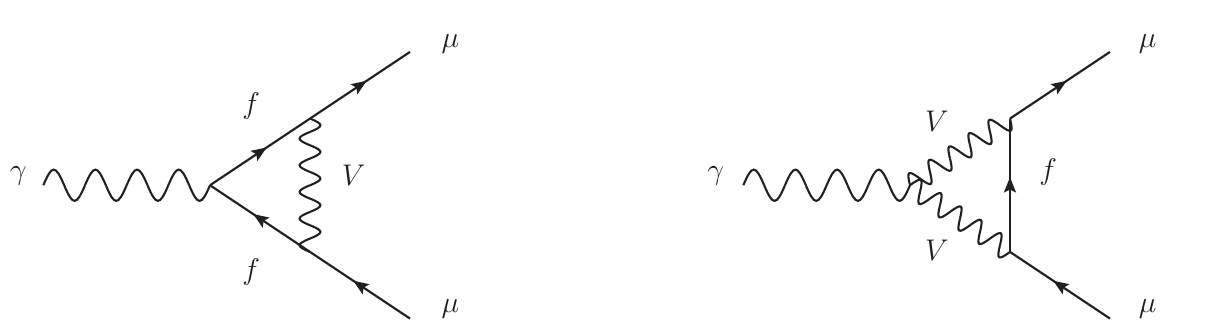}
\caption{Typical scalar (upper) and vector (lower) mediator Feynman diagrams which contribute to $(g-2)_\mu$ in the unitary gauge. Here, the diagrams are drawn by JaxoDraw \cite{Binosi:2008ig}.}\label{fig:sim:FeynSV}
\end{center}
\end{figure}

In the SM, the electroweak corrections at one-loop level have been calculated for a long time \cite{Jackiw:1972jz, Bardeen:1972vi, Leveille:1977rc}. There are $W,Z,h$ contributions, of which the Higgs contribution is negligible because of the light muon Yukawa coupling. In new physics models, there are two types of contributions. The first type originates from the modification of SM gauge interactions $Z\mu\mu,W\mu\nu_{\mu}$ and Yukawa interactions $h\mu\mu$. Because the $Z\mu\mu,W\mu\nu_{\mu}$ couplings have been constrained strictly \cite{ParticleDataGroup:2024cfk}, their effects are typically small. Although the deviation from SM $h\mu\mu$ coupling can be large \cite{ATLAS:2020fzp, CMS:2020xwi}, its contribution is suppressed by the small Yukawa coupling. The second type originates from the new scalar and vector mediated contributions. In general, there are two types of Feynman diagrams shown in Fig. \ref{fig:sim:FeynSV} at one-loop level. Here, we concentrate on the unitary gauge to avoid Feynman diagrams containing the unphysical Goldstones.

Starting from the canonical interactions in Eqs. \eqref{eqn:int:mufM} and \eqref{eqn:int:fMQED}, the new physics contributions to $(g-2)_\mu$ in $CP$ basis can be summed as \cite{Leveille:1977rc}
\begin{align}\label{eqn:contri:structure:CP}
&\Delta a_{\mu}\equiv\Delta a_{\mu}^S+\Delta a_{\mu}^V=\frac{N_Cm_{\mu}^2}{8\pi^2}\Bigg[\sum_S\frac{|y_S|^2(-Q_fI_S^f-Q_SI_S^S)+|y_P|^2(-Q_fI_P^f-Q_SI_P^S)}{m_S^2}\nonumber\\
&+\sum_V\frac{|g_V|^2(-Q_fL_V^f-Q_VL_V^V)+|g_A|^2(-Q_fL_A^f-Q_VL_A^V)}{m_V^2}\Bigg],
\end{align}
where the $\Delta a_{\mu}^S$ and $\Delta a_{\mu}^V$ label the contributions from the scalar and vector mediators. The $N_C$ is color factor, which is typically 3 for the colored loops and 1 for the colorless loops. Hereafter, the $I_{\times}^{\times}(L_{\times}^{\times})$ with superscript and subscript label the scalar (vector) loop functions. The superscript $``f\,"$ marks the $ff\gamma$ vertex mediated loop functions, while the superscript $``S(V)\,"$ marks the $SS\gamma$ ($VV\gamma$) vertex mediated loop functions. In this $CP$ basis, the subscripts $``S(P)\,"$ and $``V(A)\,"$ represent the contributions from scalar (pseudo-scalar) and vector (axial vector) interactions. Note that the pseudo-scalar part $\bar{\mu}(i\gamma^5)fS$ and axial vector part $\bar{\mu}\gamma^{\mu}\gamma^5fV_{\mu}$ can be converted into the form of scalar part $\bar{\mu}\psi S$ and vector part $\bar{\mu}\gamma^{\mu}\psi V_{\mu}$ through field redefinitions of $\psi\equiv i\gamma^5f$ and $\psi\equiv\gamma^5f$, respectively. Thus, we have the relations $I_P^{f(S)}=I_S^{f(S)}(m_f\rightarrow-m_f)$ and $L_A^{f(V)}=L_V^{f(V)}(m_f\rightarrow-m_f)$.

Similarly, the contributions in the preferred chiral basis can be expressed as \footnote{Note that the parameterization is invariant under the transformations of $y_L\leftrightarrow y_R$, $y_L\leftrightarrow (y_R)^{\ast}$, and so on, which is a consequence of the $\mc{C},\mc{P},\mc{T}$ symmetry of the dipole operator $\bar{\mu}\sigma^{\mu\nu}\mu F_{\mu\nu}$.}
\begin{align}\label{eqn:contri:structure:chiral}
&\Delta a_{\mu}=\frac{N_Cm_{\mu}^2}{8\pi^2}\Bigg\{\sum_S\frac{(|y_L|^2+|y_R|^2)(-Q_fI_{LL}^f-Q_SI_{LL}^S)+\left[y_L(y_R)^\ast+y_R(y_L)^\ast\right](-Q_fI_{LR}^f-Q_SI_{LR}^S)}{m_S^2}\nonumber\\
	&+\sum_V\frac{(|g_L|^2+|g_R|^2)(-Q_fL_{LL}^f-Q_VL_{LL}^V)+\left[g_L(g_R)^\ast+g_R(g_L)^\ast\right](-Q_fL_{LR}^f-Q_VL_{LR}^V)}{m_V^2}\Bigg\},
\end{align}
Again, $\sum_S$ and $\sum_V$ represent sum over all the possible scalar and vector mediated interactions. The subscripts different from the $CP$ basis are to identify the chiral basis. Here, the subscripts $``LL\,"$ and $``LR\,"$ represent the contributions with chiral flips in the external muon line and internal fermion line, respectively. As we can see, contributions to $(g-2)_\mu$ are determined by the Yukawa (gauge) coupling, the scalar (vector) mediator mass, and the fermion mass. In this manuscript, we do not take into account the width of particles. In Ref. \cite{Crivellin:2022gfu}, the authors describe the width effects characterized by the K\"all\'en--Lehmann representation spectral function.

Using the relations in Eqs. \eqref{eqn:sim:rep1} and  \eqref{eqn:sim:rep2}, we can derive the identities between the $CP$ and chiral basis integrals.
\begin{align}
&I_S^{f(S)}=2(I_{LL}^{f(S)}+I_{LR}^{f(S)}),\qquad I_P^{f(S)}=2(I_{LL}^{f(S)}-I_{LR}^{f(S)}),\nonumber\\
&L_V^{f(V)}=2(L_{LL}^{f(V)}+L_{LR}^{f(V)}),\qquad L_A^{f(V)}=2(L_{LL}^{f(V)}-L_{LR}^{f(V)}),
\end{align}
or equivalently,
\begin{align}
&I_{LL}^{f(S)}=\frac{I_S^{f(S)}+I_P^{f(S)}}{4},\qquad I_{LR}^{f(S)}=\frac{I_S^{f(S)}-I_P^{f(S)}}{4},\nonumber\\
&L_{LL}^{f(V)}=\frac{L_V^{f(V)}+L_A^{f(V)}}{4},\qquad L_{LR}^{f(V)}=\frac{L_V^{f(V)}-L_A^{f(V)}}{4}.
\end{align}
All the scalar and vector mediated loop integrals are functions of $(m_{\mu},m_f,m_S)$ and $(m_{\mu},m_f,m_V)$, respectively. During our computations of $(g-2)_{\mu}$, the amplitude and tensor integrals are simplified with the help of FeynCalc \cite{Mertig:1990an, Shtabovenko:2016sxi, Shtabovenko:2020gxv}. The amplitude in the SM is also checked by the FeynArts \cite{Hahn:2000kx}.

\textbf{Comments:} In Eqs. \eqref{eqn:contri:structure:CP} and \eqref{eqn:contri:structure:chiral}, we parameterize general structure of the one-loop contributions for the canonical interactions defined in Eqs. \eqref{eqn:int:mufM} and \eqref{eqn:int:fMQED}. However, they can also be applied and extended for the quasi-canonical interactions mentioned in Sec. \ref{sec:sim:quasi-cano} through proper substitutions of masses, couplings, and electric charge conservation conditions. In Sec. \ref{sec:summary:examples}, we demonstrate concrete examples on deriving the MDM results.

\subsection{Different representations of the loop functions}\label{sec:model:cont:Rep}
After integrating out the loop particles, we can obtain the results of loop functions $I_{LL(LR)}^{f(S)}$ and $L_{LL(LR)}^{f(V)}$, which can be expressed in various forms. Below, we define three distinct representations of these loop functions.

\begin{itemize}[itemindent=0em, leftmargin=10pt, listparindent=1em]

\item\textbf{Passarino-Veltman representation}

In this representation, the loop functions are expressed in terms of the Passarino-Veltman (PaVe) integrals. Then, the tensor integrals can be reduced to the scalar integrals. In this manuscript, the relevant integrals are two-point $B$ and three-point $C$ functions. In App. \ref{app:loopintegrals}, we exhibit the definitions of scalar integrals and reduction of tensor integrals.

\item\textbf{Integral representation}

In this representation, the loop functions are expressed in terms of the integrals after Feynman parameterization. It is certainly not unique, as integrands can be modified by terms such as
\begin{align}
\frac{(1-2x)[(1-x)m_S^2-x(1-x)m_{\mu}^2+xm_f^2]}{(1-x)m_S^2-x(1-x)m_{\mu}^2+xm_f^2},\quad (x^2+ax-\frac{1}{3}-\frac{a}{2})\frac{(1-x)m_S^2-x(1-x)m_{\mu}^2+xm_f^2}{(1-x)m_S^2-x(1-x)m_{\mu}^2+xm_f^2}.
\end{align}
These modifications are equivalent to the original integrands, because $\int_0^1dx(1-2x)=\int_0^1dx(x^2+ax-\frac{1}{3}-\frac{a}{2})=0$.

\item\textbf{Special function representation}

In this representation, the loop functions are expressed in terms of pre-defined functions explicitly. Although the $C_0$ function will be expressed in di-logarithm forms under normal circumstances, it can be simplified here because of the zero momentum square of the photon and two same internal propagator masses. Now, let us define the following function: \footnote{When there are singularities, the ``$i\epsilon$" prescription should be considered. We thank Chen Zhang for discussions on this point.}
\begin{align}\label{eqn:rep:f}
&f(k^2,m_0^2,m_1^2)\equiv\lim_{\epsilon\rightarrow0^+}\int_0^1dx\frac{1}{xm_0^2+(1-x)m_1^2-x(1-x)k^2-i\epsilon}\nonumber\\
=&\left\{ \begin{array}{ll}
\vspace{1ex}\frac{2}{\sqrt{\Delta}}(\mr{arctanh}\frac{\sqrt{\Delta}}{m_0^2+m_1^2-k^2}+i\pi)=\frac{1}{\sqrt{\Delta}}(\log\frac{m_0^2+m_1^2-k^2+\sqrt{\Delta}}{m_0^2+m_1^2-k^2-\sqrt{\Delta}}+2i\pi)& ,\quad\textrm{if}\quad k^2>(m_0+m_1)^2\\[2ex]
\vspace{1ex}\frac{2}{\sqrt{\Delta}}\mr{arctanh}\frac{\sqrt{\Delta}}{m_0^2+m_1^2-k^2}=\frac{1}{\sqrt{\Delta}}\log\frac{m_0^2+m_1^2-k^2+\sqrt{\Delta}}{m_0^2+m_1^2-k^2-\sqrt{\Delta}}& ,\quad\textrm{if}\quad k^2<(m_0-m_1)^2\\[2ex]
\vspace{1ex}\frac{2}{\sqrt{-\Delta}}(\arctan\frac{m_0^2-m_1^2+k^2}{\sqrt{-\Delta}}+\arctan\frac{m_1^2-m_0^2+k^2}{\sqrt{-\Delta}})=\frac{2}{\sqrt{-\Delta}}\arctan\frac{\sqrt{-\Delta}}{m_0^2+m_1^2-k^2}& ,\quad\textrm{if}\quad \makecell{(m_0-m_1)^2<k^2\\<(m_0+m_1)^2}\\[2ex]
\vspace{1ex}\frac{1}{m_0m_1} & ,\quad\textrm{if}\quad k^2=(m_0-m_1)^2\\[2ex]
\vspace{1ex}-\frac{1}{m_0m_1} & ,\quad\textrm{if}\quad k^2=(m_0+m_1)^2
\end{array} \right..
\end{align}
Here, $\Delta=k^4+m_0^4+m_1^4-2(k^2m_0^2+k^2m_1^2+m_0^2m_1^2)$. The $k^2=(m_0+m_1)^2$ is the so-called normal threshold, while the $k^2=(m_0-m_1)^2$ is the so-called pseudo-threshold \cite{Eden:1966dnq, Zwicky:2016lka}. In App. \ref{app:ftech}, we investigate the properties of $f(k^2,m_0^2,m_1^2)$ function in detail. This function agrees with that defined in Refs. \cite{Degrassi:1992ue, Albergaria:2022jne}. In Package-X \cite{Patel:2015tea, Patel:2016fam}, a similar function $\mr{DiscB}$ is defined as 
\begin{align}
\mr{DiscB}(k^2,m_0,m_1)=\frac{\Delta}{2k^2}f(k^2,m_0^2,m_1^2).
\end{align}
In App. \ref{app:functions}, we show some properties of the inverse hyperbolic and inverse trigonometric functions.

\end{itemize}

\subsection{Analytic form of the loop functions: scalar mediator case}\label{sec:model:contri:ScalarRep}
In this section, we present the explicit forms of the loop functions for scalar mediators in the PaVe representation, integral representation, and special function representation. For simplicity, we drop the $-i\epsilon$ in the integral representation, which should be remembered when encountering the branch cut. 

$\bullet$ The $I_{LL}^f(m_{\mu},m_f,m_S)$ is computed as
\begin{align}\label{eqn:con:ILLf}
&\hspace{40ex}\rotatebox{10}{$\Longrightarrow$}\quad\textbf{PaVe representation}\nonumber\\
&I_{LL}^f=\lim_{k^2\rightarrow0}m_S^2[C_{11}(m_{\mu}^2,k^2,m_{\mu}^2,m_S^2,m_f^2,m_f^2)+C_{12}(m_{\mu}^2,k^2,m_{\mu}^2,m_S^2,m_f^2,m_f^2)+C_1(m_{\mu}^2,k^2,m_{\mu}^2,m_S^2,m_f^2,m_f^2)]\nonumber\\
&=\frac{m_S^2}{16m_{\mu}^4}[(3m_S^2+m_{\mu}^2-5m_f^2)B_0(0,m_f^2,m_f^2)+2m_S^2B_0(0,m_S^2,m_S^2)+(5m_f^2-m_{\mu}^2-5m_S^2)B_0(m_{\mu}^2,m_f^2,m_S^2)\nonumber\\
	&+(3m_S^4-2m_S^2m_{\mu}^2-6m_S^2m_f^2+3m_f^4-2m_f^2m_{\mu}^2-m_{\mu}^4)C_0(m_{\mu}^2,0,m_{\mu}^2,m_S^2,m_f^2,m_f^2)+2(m_S^2-m_{\mu}^2-m_f^2)]\nonumber\\[1ex]\hline\nonumber\\[-3ex]
&=\frac{1}{2}\int_0^1dx\frac{x^2(1-x)m_S^2}{(1-x)m_S^2-x(1-x)m_{\mu}^2+xm_f^2}\hspace{30pt}\Longrightarrow\textbf{integral representation}\nonumber\\[1ex]\hline\nonumber\\[-3ex]
&=\frac{m_S^2}{4m_{\mu}^6}\Big\{[m_f^6-(3m_S^2+2m_{\mu}^2)m_f^4+(3m_S^4+m_S^2m_{\mu}^2+m_{\mu}^4)m_f^2+m_S^4m_{\mu}^2-m_S^6]f(m_{\mu}^2,m_f^2,m_S^2)\nonumber\\
	&+[-m_f^4+(2m_S^2+m_{\mu}^2)m_f^2-m_S^4]\log\frac{m_f^2}{m_S^2}+m_{\mu}^2(2m_f^2-2m_S^2-m_{\mu}^2)\Big\}.\quad\rotatebox{-10}{$\Longrightarrow$}\quad\textbf{\makecell{special function\\ representation}}
\end{align}
The first identity is valid in arbitrary dimension of spacetime, while the second identity is only valid in four dimension because we dropped the $\mc{O}(D-4)$ terms. To demonstrate each representation clearly, they are separated by the horizontal lines hereafter.

$\bullet$ The $I_{LR}^f(m_{\mu},m_f,m_S)$ is computed as
\begin{align}\label{eqn:con:ILRf}
&I_{LR}^f=\frac{m_fm_S^2}{m_{\mu}}\lim_{k^2\rightarrow0}C_1(m_{\mu}^2,k^2,m_{\mu}^2,m_S^2,m_f^2,m_f^2)\nonumber\\
&=\frac{m_fm_S^2}{4m_{\mu}^3}[B_0(m_{\mu}^2,m_f^2,m_S^2)-B_0(0,m_f^2,m_f^2)+(m_f^2-m_S^2-m_{\mu}^2)C_0(m_{\mu}^2,0,m_{\mu}^2,m_S^2,m_f^2,m_f^2)]\nonumber\\[1ex]\hline\nonumber\\[-3ex]
&=\frac{m_f}{2m_{\mu}}\int_0^1dx\frac{x^2m_S^2}{(1-x)m_S^2-x(1-x)m_{\mu}^2+xm_f^2}\nonumber\\[1ex]\hline\nonumber\\[-3ex]
&=\frac{m_fm_S^2}{4m_{\mu}^5}\Big\{[m_f^4-2m_f^2(m_S^2+m_{\mu}^2)+m_S^4+m_{\mu}^4]f(m_{\mu}^2,m_f^2,m_S^2)+(m_S^2+m_{\mu}^2-m_f^2)\log\frac{m_f^2}{m_S^2}+2m_{\mu}^2\Big\}.
\end{align}

$\bullet$ The $I_{LL}^S(m_{\mu},m_f,m_S)$ is computed as \footnote{In the first version, the variable order of $C$ functions in the $I_{LL}^S$ and $I_{LR}^S$ is $(m_{\mu}^2,m_{\mu}^2,k^2,m_S^2,m_f^2,m_S^2)$. Here, we rearrange the variable order in the form of $(m_{\mu}^2,k^2,m_{\mu}^2,m_0^2,m_1^2,m_1^2)$ uniformly. Note that the scalar integrals satisfy the permutation relation of $C_0(m_{\mu}^2,m_{\mu}^2,k^2,m_1^2,m_0^2,m_1^2)=C_0(m_{\mu}^2,k^2,m_{\mu}^2,m_0^2,m_1^2,m_1^2)=C_0(k^2,m_{\mu}^2,m_{\mu}^2,m_1^2,m_1^2,m_0^2)$, while it no longer holds for the tensor integrals.}
\begin{align}\label{eqn:con:ILLS}
&I_{LL}^S=-m_S^2\lim_{k^2\rightarrow0}[C_1(m_{\mu}^2,k^2,m_{\mu}^2,m_f^2,m_S^2,m_S^2)+C_{11}(m_{\mu}^2,k^2,m_{\mu}^2,m_f^2,m_S^2,m_S^2)+C_{12}(m_{\mu}^2,k^2,m_{\mu}^2,m_f^2,m_S^2,m_S^2)]\nonumber\\
&=\frac{m_S^2}{16m_{\mu}^4}[(5m_S^2-m_{\mu}^2-3m_f^2)B_0(0,m_S^2,m_S^2)-2m_f^2B_0(0,m_f^2,m_f^2)+(5m_f^2+m_{\mu}^2-5m_S^2)B_0(m_{\mu}^2,m_f^2,m_S^2)\nonumber\\
	&+(-3m_S^4+2m_S^2m_{\mu}^2+6m_S^2m_f^2-3m_f^4+2m_f^2m_{\mu}^2+m_{\mu}^4)C_0(m_{\mu}^2,0,m_{\mu}^2,m_f^2,m_S^2,m_S^2)+2(m_S^2+m_{\mu}^2-m_f^2)]\nonumber\\[1ex]\hline\nonumber\\[-3ex]
&=-\frac{1}{2}\int_0^1dx\frac{x(1-x)^2m_S^2}{(1-x)m_S^2-x(1-x)m_{\mu}^2+xm_f^2}\nonumber\\[1ex]\hline\nonumber\\[-3ex]
&=\frac{m_S^2}{4m_{\mu}^6}\Big\{[m_f^6-(3m_S^2+m_{\mu}^2)m_f^4+(3m_S^2-m_{\mu}^2)m_S^2m_f^2-m_S^2(m_{\mu}^2-m_S^2)^2]f(m_{\mu}^2,m_f^2,m_S^2)\nonumber\\
	&+[-m_f^4+2m_S^2m_f^2+m_S^2m_{\mu}^2-m_S^4]\log\frac{m_f^2}{m_S^2}+m_{\mu}^2(2m_f^2-2m_S^2+m_{\mu}^2)\Big\}.
\end{align}

$\bullet$ The $I_{LR}^S(m_{\mu},m_f,m_S)$ is computed as
\begin{align}\label{eqn:con:ILRS}
&I_{LR}^S=\frac{m_fm_S^2}{2m_{\mu}}\lim_{k^2\rightarrow0}[C_0(m_{\mu}^2,k^2,m_{\mu}^2,m_f^2,m_S^2,m_S^2)+2C_1(m_{\mu}^2,k^2,m_{\mu}^2,m_f^2,m_S^2,m_S^2)]\nonumber\\
&=\frac{m_fm_S^2}{4m_{\mu}^3}[B_0(m_{\mu}^2,m_f^2,m_S^2)-B_0(0,m_S^2,m_S^2)+(m_S^2+m_{\mu}^2-m_f^2)C_0(m_{\mu}^2,0,m_{\mu}^2,m_f^2,m_S^2,m_S^2)]\nonumber\\[1ex]\hline\nonumber\\[-3ex]
&=-\frac{m_f}{2m_{\mu}}\int_0^1dx\frac{x(1-x)m_S^2}{(1-x)m_S^2-x(1-x)m_{\mu}^2+xm_f^2}\nonumber\\[1ex]\hline\nonumber\\[-3ex]
&=\frac{m_fm_S^2}{4m_{\mu}^5}\Big\{[m_f^4-m_f^2(2m_S^2+m_{\mu}^2)+m_S^4-m_S^2m_{\mu}^2]f(m_{\mu}^2,m_f^2,m_S^2)+(m_S^2-m_f^2)\log\frac{m_f^2}{m_S^2}+2m_{\mu}^2\Big\}.
\end{align}

\textbf{Comments:} For the four loop integrals $I_{LL(LR)}^{f(S)}$, our results in integral representation agree with those in Refs. \cite{Leveille:1977rc, Jegerlehner:2009ry, Jegerlehner:2017gek}. In fact, the $I_{LR}^f$ can be correlated with $I_{LL}^f$ through the following relation:
\begin{small}
\begin{align}
&I_{LR}^f(m_{\mu},m_f,m_S)=\frac{m_f}{4m_{\mu}}\cdot\nonumber\\
&\frac{4m_{\mu}^2[(m_f^2-m_S^2-m_{\mu}^2)^2-2m_S^2m_{\mu}^2]I_{LL}^f(m_{\mu},m_f,m_S)+m_S^2[m_f^4-2m_f^2(2m_S^2+m_{\mu}^2)+3m_S^4+2m_S^2m_{\mu}^2+m_{\mu}^4+2m_S^4\log\frac{m_f^2}{m_S^2}]}{m_f^6-m_f^4(3m_S^2+2m_{\mu}^2)+m_f^2(3m_S^4+m_S^2m_{\mu}^2+m_{\mu}^4)+m_S^4m_{\mu}^2-m_S^6}.\nonumber\\
\end{align}
\end{small}
Moreover, the $I_{LL}^S$ and $I_{LR}^S$ can be obtained via the following relations:
\begin{align}
&I_{LL}^S(m_{\mu},m_f,m_S)=-\frac{m_S^2}{m_f^2}\cdot I_{LL}^f(m_{\mu},m_S,m_f),\nonumber\\
&I_{LR}^S(m_{\mu},m_f,m_S)=-\frac{m_S^2}{m_fm_{\mu}}\cdot I_{LL}^f(m_{\mu},m_S,m_f)-\frac{m_f}{m_{\mu}}\cdot I_{LL}^f(m_{\mu},m_f,m_S).
\end{align}

\subsection{Analytic form of the loop functions: vector mediator case}\label{sec:model:contri:VectorRep}
In this section, we present the explicit forms of the loop functions for vector mediators in the PaVe representation, integral representation, and special function representation.

$\bullet$ The $L_{LL}^f(m_{\mu},m_f,m_V)$ is computed as
\begin{align}\label{eqn:con:LLLf}
&L_{LL}^f=\lim_{k^2\rightarrow0}\Big\{m_V^2[(D-2)(C_{11}+C_{12})+(D+2)C_1+2C_0]\nonumber\\
	&-[(D+2)C_{001}+2C_{00}+m_{\mu}^2(C_{111}+3C_{112}+C_{11}+C_{12})-k^2(C_{112}+C_{12})-m_f^2(C_{11}+C_{12})]\Big\}\nonumber\\
&=-\frac{1}{16m_{\mu}^4}\Big\{[5m_f^4+m_f^2(7m_V^2-4m_{\mu}^2)-m_{\mu}^4-6m_V^4+11m_{\mu}^2m_V^2]B_0(0,m_f^2,m_f^2)-2m_V^2(m_f^2+m_{\mu}^2+2m_V^2)B_0(0,m_V^2,m_V^2)\nonumber\\
	&+[m_{\mu}^4+m_{\mu}^2(4m_f^2-9m_V^2)-5(m_f^4+m_f^2m_V^2-2m_V^4)]B_0(m_{\mu}^2,m_f^2,m_V^2)+2(m_f^2+m_{\mu}^2-m_V^2)(m_f^2+m_{\mu}^2+2m_V^2)+\nonumber\\
	&[-3m_f^6+7m_f^4m_{\mu}^2+m_f^2(3m_V^2-5m_{\mu}^2)(3m_V^2+m_{\mu}^2)+m_{\mu}^6-6m_V^6+17m_{\mu}^2m_V^4-12m_{\mu}^4m_V^2]C_0(m_{\mu}^2,0,m_{\mu}^2,m_V^2,m_f^2,m_f^2)\Big\}\nonumber\\[1ex]\hline\nonumber\\[-3ex]
&=-\frac{1}{2}\int_0^1dx\frac{x[x(1+x)m_f^2-x(1-x)m_{\mu}^2+2(1-x)(2-x)m_V^2]}{(1-x)m_V^2-x(1-x)m_{\mu}^2+xm_f^2}\nonumber\\[1ex]\hline\nonumber\\[-3ex]
&=\frac{1}{4m_{\mu}^6}\Big\{[m_f^8-m_f^6(m_V^2+3m_{\mu}^2)+m_f^4(3m_{\mu}^4+2m_{\mu}^2m_V^2-3m_V^4)+m_f^2(-m_{\mu}^6-m_{\mu}^4m_V^2-4m_{\mu}^2m_V^4+5m_V^6)\nonumber\\
	&-3m_{\mu}^4m_V^4+5m_{\mu}^2m_V^6-2m_V^8]f(m_{\mu}^2,m_f^2,m_V^2)+m_{\mu}^2[2m_f^4+m_f^2(2m_V^2-3m_{\mu}^2)-(m_{\mu}^2-2m_V^2)^2]\nonumber\\
	&+[-m_f^6+2m_f^4m_{\mu}^2+m_f^2(-m_{\mu}^4-2m_{\mu}^2m_V^2+3m_V^4)+3m_{\mu}^2m_V^4-2m_V^6]\log\frac{m_f^2}{m_V^2}\Big\}.
\end{align}
In the first identity above, the variables of $C$ functions are $(m_{\mu}^2,k^2,m_{\mu}^2,m_V^2,m_f^2,m_f^2)$.

$\bullet$ The $L_{LR}^f(m_{\mu},m_f,m_V)$ is computed as
\begin{align}\label{eqn:con:LLRf}
&L_{LR}^f=\frac{m_f}{m_{\mu}}\cdot\lim_{k^2\rightarrow0}[-m_V^2(D\cdot C_1+2C_0)+(D+2)C_{001}+2C_{00}+m_{\mu}^2(C_{111}+3C_{112})-k^2(C_{112}+C_{12})]\nonumber\\
&=\frac{m_f}{m_{\mu}}\cdot\frac{1}{8m_{\mu}^2}\Big\{(3m_f^2-3m_{\mu}^2+5m_V^2)B_0(0,m_f^2,m_f^2)-2m_V^2B_0(0,m_V^2,m_V^2)-3(m_f^2-m_{\mu}^2+m_V^2)B_0(m_{\mu}^2,m_f^2,m_V^2)\nonumber\\
	&+2(m_f^2+m_{\mu}^2-m_V^2)-[m_f^4-2m_{\mu}^2(m_f^2-4m_V^2)+4m_f^2m_V^2+m_{\mu}^4-5m_V^4]C_0(m_{\mu}^2,0,m_{\mu}^2,m_V^2,m_f^2,m_f^2)\Big\}\nonumber\\[1ex]\hline\nonumber\\[-3ex]
&=\frac{m_f}{2m_{\mu}}\int_0^1dx\frac{x[xm_f^2-x(1-2x)m_{\mu}^2+4(1-x)m_V^2]}{(1-x)m_V^2-x(1-x)m_{\mu}^2+xm_f^2}\nonumber\\[1ex]\hline\nonumber\\[-3ex]
&=\frac{m_f}{4m_{\mu}^5}\Big\{[-m_f^6+3m_f^4m_{\mu}^2+3m_f^2(-m_{\mu}^4+m_V^4)+m_{\mu}^6+3m_{\mu}^2m_V^4-2m_V^6]f(m_{\mu}^2,m_f^2,m_V^2)\nonumber\\
	&+2m_{\mu}^2(-m_f^2+2m_{\mu}^2-2m_V^2)+[m_f^4+m_f^2(-2m_{\mu}^2+m_V^2)+m_{\mu}^4+m_{\mu}^2m_V^2-2m_V^4]\log\frac{m_f^2}{m_V^2}\Big\}.
\end{align}
In the first identity above, the variables of $C$ functions are $(m_{\mu}^2,k^2,m_{\mu}^2,m_V^2,m_f^2,m_f^2)$.

$\bullet$ The $L_{LL}^V(m_{\mu},m_f,m_V)$ is computed as
\begin{align}\label{eqn:con:LLLV}
&L_{LL}^V=-\frac{1}{2m_V^2}\lim_{k^2\rightarrow0}\Big\{m_f^2(2m_V^2-k^2)C_0+[2m_V^2\big((D-6)m_V^2+m_{\mu}^2+3m_f^2\big)-k^2(m_{\mu}^2+3m_f^2-2m_V^2)]C_1\nonumber\\
	&+[2m_V^2\big((D-2)m_V^2+m_{\mu}^2+m_f^2\big)-k^2(m_{\mu}^2+m_f^2)](C_{11}+C_{12})\Big\}\nonumber\\
&=\frac{1}{16m_{\mu}^4}\Big\{-2m_f^2(m_{\mu}^2+m_f^2+2m_V^2)B_0(0,m_f^2,m_f^2)-[m_{\mu}^4+m_{\mu}^2(13m_V^2-4m_f^2)+3m_f^4+m_f^2m_V^2-10m_V^4]B_0(0,m_V^2,m_V^2)\nonumber\\
	&+[m_{\mu}^4+m_{\mu}^2(13m_V^2-2m_f^2)+5(m_f^4+m_f^2m_V^2-2m_V^4)]B_0(m_{\mu}^2,m_f^2,m_V^2)+2(m_{\mu}^4+3m_V^2m_{\mu}^2-m_f^4-m_f^2m_V^2+2m_V^4)\nonumber\\
	&+[m_{\mu}^6-m_{\mu}^4(5m_f^2+12m_V^2)+m_{\mu}^2(7m_f^4-12m_f^2m_V^2+17m_V^4)-3(m_f^6-3m_f^2m_V^4+2m_V^6)]C_0(m_{\mu}^2,0,m_{\mu}^2,m_f^2,m_V^2,m_V^2)\Big\}\nonumber\\[1ex]\hline\nonumber\\[-3ex]
&=\frac{1}{2}\int_0^1dx\frac{(1-x)[x(1+x)m_f^2-x(1-x)m_{\mu}^2+2(1-x)(2-x)m_V^2]}{(1-x)m_V^2-x(1-x)m_{\mu}^2+xm_f^2}\nonumber\\[1ex]\hline\nonumber\\[-3ex]
&=\frac{1}{4m_{\mu}^6}\Big\{[m_f^8-m_f^6(m_V^2+2m_{\mu}^2)+m_f^4(m_{\mu}^4+2m_{\mu}^2m_V^2-3m_V^4)+m_f^2m_V^4(5m_V^2-7m_{\mu}^2)\nonumber\\
	&+m_V^2(3m_{\mu}^2-2m_V^2)(m_{\mu}^2-m_V^2)^2]f(m_{\mu}^2,m_f^2,m_V^2)+m_{\mu}^2[2m_f^4+m_f^2(2m_V^2-m_{\mu}^2)+m_{\mu}^4+8m_{\mu}^2m_V^2-4m_V^4]\nonumber\\
	&+[-m_f^6+m_f^4m_{\mu}^2+3m_f^2m_V^2(-m_{\mu}^2+m_V^2)-3m_{\mu}^4m_V^2+5m_{\mu}^2m_V^4-2m_V^6]\log\frac{m_f^2}{m_V^2}\Big\}.
\end{align}
In the first identity above, the variables of $C$ functions are $(m_{\mu}^2,k^2,m_{\mu}^2,m_f^2,m_V^2,m_V^2)$.

$\bullet$ The $L_{LR}^V(m_{\mu},m_f,m_V)$ is computed as
\begin{align}\label{eqn:con:LLRV}
&L_{LR}^V=\frac{m_f}{4m_{\mu}m_V^2}\cdot\lim_{k^2\rightarrow0}\Big\{[2m_V^2\big((4-D)m_V^2+m_{\mu}^2+m_f^2\big)-k^2(m_{\mu}^2+m_f^2)]C_0+4m_{\mu}^2(2m_V^2-k^2)(C_{11}+C_{12})\nonumber\\
	&+[4m_V^2(-Dm_V^2+3m_{\mu}^2+m_f^2)-2k^2(3m_{\mu}^2+m_f^2-2m_V^2)]C_1\Big\}\nonumber\\
&=\frac{m_f}{m_{\mu}}\cdot\frac{1}{8m_{\mu}^2}\Big\{2m_f^2B_0(0,m_f^2,m_f^2)+(3m_V^2+m_f^2-m_{\mu}^2)B_0(0,m_V^2,m_V^2)+(m_{\mu}^2-3m_f^2-3m_V^2)B_0(m_{\mu}^2,m_f^2,m_V^2)\nonumber\\
	&+[m_{\mu}^4+2m_{\mu}^2(4m_V^2-m_f^2)+m_f^4+4m_f^2m_V^2-5m_V^4]C_0(m_{\mu}^2,0,m_{\mu}^2,m_f^2,m_V^2,m_V^2)+2m_f^2-2m_{\mu}^2-2m_V^2\Big\}\nonumber\\[1ex]\hline\nonumber\\[-3ex]
&=-\frac{m_f}{2m_{\mu}}\int_0^1dx\frac{(1-x)[xm_f^2+x(2x-1)m_{\mu}^2+4(1-x)m_V^2]}{(1-x)m_V^2-x(1-x)m_{\mu}^2+xm_f^2}\nonumber\\
&=-\frac{m_f}{2m_{\mu}}\int_0^1dx\frac{x^2m_f^2+(3-2x)(1-x)m_V^2}{(1-x)m_V^2-x(1-x)m_{\mu}^2+xm_f^2}\nonumber\\[1ex]\hline\nonumber\\[-3ex]
&=\frac{m_f}{4m_{\mu}^5}\Big\{[-m_f^6+2m_f^4m_{\mu}^2+m_f^2(-m_{\mu}^4-m_{\mu}^2m_V^2+3m_V^4)-3m_{\mu}^4m_V^2+5m_{\mu}^2m_V^4-2m_V^6]f(m_{\mu}^2,m_f^2,m_V^2)\nonumber\\
	&-2m_{\mu}^2(m_f^2+2m_V^2)+[m_f^4+m_f^2(-m_{\mu}^2+m_V^2)+3m_{\mu}^2m_V^2-2m_V^4]\log\frac{m_f^2}{m_V^2}\Big\}.
\end{align}
In the first identity above, the variables of $C$ functions are $(m_{\mu}^2,k^2,m_{\mu}^2,m_f^2,m_V^2,m_V^2)$.\\

\textbf{Comments:} In this work, we perform the calculations in the unitary gauge. However, the contributions are gauge independent, as confirmed by studies in other gauges, such as Feynman gauge \cite{Fujikawa:1972fe, Athron:2021iuf}. For the four loop integrals $L_{LL(LR)}^{f(V)}$, our results in the integral representation agree with those in Refs. \cite{Leveille:1977rc}. Our results of $L_{LL(LR)}^{f}$ in the integral representation also agree with those in Refs. \cite{Jegerlehner:2009ry, Jegerlehner:2017gek}, while our results of $L_{LL(LR)}^{V}$ in the integral representation correspond to $Q_V=2x^2(1+x-2\epsilon)+\lambda^2(1-\epsilon)^2Q_S$ and $Q_A=2x^2(1+x+2\epsilon)+\lambda^2(1+\epsilon)^2Q_P$ below Eq. (7.14) in Ref. \cite{Jegerlehner:2017gek}, namely, the $Q_V$ and $Q_A$ below Eq. (264) in \textit{Physics Reports} paper \cite{Jegerlehner:2009ry}. This sign difference in front of the $\lambda^2\equiv m_{\mu}^2/m_V^2$ usually does not matter because the muon mass is typically much less than the charged gauge boson mass. However, this difference can be sizable in certain new physics models where the charged vector boson mass is light.

\subsection{Singularities of the loop functions}\label{sec:contri:singularities}
In Sec. \ref{sec:model:contri:ScalarRep} and Sec. \ref{sec:model:contri:VectorRep}, we have presented the analytic forms of the loop functions for scalar and vector mediator cases in three representations. In the PaVe representation, the muon MDM contributions are expressed in terms of two-point and three-point Passarino-Veltman scalar integrals. The relevant forms are $B_0(m_{\mu}^2,m_0^2,m_1^2)$ and $C_0(m_{\mu}^2,0,m_{\mu}^2,m_0^2,m_1^2,m_1^2)$. Their Landau singularities are $m_{\mu}^2=(m_0+m_1)^2$, $m_0=0$, and $m_1=0$, which are investigated detailedly in App. \ref{app:singularity}. The singularities can also be verified in the integral representation. According to the muon MDM contributions, all the loop functions contain the denominator of $xm_0^2-x(1-x)m_{\mu}^2+(1-x)m_1^2$. For $m_0=0$, the denominator is gone at $x=1$. For $m_1=0$, the denominator is gone at $x=0$. It is the so called end-point singularity. When analyzing zero points of the quadratic polynomial, we find that solutions for $x$ exist in the interval $(0,1)$ when $m_{\mu}^2>(m_0+m_1)^2$. As to the special function representation, we can identify the singularities of pre-defined function $f(k^2,m_0^2,m_1^2)$ explicitly, which are discussed in detail in App. \ref{app:ftech}.

\section{Expansion of the loop functions: scalar mediator case}\label{sec:expansionS}

In Sec. \ref{sec:model:contri:ScalarRep}, we have exhibited the analytic forms of loop functions for the scalar mediator case, including the PaVe representation, integral representation, and special function representation. Although the analytic formulae are valid for all the mass scales, they are quite lengthy and inconvenient to adopt when evaluating the contributions in new physics models. Based on the analytic forms, this section is dedicated to the expansion of loop functions. These expansion results are not only simple in practical calculations, but also enough to capture and understand the effects of new physics.

There are three mass scales $m_{\mu}$, $m_f$, and $m_S$ in the loop functions, while one of the mass scale can be very light, very heavy, or degenerate with the other scales in specific models. First, there can be a large hierarchy between one mass scale and the other two mass scales, for which we consider six hierarchical scenarios: 
\begin{align}
&m_\mu\ll m_f,m_S;\qquad m_S\ll m_f,m_\mu;\qquad m_f\ll m_\mu,m_S; \nonumber\\
&m_f,m_S\ll m_\mu;\qquad m_f,m_\mu\ll m_S;\qquad m_\mu,m_S\ll m_f. 
\end{align}
Second, we also study the four degenerate scenarios:
\begin{align}
m_f=m_\mu,\qquad m_\mu=m_S,\qquad m_f=m_S,\qquad m_f=m_\mu=m_S.
\end{align}

The scenario of $m_\mu\ll m_f,m_S$ is the most common situation, because current experimental constraints push the new physics scale to almost TeV regardless of the spin. The scenario of $m_S\ll m_f,m_\mu$ is available to axion contributions. The scenario of $m_f\ll m_\mu,m_S$ is applicable to neutrino contributions. The scenario of $m_f,m_S\ll m_\mu$ is less to encounter. The scenario of $m_f,m_\mu\ll m_S$ is suitable to the contributions of heavy scalar. The scenario of $m_\mu,m_S\ll m_f$ is suitable to the contributions of heavy fermion. For the degenerate scenario of $m_f=m_\mu$, it means that the internal fermion line is just muon. For the degenerate scenarios of $m_\mu=m_S$, $m_f=m_S$, and $m_f=m_\mu=m_S$, they are also less to encounter. Of course, not all of the scenarios widely appear in new physics models. It is not the priority of this manuscript to check whether each scenario is physical accessible and corresponds to a well realized ultraviolet (UV) complete model. However, we keep the whole scenarios for completeness.
\subsection{Hierarchical mass expansion for the $I_{LL}^f$ and $I_{LR}^f$}\label{sec:expansion:If}
\subsubsection{Scenario of $m_\mu\ll m_f,m_S$}\label{sec:expansion:Ifhie:mmullmfmS}

Up to $m_{\mu}^4$, the integral $I_{LL}^f$ can be expanded as
\begin{align}\label{eqn:expansion:IfLL:mmullmfmS}
&I_{LL}^f=\frac{m_S^2}{12(m_S^2-m_f^2)^4}(2m_S^6+3m_S^4m_f^2-6m_S^2m_f^4+m_f^6+6m_S^4m_f^2\log\frac{m_f^2}{m_S^2})\nonumber\\
&+\frac{m_\mu^2m_S^2}{24(m_S^2-m_f^2)^6}\Big[3m_S^8+44m_S^6m_f^2-36m_S^4m_f^4-12m_S^2m_f^6+m_f^8+12m_S^4m_f^2(3m_f^2+2m_S^2)\log\frac{m_f^2}{m_S^2}\Big]\nonumber\\
&+\frac{m_\mu^4m_S^2}{40(m_S^2-m_f^2)^8}\big[4m_S^{10}+155m_S^8m_f^2+80m_S^6m_f^4-220m_S^4m_f^6-20m_S^2m_f^8+m_f^{10}\nonumber\\
	&+60m_S^4m_f^2(m_S^4+4m_S^2m_f^2+2m_f^4)\log\frac{m_f^2}{m_S^2}\Big]+\mathcal{O}(m_\mu^6).
\end{align}
Up to $m_{\mu}^3$, the integral $I_{LR}^f$ can be expanded as
\begin{align}\label{eqn:expansion:IfLR:mmullmfmS}
&I_{LR}^f=\frac{m_f}{m_\mu}\Big\{\frac{m_S^2}{4(m_S^2-m_f^2)^3}(-3m_S^4+4m_S^2m_f^2-m_f^4-2m_S^4\log\frac{m_f^2}{m_S^2})\nonumber\\
&+\frac{m_\mu^2m_S^2}{12(m_S^2-m_f^2)^5}\Big[-17m_S^6+9m_S^4m_f^2+9m_S^2m_f^4-m_f^6-6m_S^4(3m_f^2+m_S^2)\log\frac{m_f^2}{m_S^2}\Big]\nonumber\\
&+\frac{m_\mu^4m_S^2}{24(m_S^2-m_f^2)^7}\big[-43m_S^8-80m_S^6m_f^2+108m_S^4m_f^4+16m_S^2m_f^6-m_f^8\nonumber\\
	&-12m_S^4(m_S^4+8m_S^2m_f^2+6m_f^4)\log\frac{m_f^2}{m_S^2}\Big]\Big\}+\mathcal{O}(m_\mu^5).
\end{align}
This coincides with those results in the supersymmetry models \cite{Moroi:1995yh, Martin:2001st}.\footnote{The $\frac{1}{2}$ factor for $F_2^C(x)$ is dropped in Stockinger's paper \cite{Stockinger:2006zn}.}

Especially, we have the following approximations:
\begin{align}\label{eqn:expansion:IfLL:mmullmfmS:esp}
I_{LL}^f\approx\left\{\begin{array}{ll}
\frac{1}{6}(1+\frac{11m_f^2}{2m_S^2}-\frac{6m_f^2}{m_S^2}\log\frac{m_S}{m_f}),& \mathrm{for}~m_\mu\ll m_f\ll m_S\vspace{1ex}\\
\frac{m_S^2}{12m_f^2}(1-\frac{2m_S^2}{m_f^2}),& \mathrm{for}~m_\mu\ll m_S\ll m_f
\end{array}\right.
\end{align}
and
\begin{align}\label{eqn:expansion:IfLR:mmullmfmS:esp}
I_{LR}^f\approx\left\{\begin{array}{ll}
\frac{m_f}{m_\mu}\Big[\log\frac{m_S}{m_f}-\frac{3}{4}+\frac{m_f^2}{m_S^2}(3\log\frac{m_S}{m_f}-\frac{5}{4})\Big],& \mathrm{for}~m_\mu\ll m_f\ll m_S\vspace{1ex}\\
\frac{m_S^2}{4m_fm_\mu}(1-\frac{m_S^2}{m_f^2}),& \mathrm{for}~m_\mu\ll m_S\ll m_f
\end{array}\right. .
\end{align}

For the scenario of $m_\mu\ll m_f=m_S$, please refer to Eqs. \eqref{eqn:expansion:IfLL:mfmS:esp} and \eqref{eqn:expansion:IfLR:mfmS:esp}.
\subsubsection{Scenario of $m_S\ll m_f,m_\mu$}\label{sec:expansion:Ifhie:mSllmfmmu}

Up to $m_S^6$, the integrals $I_{LL}^f$ and $I_{LR}^f$ can be expanded as \footnote{If $m_\mu>m_f+m_S$, there can be imaginary part for the contributions to $a_{\mu}$ because of the threshold effects. Technically speaking, it is caused by the branch cut. When the loop particles are light, the EFT framework breaks down. It is not proper to integrate out the loop particles and induce the operator $\bar{\mu}\sigma^{\mu\nu}\mu F_{\mu\nu}$. The physical understanding and observable effect of such imaginary part are still under investigation.}
\begin{align}\label{eqn:expansion:IfLL:mSllmfmmu}
&I_{LL}^f=\left\{\begin{array}{ll}
\frac{m_S^2}{4m_\mu^6}(1+\frac{2m_S^2}{m_\mu^2-m_f^2})\big[2m_f^2m_\mu^2-m_\mu^4+2m_f^2(m_f^2-m_\mu^2)\log\frac{m_f^2-m_\mu^2}{m_f^2}\big]\vspace{1ex}\\
+\frac{m_S^6}{4m_\mu^6(m_\mu^2-m_f^2)^3}\big[m_f^2m_\mu^2(5m_\mu^2-2m_f^2)-2m_\mu^6\log\frac{m_f^2}{m_S^2}\vspace{1ex}\\
+(-2m_f^6+6m_f^4m_\mu^2-6m_f^2m_\mu^4-2m_\mu^6)\log\frac{m_f^2-m_\mu^2}{m_f^2}\big]+\mathcal{O}(m_S^8),& \mathrm{for}~m_\mu<m_f\vspace{2ex}\\
\frac{m_S^2}{4m_\mu^6}(1+\frac{2m_S^2}{m_\mu^2-m_f^2})\big[2m_f^2m_\mu^2-m_\mu^4+2m_f^2(m_f^2-m_\mu^2)(\log\frac{m_\mu^2-m_f^2}{m_f^2}-i\pi)\big]\vspace{1ex}\\
+\frac{m_S^6}{4m_\mu^6(m_\mu^2-m_f^2)^3}\big[m_f^2m_\mu^2(5m_\mu^2-2m_f^2)-2m_\mu^6\log\frac{m_f^2}{m_S^2}\vspace{1ex}\\
+(-2m_f^6+6m_f^4m_\mu^2-6m_f^2m_\mu^4-2m_\mu^6)(\log\frac{m_\mu^2-m_f^2}{m_f^2}-i\pi)\big]+\mathcal{O}(m_S^8),& \mathrm{for}~m_\mu>m_f
\end{array}\right.
\end{align}
and
\begin{align}\label{eqn:expansion:IfLR:mSllmfmmu}
&I_{LR}^f=\left\{\begin{array}{ll}
\frac{m_f}{m_\mu}\Big\{\frac{m_S^2}{2m_\mu^4}(1+\frac{m_S^2}{m_\mu^2-m_f^2})\big[m_\mu^2+(m_f^2-m_\mu^2)\log\frac{m_f^2-m_\mu^2}{m_f^2}\big]\vspace{1ex}\\
+\frac{m_S^6}{4(m_\mu^2-m_f^2)^3}(3-4\log\frac{m_f^2-m_\mu^2}{m_f^2}-2\log\frac{m_f^2}{m_S^2})\Big\}+\mathcal{O}(m_S^8),& \hspace{5ex}\mathrm{for}~m_\mu<m_f\vspace{2ex}\\
\frac{m_f}{m_\mu}\Big\{\frac{m_S^2}{2m_\mu^4}(1+\frac{m_S^2}{m_\mu^2-m_f^2})\big[m_\mu^2+(m_f^2-m_\mu^2)(\log\frac{m_\mu^2-m_f^2}{m_f^2}-i\pi)\big]\vspace{1ex}\\
+\frac{m_S^6}{4(m_\mu^2-m_f^2)^3}(3-4\log\frac{m_\mu^2-m_f^2}{m_f^2}-2\log\frac{m_f^2}{m_S^2}+4i\pi))\Big\}+\mathcal{O}(m_S^8),&\hspace{5ex} \mathrm{for}~m_\mu>m_f\\
\end{array}\right..
\end{align}
Especially, we have the following approximations:
\begin{align}\label{eqn:expansion:IfLL:mSllmfmmu:esp}
I_{LL}^f\approx\left\{\begin{array}{ll}
-\frac{m_S^2}{4m_\mu^2}\big[1-\frac{2m_f^2}{m_\mu^2}(1-\log\frac{m_\mu^2}{m_f^2}+i\pi)\big],& \mathrm{for}~m_S\ll m_f\ll m_\mu\vspace{1ex}\\
\frac{m_S^2}{12m_f^2}(1+\frac{m_\mu^2}{2m_f^2}),& \mathrm{for}~m_S\ll m_\mu\ll m_f
\end{array}\right.
\end{align}
and
\begin{align}\label{eqn:expansion:IfLR:mSllmfmmu:esp}
I_{LR}^f\approx\left\{\begin{array}{ll}
\frac{m_S^2m_f}{2m_\mu^3}\big[1-\log\frac{m_\mu^2}{m_f^2}+i\pi+\frac{m_f^2}{m_\mu^2}(1+\log\frac{m_\mu^2}{m_f^2}-i\pi)\big],& \mathrm{for}~m_S\ll m_f\ll m_\mu\vspace{1ex}\\
\frac{m_S^2}{4m_fm_\mu}(1+\frac{m_\mu^2}{3m_f^2}),& \mathrm{for}~m_S\ll m_\mu\ll m_f
\end{array}\right. .
\end{align}

For the scenario of $m_S\ll m_f=m_\mu$, please refer to Eqs. \eqref{eqn:expansion:IfLL:mfmmu:esp} and \eqref{eqn:expansion:IfLR:mfmmu:esp}.
\subsubsection{Scenario of $m_f\ll m_\mu,m_S$}\label{sec:expansion:Ifhie:mfllmmumS}
Up to $m_f^4$, the integral $I_{LL}^f$ can be expanded as
\begin{align}\label{eqn:expansion:IfLL:mfllmmumS}
&I_{LL}^f=\left\{\begin{array}{ll}
-\frac{m_S^2}{4m_{\mu}^6}\big[m_{\mu}^4+2m_S^2m_{\mu}^2+2m_S^4\log\frac{m_S^2-m_{\mu}^2}{m_S^2}\big]+\frac{m_f^2m_S^2}{2m_{\mu}^6(m_S^2-m_\mu^2)^2}\big[m_{\mu}^2(m_\mu^4-2m_S^2m_{\mu}^2+2m_S^4)\vspace{1ex}\\
+(2m_S^6-3m_S^4m_{\mu}^2-m_\mu^6)\log\frac{m_S^2-m_{\mu}^2}{m_S^2}-m_\mu^6\log\frac{m_S^2}{m_f^2}\big]\vspace{1ex}\\
+\frac{m_f^4m_S^2}{4m_{\mu}^6(m_S^2-m_\mu^2)^4}\big[m_\mu^2(2m_\mu^6+7m_S^4m_{\mu}^2-2m_S^6)+2m_\mu^6(m_\mu^2-4m_S^2)\log\frac{m_S^2}{m_f^2}\vspace{1ex}\\
+(2m_\mu^8-8m_S^2m_\mu^6-12m_S^4m_{\mu}^4+8m_S^6m_{\mu}^2-2m_S^8)\log\frac{m_S^2-m_{\mu}^2}{m_S^2}\big]
+\mathcal{O}(m_f^6),& \hspace{-13ex}\mathrm{for}~m_\mu<m_S\vspace{3ex}\\
-\frac{m_S^2}{4m_{\mu}^6}\big[m_{\mu}^4+2m_S^2m_{\mu}^2+2m_S^4(\log\frac{m_{\mu}^2-m_S^2}{m_S^2}-i\pi)\big]+\frac{m_f^2m_S^2}{2m_{\mu}^6(m_S^2-m_\mu^2)^2}\big[m_{\mu}^2(m_\mu^4-2m_S^2m_{\mu}^2+2m_S^4)\vspace{1ex}\\
+(2m_S^6-3m_S^4m_{\mu}^2-m_\mu^6)(\log\frac{m_{\mu}^2-m_S^2}{m_S^2}-i\pi)-m_\mu^6\log\frac{m_S^2}{m_f^2}\big]\vspace{1ex}\\
+\frac{m_f^4m_S^2}{4m_{\mu}^6(m_S^2-m_\mu^2)^4}\big[m_\mu^2(2m_\mu^6+7m_S^4m_{\mu}^2-2m_S^6)+2m_\mu^6(m_\mu^2-4m_S^2)\log\frac{m_S^2}{m_f^2}\vspace{1ex}\\
+(2m_\mu^8-8m_S^2m_\mu^6-12m_S^4m_{\mu}^4+8m_S^6m_{\mu}^2-2m_S^8)(\log\frac{m_{\mu}^2-m_S^2}{m_S^2}-i\pi)\big]+\mathcal{O}(m_f^6),& \hspace{-13ex}\mathrm{for}~m_\mu>m_S
\end{array}\right..
\end{align}
Up to $m_f^5$, the integral $I_{LR}^f$ can be expanded as
\begin{align}\label{eqn:expansion:IfLR:mfllmmumS}
&I_{LR}^f=\left\{\begin{array}{ll}
\frac{m_f}{m_\mu}\Big\{
\frac{m_S^2}{2m_{\mu}^4(m_S^2-m_{\mu}^2)}\big[m_{\mu}^2(m_S^2-m_{\mu}^2)+(m_S^4+m_\mu^4)\log\frac{m_S^2-m_{\mu}^2}{m_S^2}+m_{\mu}^4\log\frac{m_S^2}{m_f^2}\big]\vspace{1ex}\\
+\frac{m_f^2m_S^2}{2m_{\mu}^4(m_S^2-m_\mu^2)^3}\big[-m_{\mu}^2(m_S^4+m_{\mu}^4)+(-m_{\mu}^6+3m_S^2m_{\mu}^4+3m_S^4m_{\mu}^2-m_S^6)\log\frac{m_S^2-m_{\mu}^2}{m_S^2}\vspace{1ex}\\
+m_\mu^4(3m_S^2-m_\mu^2)\log\frac{m_S^2}{m_f^2}\big]+\frac{m_f^4m_S^2}{4(m_S^2-m_\mu^2)^5}\big[m_{\mu}^4-8m_S^2m_{\mu}^2-7m_S^4+24m_S^4\log\frac{m_S^2-m_{\mu}^2}{m_S^2}\vspace{1ex}\\
+12m_S^4\log\frac{m_S^2}{m_f^2}\big]\Big\}+\mathcal{O}(m_f^7),& \hspace{-15ex}\mathrm{for}~m_\mu<m_S\vspace{2ex}\\
\frac{m_f}{m_\mu}\Big\{
\frac{m_S^2}{2m_{\mu}^4(m_S^2-m_{\mu}^2)}\big[m_{\mu}^2(m_S^2-m_{\mu}^2)+(m_S^4+m_\mu^4)(\log\frac{m_{\mu}^2-m_S^2}{m_S^2}-i\pi)+m_{\mu}^4\log\frac{m_S^2}{m_f^2}\big]\vspace{1ex}\\
+\frac{m_f^2m_S^2}{2m_{\mu}^4(m_S^2-m_\mu^2)^3}\big[-m_{\mu}^2(m_S^4+m_{\mu}^4)+(-m_{\mu}^6+3m_S^2m_{\mu}^4+3m_S^4m_{\mu}^2-m_S^6)(\log\frac{m_{\mu}^2-m_S^2}{m_S^2}-i\pi)\vspace{1ex}\\
+m_\mu^4(3m_S^2-m_\mu^2)\log\frac{m_S^2}{m_f^2}\big]+\frac{m_f^4m_S^2}{4(m_S^2-m_\mu^2)^5}\big[m_{\mu}^4-8m_S^2m_{\mu}^2-7m_S^4+24m_S^4(\log\frac{m_{\mu}^2-m_S^2}{m_S^2}-i\pi)\vspace{1ex}\\
+12m_S^4\log\frac{m_S^2}{m_f^2}\big]\Big\}+\mathcal{O}(m_f^7),& \hspace{-15ex}\mathrm{for}~m_\mu>m_S
\end{array}\right..
\end{align}
Especially, we have the following approximations:
\begin{align}\label{eqn:expansion:IfLL:mfllmmumS:esp}
I_{LL}^f\approx\left\{\begin{array}{ll}
\frac{1}{6}(1+\frac{3m_{\mu}^2}{4m_S^2}),& \mathrm{for}~m_f\ll m_\mu\ll m_S\\
-\frac{m_S^2}{4m_{\mu}^2}(1+\frac{2m_S^2}{m_{\mu}^2}),& \mathrm{for}~m_f\ll m_S\ll m_\mu
\end{array}\right.
\end{align}
and
\begin{align}\label{eqn:expansion:IfLR:mfllmmumS:esp}
I_{LR}^f\approx\left\{\begin{array}{ll}
\frac{m_f}{4m_{\mu}}\big[-3+2\log\frac{m_S^2}{m_f^2}+\frac{m_{\mu}^2}{m_S^2}(2\log\frac{m_S^2}{m_f^2}-\frac{17}{3})\big],& \mathrm{for}~m_f\ll m_\mu\ll m_S\\
\frac{m_S^2m_f}{2m_{\mu}^3}(1+\frac{m_S^2}{m_{\mu}^2})(1-\log\frac{m_{\mu}^2}{m_f^2}+i\pi),& \mathrm{for}~m_f\ll m_S\ll m_\mu
\end{array}\right..
\end{align}

For the scenario of $m_f\ll m_\mu=m_S$, please refer to Eqs. \eqref{eqn:expansion:IfLL:mmumS:esp} and \eqref{eqn:expansion:IfLR:mmumS:esp}.
\subsubsection{Scenario of $m_f,m_S\ll m_\mu$}

Up to $1/m_{\mu}^6$, the integral $I_{LL}^f$ can be expanded as
\begin{align}\label{eqn:expansion:IfLL:mfmSllmmu}
&I_{LL}^f=-\frac{m_S^2}{4m_\mu^2}-\frac{m_S^2}{2m_\mu^4}(m_S^2-m_f^2-i\pi m_f^2+m_f^2\log\frac{m_\mu^2}{m_f^2})\nonumber\\
&+\frac{m_S^2}{2m_\mu^6}\Big[m_f^2(m_f^2+m_S^2)+m_f^2(m_f^2-2m_S^2)(\log\frac{m_\mu^2}{m_f^2}-i\pi)-m_S^4(\log\frac{m_\mu^2}{m_S^2}-i\pi)\Big]+\mathcal{O}(\frac{1}{m_\mu^8}).
\end{align}
Up to $1/m_{\mu}^7$, the integral $I_{LR}^f$ can be expanded as
\begin{align}
&I_{LR}^f=\frac{m_f}{m_\mu}\Big\{\frac{m_S^2}{2m_\mu^2}(1-\log\frac{m_\mu^2}{m_f^2}+i\pi)+\frac{m_S^2}{2m_\mu^4}\big[m_f^2+m_S^2+(m_f^2-m_S^2)(\log\frac{m_\mu^2}{m_f^2}-i\pi)\big]\nonumber\\
&+\frac{m_S^2}{4m_\mu^6}\big[3m_S^4+4m_f^2m_S^2-m_f^4-4m_S^4(\log\frac{m_\mu^2}{m_fm_S}-i\pi)\big]\Big\}+\mathcal{O}(\frac{1}{m_\mu^9}).
\end{align}

For the scenario of $m_f\ll m_S\ll m_\mu$, the approximations have been given in Eqs. \eqref{eqn:expansion:IfLL:mfllmmumS:esp} and \eqref{eqn:expansion:IfLR:mfllmmumS:esp}. For the scenario of $m_S\ll m_f\ll m_\mu$, the approximations have been given in Eqs. \eqref{eqn:expansion:IfLL:mSllmfmmu:esp} and \eqref{eqn:expansion:IfLR:mSllmfmmu:esp}. For the scenario of $m_f=m_S\ll m_\mu$, please refer to Eqs. \eqref{eqn:expansion:IfLL:mfmS:esp} and \eqref{eqn:expansion:IfLR:mfmS:esp}.
\subsubsection{Scenario of $m_f,m_\mu\ll m_S$}

Up to $1/m_S^4$, the integrals $I_{LL}^f$ and $I_{LR}^f$ can be expanded as
\begin{align}
&I_{LL}^f=\frac{1}{6}+\frac{1}{24m_S^2}(3m_\mu^2+22m_f^2-12m_f^2\log\frac{m_S^2}{m_f^2})\nonumber\\
&+\frac{1}{60m_S^4}\big[130m_f^4+155m_f^2m_\mu^2+6m_\mu^4-60m_f^2(2m_f^2+m_\mu^2)\log\frac{m_S^2}{m_f^2}\big]+\mathcal{O}(\frac{1}{m_S^6})
\end{align}
and
\begin{align}
&I_{LR}^f=\frac{m_f}{m_\mu}\Big\{\frac{1}{4}(2\log\frac{m_S^2}{m_f^2}-3)+\frac{1}{12m_S^2}\big[-17m_\mu^2-15m_f^2+(6m_\mu^2+18m_f^2)\log\frac{m_S^2}{m_f^2}\big]\nonumber\\
&+\frac{1}{24m_S^4}\big[-42m_f^4-152m_f^2m_\mu^2-43m_\mu^4+(72m_f^4+96m_f^2m_\mu^2+12m_\mu^4)\log\frac{m_S^2}{m_f^2}\big]\Big\}+\mathcal{O}(\frac{1}{m_S^6}).
\end{align}

For the scenario of $m_f\ll m_\mu\ll m_S$, the approximations have been given in Eqs. \eqref{eqn:expansion:IfLL:mfllmmumS:esp} and \eqref{eqn:expansion:IfLR:mfllmmumS:esp}. For the scenario of $m_\mu\ll m_f\ll m_S$, the approximations have been given in Eqs. \eqref{eqn:expansion:IfLL:mmullmfmS:esp} and \eqref{eqn:expansion:IfLR:mmullmfmS:esp}. For the scenario of $m_f=m_\mu\ll m_S$, please refer to Eqs. \eqref{eqn:expansion:IfLL:mfmmu:esp} and \eqref{eqn:expansion:IfLR:mfmmu:esp}.
\subsubsection{Scenario of $m_\mu,m_S\ll m_f$}
Up to $1/m_f^6$, the integral $I_{LL}^f$ can be expanded as
\begin{align}
&I_{LL}^f=\frac{m_S^2}{12m_f^2}\Big[1+\frac{1}{2m_f^2}(m_\mu^2-4m_S^2)+\frac{1}{10m_f^4}(3m_\mu^4-30m_S^2m_\mu^2-110m_S^4+60m_S^4\log\frac{m_f^2}{m_S^2})\Big]+\mathcal{O}(\frac{1}{m_f^8}).
\end{align}
Up to $1/m_f^5$, the integral $I_{LR}^f$ can be expanded as
\begin{align}\label{eqn:expansion:IfLR:mmumSllmf}
&I_{LR}^f=\frac{m_S^2}{4m_\mu m_f}\Big[1+\frac{1}{3m_f^2}(m_\mu^2-3m_S^2)+\frac{1}{6m_f^4}(m_\mu^4-8m_S^2m_\mu^2-18m_S^4+12m_S^4\log\frac{m_f^2}{m_S^2})\Big]+\mathcal{O}(\frac{1}{m_f^7}).
\end{align}

For the scenario of $m_\mu\ll m_S\ll m_f$, the approximations have been given in Eqs. \eqref{eqn:expansion:IfLL:mmullmfmS:esp} and \eqref{eqn:expansion:IfLR:mmullmfmS:esp}. For the scenario of $m_S\ll m_\mu\ll m_f$, the approximations have been given in Eqs. \eqref{eqn:expansion:IfLL:mSllmfmmu:esp} and \eqref{eqn:expansion:IfLR:mSllmfmmu:esp}. For the scenario of $m_\mu=m_S\ll m_f$, please refer to Eqs. \eqref{eqn:expansion:IfLL:mmumS:esp} and \eqref{eqn:expansion:IfLR:mmumS:esp}.
\subsection{Degenerate mass case for the $I_{LL}^f$ and $I_{LR}^f$}\label{sec:expansion:Ifdeg}
\subsubsection{Scenario of $m_f=m_\mu$}\label{sec:expansion:Ifdeg:mfmmu}
For $m_f=m_\mu$, the integrals $I_{LL}^f$ and $I_{LR}^f$ can be calculated as
\begin{small}
\begin{align}\label{eqn:expansion:IfLL:mfmmu}
&I_{LL}^f=\nonumber\\
&\left\{\begin{array}{ll}
\frac{m_S^2}{4m_{\mu}^2}\Big[1-\frac{2m_S^2}{m_{\mu}^2}+2(-\frac{2m_S^2}{m_{\mu}^2}+\frac{m_S^4}{m_{\mu}^4})\log\frac{m_S}{m_{\mu}}-\frac{2m_S}{\sqrt{m_S^2-4m_{\mu}^2}}(2-\frac{4m_S^2}{m_{\mu}^2}+\frac{m_S^4}{m_{\mu}^4})\log\frac{m_S+\sqrt{m_S^2-4m_{\mu}^2}}{2m_{\mu}}\Big],&\mathrm{for}~2m_\mu<m_S\\
\frac{m_S^2}{4m_{\mu}^2}\Big[1-\frac{2m_S^2}{m_{\mu}^2}+2(-\frac{2m_S^2}{m_{\mu}^2}+\frac{m_S^4}{m_{\mu}^4})\log\frac{m_S}{m_{\mu}}-\frac{2m_S}{\sqrt{4m_{\mu}^2-m_S^2}}(2-\frac{4m_S^2}{m_{\mu}^2}+\frac{m_S^4}{m_{\mu}^4})\arctan\sqrt{\frac{4m_{\mu}^2}{m_S^2}-1}\Big],&\mathrm{for}~2m_\mu>m_S\\
16\log 2-11,&\mathrm{for}~2m_\mu=m_S
\end{array}\right.
\end{align}
\end{small}
and
\begin{small}
\begin{align}\label{eqn:expansion:IfLR:mfmmu}
&I_{LR}^f=\nonumber\\
&\left\{\begin{array}{ll}
\frac{m_S^2}{4m_{\mu}^2}\Big[2-\frac{2m_S^2}{m_{\mu}^2}\log\frac{m_S}{m_{\mu}}-\frac{4m_S}{\sqrt{m_S^2-4m_{\mu}^2}}(1-\frac{m_S^2}{2m_{\mu}^2})\log\frac{m_S+\sqrt{m_S^2-4m_{\mu}^2}}{2m_{\mu}}\Big],&\mathrm{for}~2m_\mu<m_S\\
\frac{m_S^2}{4m_{\mu}^2}\Big[2-\frac{2m_S^2}{m_{\mu}^2}\log\frac{m_S}{m_{\mu}}-\frac{4m_S}{\sqrt{4m_{\mu}^2-m_S^2}}(1-\frac{m_S^2}{2m_{\mu}^2})\arctan\sqrt{\frac{4m_{\mu}^2}{m_S^2}-1}\Big],&\mathrm{for}~2m_\mu>m_S\\
6-8\log 2,&\mathrm{for}~2m_\mu=m_S
\end{array}\right..
\end{align}
\end{small}
Then, they can be expanded as
\begin{align}\label{eqn:expansion:IfLL:mfmmu:esp}
I_{LL}^f\approx\left\{\begin{array}{ll}
\frac{1}{6}+\frac{m_{\mu}^2}{m_S^2}(\frac{25}{24}-\log\frac{m_S}{m_{\mu}})+\frac{m_{\mu}^4}{m_S^4}(\frac{97}{20}-6\log\frac{m_S}{m_{\mu}}),& \mathrm{for}~m_\mu\ll m_S\vspace{1ex}\\
\frac{m_S^2}{4m_{\mu}^2}-\frac{\pi m_S^3}{4m_{\mu}^3}+\frac{m_S^4}{4m_{\mu}^4}(4\log\frac{m_{\mu}}{m_S}-1)+\frac{15\pi m_S^5}{32m_{\mu}^5}-\frac{m_S^6}{24m_{\mu}^6}(12\log\frac{m_{\mu}}{m_S}+11),& \mathrm{for}~m_S\ll m_\mu
\end{array}\right.
\end{align}
and
\begin{align}\label{eqn:expansion:IfLR:mfmmu:esp}
I_{LR}^f\approx\left\{\begin{array}{ll}
\log\frac{m_S}{m_{\mu}}-\frac{3}{4}+\frac{m_{\mu}^2}{m_S^2}(4\log\frac{m_S}{m_{\mu}}-\frac{8}{3})+\frac{m_{\mu}^4}{m_S^4}(15\log\frac{m_S}{m_{\mu}}-\frac{79}{8}),& \mathrm{for}~m_\mu\ll m_S\vspace{1ex}\\
\frac{m_S^2}{2m_{\mu}^2}-\frac{\pi m_S^3}{4m_{\mu}^3}+\frac{m_S^4}{4m_{\mu}^4}(2\log\frac{m_{\mu}}{m_S}+1)+\frac{3\pi m_S^5}{32m_{\mu}^5}-\frac{m_S^6}{12m_{\mu}^6},& \mathrm{for}~m_S\ll m_\mu
\end{array}\right..
\end{align}

\subsubsection{Scenario of $m_\mu=m_S$}\label{sec:expansion:Ifdeg:mmumS}
For $m_\mu=m_S$, the integrals $I_{LL}^f$ and $I_{LR}^f$ can be calculated as
\begin{align}\label{eqn:expansion:IfLL:mmumS}
I_{LL}^f=\left\{\begin{array}{ll}
-\frac{3}{4}+\frac{m_f^2}{2m_{\mu}^2}-\frac{1}{2}(1-\frac{3m_f^2}{m_{\mu}^2}+\frac{m_f^4}{m_{\mu}^4})\log\frac{m_f}{m_{\mu}}+\frac{m_f}{2\sqrt{m_f^2-4m_{\mu}^2}}(5-\frac{5m_f^2}{m_{\mu}^2}+\frac{m_f^4}{m_{\mu}^4})\log\frac{m_f+\sqrt{m_f^2-4m_{\mu}^2}}{2m_{\mu}},& \mathrm{for}~m_f>2m_{\mu}\\
-\frac{3}{4}+\frac{m_f^2}{2m_{\mu}^2}-\frac{1}{2}(1-\frac{3m_f^2}{m_{\mu}^2}+\frac{m_f^4}{m_{\mu}^4})\log\frac{m_f}{m_{\mu}}+\frac{m_f}{2\sqrt{4m_{\mu}^2-m_f^2}}(5-\frac{5m_f^2}{m_{\mu}^2}+\frac{m_f^4}{m_{\mu}^4})\arctan\sqrt{\frac{4m_{\mu}^2}{m_f^2}-1},& \mathrm{for}~m_f<2m_\mu\\
\frac{1}{4}(7-10\log2),& \mathrm{for}~m_f=2m_\mu
\end{array}\right.
\end{align}
and
\begin{align}\label{eqn:expansion:IfLR:mmumS}
I_{LR}^f=\left\{\begin{array}{ll}
\frac{m_f}{2m_{\mu}}[1+(2-\frac{m_f^2}{m_{\mu}^2})\log\frac{m_f}{m_{\mu}}]+\frac{m_{\mu}}{2\sqrt{m_f^2-4m_{\mu}^2}}(2-\frac{4m_f^2}{m_{\mu}^2}+\frac{m_f^4}{m_{\mu}^4})\log\frac{m_f+\sqrt{m_f^2-4m_{\mu}^2}}{2m_{\mu}},& \mathrm{for}~m_f>2m_{\mu}\\
\frac{m_f}{2m_{\mu}}[1+(2-\frac{m_f^2}{m_{\mu}^2})\log\frac{m_f}{m_{\mu}}]+\frac{m_{\mu}}{2\sqrt{4m_{\mu}^2-m_f^2}}(2-\frac{4m_f^2}{m_{\mu}^2}+\frac{m_f^4}{m_{\mu}^4})\arctan\sqrt{\frac{4m_{\mu}^2}{m_f^2}-1},& \mathrm{for}~m_f<2m_\mu\\
\frac{3}{2}-2\log2,& \mathrm{for}~m_f=2m_\mu
\end{array}\right. .
\end{align}
Then, they can be expanded as
\begin{align}\label{eqn:expansion:IfLL:mmumS:esp}
I_{LL}^f\approx\left\{\begin{array}{ll}
\frac{m_{\mu}^2}{12m_f^2}\big[1-\frac{3m_{\mu}^2}{2m_f^2}+\frac{m_{\mu}^4}{10m_f^4}(120\log\frac{m_f}{m_{\mu}}-137)\big],& \mathrm{for}~m_\mu\ll m_f\vspace{1ex}\\
\frac{1}{2}\log\frac{m_{\mu}}{m_f}-\frac{3}{4}+\frac{5\pi m_f}{8m_{\mu}}-\frac{m_f^2}{8m_{\mu}^2}(12\log\frac{m_{\mu}}{m_f}+1)-\frac{35\pi m_f^3}{64m_{\mu}^3}+\frac{m_f^4}{48m_{\mu}^4}(24\log\frac{m_{\mu}}{m_f}+25),& \mathrm{for}~m_f\ll m_\mu
\end{array}\right.
\end{align}
and
\begin{align}\label{eqn:expansion:IfLR:mmumS:esp}
I_{LR}^f\approx\left\{\begin{array}{ll}
\frac{m_{\mu}}{4m_f}\big[1-\frac{2m_{\mu}^2}{3m_f^2}+\frac{m_{\mu}^4}{6m_f^4}(24\log\frac{m_f}{m_{\mu}}-25)\big],& \mathrm{for}~m_\mu\ll m_f\vspace{1ex}\\
\frac{\pi}{4}+\frac{m_f}{4m_{\mu}}(1-4\log\frac{m_{\mu}}{m_f})-\frac{15\pi m_f^2}{32m_{\mu}^2}+\frac{m_f^3}{24m_{\mu}^3}(12\log\frac{m_{\mu}}{m_f}+11)+\frac{35\pi m_f^4}{512m_{\mu}^4}-\frac{m_f^5}{20m_{\mu}^5},& \mathrm{for}~m_f\ll m_\mu
\end{array}\right..
\end{align}

\subsubsection{Scenario of $m_f=m_S$}\label{sec:expansion:Ifdeg:mfmS}
For $m_f=m_S$, the integrals $I_{LL}^f$ and $I_{LR}^f$ can be calculated as
\begin{align}\label{eqn:expansion:IfLL:mfmS}
I_{LL}^f=\left\{\begin{array}{ll}
-\frac{m_S^2}{4m_{\mu}^2}(1-\frac{4m_S^2}{m_{\mu}\sqrt{4m_S^2-m_{\mu}^2}}\arctan\frac{m_{\mu}}{\sqrt{4m_S^2-m_{\mu}^2}}),&\mathrm{for}~m_\mu<2m_S\\
-\frac{m_S^2}{4m_{\mu}^2}\big[1+\frac{4m_S^2}{m_{\mu}\sqrt{m_{\mu}^2-4m_S^2}}(\log\frac{m_{\mu}+\sqrt{m_{\mu}^2-4m_S^2}}{2m_S}-\frac{i\pi}{2})\big],&\mathrm{for}~m_\mu>2m_S\\
-\frac{1}{8},&\mathrm{for}~m_\mu=2m_S
\end{array}\right.
\end{align}
and
\begin{small}\label{eqn:expansion:IfLR:mfmS}
\begin{align}
I_{LR}^f=\left\{\begin{array}{ll}
\frac{m_S^3}{2m_{\mu}^3}(1+\frac{2m_{\mu}^2-4m_S^2}{m_{\mu}\sqrt{4m_S^2-m_{\mu}^2}}\arctan\frac{m_{\mu}}{\sqrt{4m_S^2-m_{\mu}^2}}),&\mathrm{for}~m_\mu<2m_S\\
\frac{m_S^3}{2m_{\mu}^3}\big[1+\frac{4m_S^2-2m_{\mu}^2}{m_{\mu}\sqrt{m_{\mu}^2-4m_S^2}}(\log\frac{m_{\mu}+\sqrt{m_{\mu}^2-4m_S^2}}{2m_S}-\frac{i\pi}{2})\big],&\mathrm{for}~m_\mu>2m_S\\
0,&\mathrm{for}~m_\mu=2m_S
\end{array}\right..
\end{align}
\end{small}
Then, they can be expanded as
\begin{align}\label{eqn:expansion:IfLL:mfmS:esp}
I_{LL}^f\approx\left\{\begin{array}{ll}
\frac{1}{24}(1+\frac{m_{\mu}^2}{5m_S^2}+\frac{3m_{\mu}^4}{70m_S^4}),& \mathrm{for}~m_\mu\ll m_S\vspace{1ex}\\
-\frac{m_S^2}{4m_{\mu}^2}\big[1+\frac{2m_S^2}{m_{\mu}^2}(2\log\frac{m_{\mu}}{m_S}-i\pi)+\frac{4m_S^4}{m_{\mu}^4}(2\log\frac{m_{\mu}}{m_S}-i\pi-1)\big],& \mathrm{for}~m_S\ll m_\mu
\end{array}\right.
\end{align}
and
\begin{align}\label{eqn:expansion:IfLR:mfmS:esp}
I_{LR}^f\approx\left\{\begin{array}{ll}
\frac{m_S}{6m_{\mu}}(1+\frac{3m_{\mu}^2}{20m_S^2}+\frac{m_{\mu}^4}{35m_S^4}),& \mathrm{for}~m_\mu\ll m_S\vspace{1ex}\\
\frac{m_S^3}{2m_{\mu}^3}(1-2\log\frac{m_{\mu}}{m_S}+i\pi)+\frac{m_S^5}{m_{\mu}^5}+\frac{m_S^7}{2m_{\mu}^7}(3-4\log\frac{m_{\mu}}{m_S}+2i\pi),& \mathrm{for}~m_S\ll m_\mu
\end{array}\right..
\end{align}

\subsubsection{Scenario of $m_f=m_\mu=m_S$}
In this case, we have $I_{LL}^f=\frac{2\sqrt{3}\pi-9}{36}$ and
$I_{LR}^f=\frac{9-\sqrt{3}\pi}{18}$.

\subsection{Hierarchical mass expansion for the $I_{LL}^S$ and $I_{LR}^S$}\label{sec:expansion:IS}
\subsubsection{Scenario of $m_\mu\ll m_f,m_S$}\label{sec:expansion:IShie:mmullmfmS}
Up to $m_{\mu}^4$, the integral $I_{LL}^S$ can be expanded as
\begin{align}\label{eqn:expansion:ISLL:mmullmfmS}
&I_{LL}^S=-\frac{m_S^2}{12(m_S^2-m_f^2)^4}(m_S^6-6m_S^4m_f^2+3m_S^2m_f^4+2m_f^6-6m_S^2m_f^4\log\frac{m_f^2}{m_S^2})\nonumber\\
&+\frac{m_\mu^2m_S^2}{24(m_S^2-m_f^2)^6}\Big[-m_S^8+12m_S^6m_f^2+36m_S^4m_f^4-44m_S^2m_f^6-3m_f^8+12m_S^2m_f^4(2m_f^2+3m_S^2)\log\frac{m_f^2}{m_S^2}\Big]\nonumber\\
&+\frac{m_\mu^4m_S^2}{40(m_S^2-m_f^2)^8}\big[-m_S^{10}+20m_S^8m_f^2+220m_S^6m_f^4-80m_S^4m_f^6-155m_S^2m_f^8-4m_f^{10}\nonumber\\
	&+60m_S^2m_f^4(2m_S^4+4m_S^2m_f^2+m_f^4)\log\frac{m_f^2}{m_S^2}\Big]+\mathcal{O}(m_\mu^6).
\end{align}
Up to $m_{\mu}^3$, the integral $I_{LR}^S$ can be expanded as
\begin{align}\label{eqn:expansion:ISLR:mmullmfmS}
&I_{LR}^S=\frac{m_f}{m_\mu}\Big\{\frac{m_S^2}{4(m_S^2-m_f^2)^3}(-m_S^4+m_f^4-2m_S^2m_f^2\log\frac{m_f^2}{m_S^2})\nonumber\\
&+\frac{m_\mu^2m_S^2}{6(m_S^2-m_f^2)^5}\big[-m_S^6-9m_S^4m_f^2+9m_S^2m_f^4+m_f^6-6m_S^2m_f^2(m_f^2+m_S^2)\log\frac{m_f^2}{m_S^2}\big]\nonumber\\
&+\frac{m_\mu^4m_S^2}{8(m_S^2-m_f^2)^7}\big[-m_S^8-28m_S^6m_f^2+28m_S^2m_f^6+m_f^8\nonumber\\
	&-12m_S^2m_f^2(m_S^4+3m_S^2m_f^2+m_f^4)\log\frac{m_f^2}{m_S^2}\big]\Big\}+\mathcal{O}(m_\mu^5).
\end{align}
Especially, we have the following approximations:
\begin{align}\label{eqn:expansion:ISLL:mmullmfmS:esp}
I_{LL}^S\approx\left\{\begin{array}{ll}
-\frac{1}{12}(1-\frac{2m_f^2}{m_S^2}),& \mathrm{for}~m_\mu\ll m_f\ll m_S\vspace{1ex}\\
-\frac{m_S^2}{6m_f^2}\big[1+\frac{m_S^2}{2m_f^2}(11-12\log\frac{m_f}{m_S})\big],& \mathrm{for}~m_\mu\ll m_S\ll m_f
\end{array}\right.
\end{align}
and
\begin{align}\label{eqn:expansion:ISLR:mmullmfmS:esp}
I_{LR}^S\approx\left\{\begin{array}{ll}
-\frac{m_f}{4m_{\mu}}\Big[1+\frac{m_f^2}{m_S^2}(3-4\log\frac{m_S}{m_f})\Big],& \mathrm{for}~m_\mu\ll m_f\ll m_S\vspace{1ex}\\
-\frac{m_S^2}{4m_fm_{\mu}}\Big[1+\frac{m_S^2}{m_f^2}(3-4\log\frac{m_f}{m_S})\Big],& \mathrm{for}~m_\mu\ll m_S\ll m_f
\end{array}\right. .
\end{align}

For the scenario of $m_\mu\ll m_f=m_S$, please refer to Eqs. \eqref{eqn:expansion:ISLL:mfmS:esp} and \eqref{eqn:expansion:ISLR:mfmS:esp}.
\subsubsection{Scenario of $m_S\ll m_f,m_\mu$}\label{sec:expansion:IShie:mSllmfmmu}
Up to $m_S^6$, the integrals $I_{LL}^S$ and $I_{LR}^S$ can be expanded as
\begin{align}\label{eqn:expansion:ISLL:mSllmfmmu}
&I_{LL}^S=\left\{\begin{array}{ll}
\frac{m_S^2}{4m_\mu^6}\big[2m_f^2m_\mu^2+m_\mu^4+2m_f^4\log\frac{m_f^2-m_\mu^2}{m_f^2}\big]+\frac{m_S^4}{2m_\mu^6(m_\mu^2-m_f^2)^2}\big[m_\mu^2(-2m_f^4+2m_f^2m_\mu^2-m_\mu^4)\vspace{1ex}\\
+(m_\mu^6+3m_f^4m_\mu^2-2m_f^6)\log\frac{m_f^2-m_\mu^2}{m_f^2}+m_\mu^6\log\frac{m_f^2}{m_S^2}\big]\vspace{1ex}\\
+\frac{m_S^6}{4m_\mu^6(m_\mu^2-m_f^2)^4}\big[m_\mu^2(2m_f^6-7m_f^4m_\mu^2-2m_\mu^6)+2m_\mu^6(4m_f^2-m_\mu^2)\log\frac{m_f^2}{m_S^2}\vspace{1ex}\\
+(2m_f^8-8m_f^6m_\mu^2+12m_f^4m_\mu^4+8m_f^2m_\mu^6-2m_\mu^8)\log\frac{m_f^2-m_\mu^2}{m_f^2}\big]+\mathcal{O}(m_S^8),& \hspace{-10ex}\mathrm{for}~m_\mu<m_f\vspace{2ex}\\
\frac{m_S^2}{4m_\mu^6}\big[2m_f^2m_\mu^2+m_\mu^4+2m_f^4(\log\frac{m_\mu^2-m_f^2}{m_f^2}-i\pi)\big]+\frac{m_S^4}{2m_\mu^6(m_\mu^2-m_f^2)^2}\big[m_\mu^2(-2m_f^4+2m_f^2m_\mu^2-m_\mu^4)\vspace{1ex}\\
+(m_\mu^6+3m_f^4m_\mu^2-2m_f^6)(\log\frac{m_\mu^2-m_f^2}{m_f^2}-i\pi)+m_\mu^6\log\frac{m_f^2}{m_S^2}\big]\vspace{1ex}\\
+\frac{m_S^6}{4m_\mu^6(m_\mu^2-m_f^2)^4}\big[m_\mu^2(2m_f^6-7m_f^4m_\mu^2-2m_\mu^6)+2m_\mu^6(4m_f^2-m_\mu^2)\log\frac{m_f^2}{m_S^2}\vspace{1ex}\\
+(2m_f^8-8m_f^6m_\mu^2+12m_f^4m_\mu^4+8m_f^2m_\mu^6-2m_\mu^8)(\log\frac{m_\mu^2-m_f^2}{m_f^2}-i\pi)\big]+\mathcal{O}(m_S^8),& \hspace{-10ex}\mathrm{for}~m_\mu>m_f
\end{array}\right.
\end{align}
and
\begin{align}\label{eqn:expansion:ISLR:mSllmfmmu}
&I_{LR}^S=\left\{\begin{array}{ll}
\frac{m_f}{m_\mu}\Big\{\frac{m_S^2}{2m_\mu^4}\big[m_\mu^2+m_f^2\log\frac{m_f^2-m_\mu^2}{m_f^2}\big]+\frac{m_S^4}{2m_\mu^4(m_\mu^2-m_f^2)^2}\big[-m_f^2m_\mu^2+m_\mu^4\log\frac{m_f^2}{m_S^2}\vspace{1ex}\\
+(m_\mu^4+2m_f^2m_\mu^2-m_f^4)\log\frac{m_f^2-m_\mu^2}{m_f^2}\big]+\frac{m_S^6}{4(m_\mu^2-m_f^2)^4}\big[-5m_f^2-2m_\mu^2\vspace{1ex}\\
+12m_f^2\log\frac{m_f^2-m_\mu^2}{m_fm_S}\big]\Big\}+\mathcal{O}(m_S^8),& \hspace{-15ex}\mathrm{for}~m_\mu<m_f\vspace{2ex}\\
\frac{m_f}{m_\mu}\Big\{\frac{m_S^2}{2m_\mu^4}\big[m_\mu^2+m_f^2(\log\frac{m_\mu^2-m_f^2}{m_f^2}-i\pi)\big]+\frac{m_S^4}{2m_\mu^4(m_\mu^2-m_f^2)^2}\big[-m_f^2m_\mu^2+m_\mu^4\log\frac{m_f^2}{m_S^2}\vspace{1ex}\\
+(m_\mu^4+2m_f^2m_\mu^2-m_f^4)(\log\frac{m_\mu^2-m_f^2}{m_f^2}-i\pi)\big]+\frac{m_S^6}{4(m_\mu^2-m_f^2)^4}\big[-5m_f^2-2m_\mu^2\vspace{1ex}\\
+12m_f^2(\log\frac{m_\mu^2-m_f^2}{m_fm_S}-i\pi)\big]\Big\}+\mathcal{O}(m_S^8),& \hspace{-15ex}\mathrm{for}~m_\mu>m_f
\end{array}\right..
\end{align}

Especially, we have the following approximations:
\begin{align}\label{eqn:expansion:ISLL:mSllmfmmu:esp}
I_{LL}^S\approx\left\{\begin{array}{ll}
\frac{m_S^2}{4m_\mu^2}(1+\frac{2m_f^2}{m_\mu^2}),& \mathrm{for}~m_S\ll m_f\ll m_\mu\\
-\frac{m_S^2}{6m_f^2}(1+\frac{3m_\mu^2}{4m_f^2}),& \mathrm{for}~m_S\ll m_\mu\ll m_f
\end{array}\right.
\end{align}
and
\begin{align}\label{eqn:expansion:ISLR:mSllmfmmu:esp}
I_{LR}^S\approx\left\{\begin{array}{ll}
\frac{m_S^2m_f}{2m_\mu^3}\left[1+\frac{m_f^2}{m_\mu^2}(\log\frac{m_\mu^2}{m_f^2}-i\pi)\right],& \mathrm{for}~m_S\ll m_f\ll m_\mu\\
-\frac{m_S^2}{4m_fm_\mu}(1+\frac{2m_\mu^2}{3m_f^2}),& \mathrm{for}~m_S\ll m_\mu\ll m_f
\end{array}\right. .
\end{align}

For the scenario of $m_S\ll m_f=m_\mu$, please refer to Eqs. \eqref{eqn:expansion:ISLL:mfmmu:esp} and \eqref{eqn:expansion:ISLR:mfmmu:esp}.
\subsubsection{Scenario of $m_f\ll m_\mu,m_S$}\label{sec:expansion:IShie:mfllmmumS}
Up to $m_f^4$, the integral $I_{LL}^S$ can be expanded as
\begin{align}\label{eqn:expansion:ISLL:mfllmmumS}
&I_{LL}^S=\left\{\begin{array}{ll}
\frac{m_S^2}{4m_{\mu}^6}\big[m_{\mu}^4-2m_S^2m_{\mu}^2+2m_S^2(m_{\mu}^2-m_S^2)\log\frac{m_S^2-m_{\mu}^2}{m_S^2}\big]+\frac{m_f^2m_S^2}{2m_{\mu}^6(m_S^2-m_\mu^2)}\big[2m_S^2m_{\mu}^2-m_{\mu}^4\vspace{1ex}\\
+2m_S^2(m_S^2-m_{\mu}^2)\log\frac{m_S^2-m_{\mu}^2}{m_S^2}\big]+\frac{m_f^4m_S^2}{4m_{\mu}^6(m_S^2-m_\mu^2)^3}\big[m_S^2m_\mu^2(5m_\mu^2-2m_S^2)-2m_\mu^6\log\frac{m_S^2}{m_f^2}\vspace{1ex}\\
+(-2m_\mu^6-6m_S^2m_\mu^4+6m_S^4m_{\mu}^2-2m_S^6)\log\frac{m_S^2-m_{\mu}^2}{m_S^2}\big]
+\mathcal{O}(m_f^6),& \hspace{-15ex}\mathrm{for}~m_\mu<m_S\vspace{3ex}\\
\frac{m_S^2}{4m_{\mu}^6}\big[m_{\mu}^4-2m_S^2m_{\mu}^2+2m_S^2(m_{\mu}^2-m_S^2)(\log\frac{m_{\mu}^2-m_S^2}{m_S^2}-i\pi)\big]+\frac{m_f^2m_S^2}{2m_{\mu}^6(m_S^2-m_\mu^2)}\big[2m_S^2m_{\mu}^2-m_{\mu}^4\vspace{1ex}\\
+2m_S^2(m_S^2-m_{\mu}^2)(\log\frac{m_{\mu}^2-m_S^2}{m_S^2}-i\pi)\big]+\frac{m_f^4m_S^2}{4m_{\mu}^6(m_S^2-m_\mu^2)^3}\big[m_S^2m_\mu^2(5m_\mu^2-2m_S^2)-2m_\mu^6\log\frac{m_S^2}{m_f^2}\vspace{1ex}\\
+(-2m_\mu^6-6m_S^2m_\mu^4+6m_S^4m_{\mu}^2-2m_S^6)(\log\frac{m_{\mu}^2-m_S^2}{m_S^2}-i\pi)\big]
+\mathcal{O}(m_f^6),& \hspace{-15ex}\mathrm{for}~m_\mu>m_S
\end{array}\right..
\end{align}
Up to $m_f^5$, the integral $I_{LR}^S$ can be expanded as
\begin{align}\label{eqn:expansion:ISLR:mfllmmumS}
&I_{LR}^S=\left\{\begin{array}{ll}
\frac{m_f}{m_{\mu}}\Big\{\frac{m_S^2}{2m_{\mu}^4}\big[m_{\mu}^2+m_S^2\log\frac{m_S^2-m_{\mu}^2}{m_S^2}\big]+\frac{m_f^2m_S^2}{2m_{\mu}^4(m_S^2-m_\mu^2)^2}\big[-m_S^2m_{\mu}^2+m_\mu^4\log\frac{m_S^2}{m_f^2}\vspace{1ex}\\
+(m_{\mu}^4+2m_S^2m_{\mu}^2-m_S^4)\log\frac{m_S^2-m_{\mu}^2}{m_S^2}\big]+\frac{m_f^4m_S^2}{4(m_S^2-m_\mu^2)^4}\big[-2m_{\mu}^2-5m_S^2\vspace{1ex}\\
+12m_S^2\log\frac{m_S^2-m_{\mu}^2}{m_Sm_f}\big]\Big\}+\mathcal{O}(m_f^7),& \hspace{-15ex}\mathrm{for}~m_\mu<m_S\vspace{2ex}\\
\frac{m_f}{m_{\mu}}\Big\{\frac{m_S^2}{2m_{\mu}^4}\big[m_{\mu}^2+m_S^2(\log\frac{m_{\mu}^2-m_S^2}{m_S^2}-i\pi)\big]+\frac{m_f^2m_S^2}{2m_{\mu}^4(m_S^2-m_\mu^2)^2}\big[-m_S^2m_{\mu}^2+m_\mu^4\log\frac{m_S^2}{m_f^2}\vspace{1ex}\\
+(m_{\mu}^4+2m_S^2m_{\mu}^2-m_S^4)(\log\frac{m_{\mu}^2-m_S^2}{m_S^2}-i\pi)\big]+\frac{m_f^4m_S^2}{4(m_S^2-m_\mu^2)^4}\big[-2m_{\mu}^2-5m_S^2\vspace{1ex}\\
+12m_S^2(\log\frac{m_{\mu}^2-m_S^2}{m_Sm_f}-i\pi)\big]\Big\}+\mathcal{O}(m_f^7),& \hspace{-15ex}\mathrm{for}~m_\mu>m_S
\end{array}\right..
\end{align}

Especially, we have the following approximations:
\begin{align}\label{eqn:expansion:ISLL:mfllmmumS:esp}
I_{LL}^S\approx\left\{\begin{array}{ll}
-\frac{1}{12}(1+\frac{m_{\mu}^2}{2m_S^2}),& \mathrm{for}~m_f\ll m_\mu\ll m_S\\
\frac{m_S^2}{4m_{\mu}^2}\left[1+\frac{2m_S^2}{m_{\mu}^2}(\log\frac{m_{\mu}^2}{m_S^2}-i\pi-1)\right],& \mathrm{for}~m_f\ll m_S\ll m_\mu
\end{array}\right.
\end{align}
and
\begin{align}\label{eqn:expansion:ISLR:mfllmmumS:esp}
I_{LR}^S\approx\left\{\begin{array}{ll}
-\frac{m_f}{4m_{\mu}}(1+\frac{2m_{\mu}^2}{3m_S^2}),& \mathrm{for}~m_f\ll m_\mu\ll m_S\\
\frac{m_S^2m_f}{2m_{\mu}^3}[1+\frac{m_S^2}{m_{\mu}^2}(\log\frac{m_{\mu}^2}{m_S^2}-i\pi)],& \mathrm{for}~m_f\ll m_S\ll m_\mu
\end{array}\right..
\end{align}

For the scenario of $m_f\ll m_\mu=m_S$, please refer to Eqs. \eqref{eqn:expansion:ISLL:mmumS:esp} and \eqref{eqn:expansion:ISLR:mmumS:esp}.
\subsubsection{Scenario of $m_f,m_S\ll m_\mu$}

Up to $1/m_{\mu}^6$, the integral $I_{LL}^S$ can be expanded as
\begin{align}
&I_{LL}^S=\frac{m_S^2}{4m_\mu^2}+\frac{m_S^2}{2m_\mu^4}\big[m_f^2-m_S^2+ m_S^2(\log\frac{m_\mu^2}{m_S^2}-i\pi)\big]\nonumber\\
&+\frac{m_S^2}{2m_\mu^6}\big[-m_S^2(m_f^2+m_S^2)+m_f^4(\log\frac{m_\mu^2}{m_f^2}-i\pi)+m_S^2(2m_f^2-m_S^2)(\log\frac{m_\mu^2}{m_S^2}-i\pi)\big]+\mathcal{O}(\frac{1}{m_\mu^8}).
\end{align}
Up to $1/m_{\mu}^7$, the integral $I_{LR}^S$ can be expanded as
\begin{align}
&I_{LR}^S=\frac{m_f}{m_\mu}\Big\{\frac{m_S^2}{2m_\mu^2}+\frac{m_S^2}{2m_\mu^4}[m_S^2(\log\frac{m_\mu^2}{m_S^2}-i\pi)+m_f^2(\log\frac{m_\mu^2}{m_f^2}-i\pi)]\nonumber\\
&+\frac{m_S^2}{2m_\mu^6}[-(m_f^2+m_S^2)^2+4m_f^2m_S^2(\log\frac{m_\mu^2}{m_fm_S}-i\pi)]\Big\}+\mathcal{O}(\frac{1}{m_\mu^9}).
\end{align}

For the scenario of $m_f\ll m_S\ll m_\mu$, the approximations have been given in Eqs. \eqref{eqn:expansion:ISLL:mfllmmumS:esp} and \eqref{eqn:expansion:ISLR:mfllmmumS:esp}. For the scenario of $m_S\ll m_f\ll m_\mu$, the approximations have been given in Eqs. \eqref{eqn:expansion:ISLL:mSllmfmmu:esp} and \eqref{eqn:expansion:ISLR:mSllmfmmu:esp}. For the scenario of $m_f=m_S\ll m_\mu$, please refer to Eqs. \eqref{eqn:expansion:ISLL:mfmS:esp} and \eqref{eqn:expansion:ISLR:mfmS:esp}.

\subsubsection{Scenario of $m_f,m_\mu\ll m_S$}

Up to $1/m_S^4$, the integrals $I_{LL}^S$ and $I_{LR}^S$ can be expanded as
\begin{align}
&I_{LL}^S=-\frac{1}{12}+\frac{4m_f^2-m_\mu^2}{24m_S^2}+\frac{1}{120m_S^4}(110m_f^4+30m_f^2m_\mu^2-3m_\mu^4-60m_f^4\log\frac{m_S^2}{m_f^2})+\mathcal{O}(\frac{1}{m_S^6})
\end{align}
and
\begin{align}
&I_{LR}^S=\frac{m_f}{m_\mu}\Big\{-\frac{1}{4}+\frac{1}{12m_S^2}(6m_f^2\log\frac{m_S^2}{m_f^2}-2m_\mu^2-9m_f^2)\nonumber\\
&+\frac{1}{24m_S^4}\big[-30m_f^4-56m_f^2m_\mu^2-3m_\mu^4+12m_f^2(3m_f^2+2m_\mu^2)\log\frac{m_S^2}{m_f^2}\big]\Big\}+\mathcal{O}(\frac{1}{m_S^6}).
\end{align}

For the scenario of $m_f\ll m_\mu\ll m_S$, the approximations have been given in Eqs. \eqref{eqn:expansion:ISLL:mfllmmumS:esp} and \eqref{eqn:expansion:ISLR:mfllmmumS:esp}. For the scenario of $m_\mu\ll m_f\ll m_S$, the approximations have been given in Eqs. \eqref{eqn:expansion:ISLL:mmullmfmS:esp} and \eqref{eqn:expansion:ISLR:mmullmfmS:esp}. For the scenario of $m_f=m_\mu\ll m_S$, please refer to Eqs. \eqref{eqn:expansion:ISLL:mfmmu:esp} and \eqref{eqn:expansion:ISLR:mfmmu:esp}.

\subsubsection{Scenario of $m_\mu,m_S\ll m_f$}

Up to $1/m_f^6$, the integral $I_{LL}^S$ can be expanded as
\begin{align}
&I_{LL}^S=-\frac{m_S^2}{6m_f^2}+\frac{m_S^2}{24m_f^4}(12m_S^2\log\frac{m_f^2}{m_S^2}-3m_\mu^2-22m_S^2)\nonumber\\
&+\frac{m_S^2}{60m_f^6}\big[-6m_\mu^4-155m_S^2m_\mu^2-130m_S^4+60m_S^2(2m_S^2+m_\mu^2)\log\frac{m_f^2}{m_S^2}\big]+\mathcal{O}(\frac{1}{m_f^8}).
\end{align}
Up to $1/m_f^5$, the integral $I_{LR}^S$ can be expanded as
\begin{align}
&I_{LR}^S=-\frac{m_S^2}{4m_\mu m_f}+\frac{m_S^2}{12m_\mu m_f^3}(6m_S^2\log\frac{m_f^2}{m_S^2}-2m_\mu^2-9m_S^2)\nonumber\\
&+\frac{m_S^2}{24m_\mu m_f^5}\big[-3m_\mu^4-56m_S^2m_\mu^2-30m_S^4+12m_S^2(2m_\mu^2+3m_S^2)\log\frac{m_f^2}{m_S^2}\big]+\mathcal{O}(\frac{1}{m_f^7}).
\end{align}

For the scenario of $m_\mu\ll m_S\ll m_f$, the approximations have been given in Eqs. \eqref{eqn:expansion:ISLL:mmullmfmS:esp} and \eqref{eqn:expansion:ISLR:mmullmfmS:esp}. For the scenario of $m_S\ll m_\mu\ll m_f$, the approximations have been given in Eqs. \eqref{eqn:expansion:ISLL:mSllmfmmu:esp} and \eqref{eqn:expansion:ISLR:mSllmfmmu:esp}. For the scenario of $m_\mu=m_S\ll m_f$, please refer to Eqs. \eqref{eqn:expansion:ISLL:mmumS:esp} and \eqref{eqn:expansion:ISLR:mmumS:esp}.

\subsection{Degenerate mass case for the $I_{LL}^S$ and $I_{LR}^S$}\label{sec:expansion:ISdeg}
\subsubsection{Scenario of $m_f=m_\mu$}\label{sec:expansion:ISdeg:mfmu}

For $m_f=m_\mu$, the integrals $I_{LL}^S$ and $I_{LR}^S$ can be calculated as
\begin{small}\label{eqn:expansion:ISLL:mfmmu}
\begin{align}
&I_{LL}^S=\nonumber\\
&\left\{\begin{array}{ll}
\frac{m_S^2}{4m_{\mu}^2}\Big[3-\frac{2m_S^2}{m_{\mu}^2}+2(1-\frac{3m_S^2}{m_{\mu}^2}+\frac{m_S^4}{m_{\mu}^4})\log\frac{m_S}{m_{\mu}}-\frac{2m_S}{\sqrt{m_S^2-4m_{\mu}^2}}(5-\frac{5m_S^2}{m_{\mu}^2}+\frac{m_S^4}{m_{\mu}^4})\log\frac{m_S+\sqrt{m_S^2-4m_{\mu}^2}}{2m_{\mu}}\Big],&\mathrm{for}~m_\mu<\frac{m_S}{2}\\
\frac{m_S^2}{4m_{\mu}^2}\Big[3-\frac{2m_S^2}{m_{\mu}^2}+2(1-\frac{3m_S^2}{m_{\mu}^2}+\frac{m_S^4}{m_{\mu}^4})\log\frac{m_S}{m_{\mu}}-\frac{2m_S}{\sqrt{4m_{\mu}^2-m_S^2}}(5-\frac{5m_S^2}{m_{\mu}^2}+\frac{m_S^4}{m_{\mu}^4})\arctan\sqrt{\frac{4m_{\mu}^2}{m_S^2}-1}\Big],&\mathrm{for}~m_\mu>\frac{m_S}{2}\\
10\log 2-7,&\mathrm{for}~m_\mu=\frac{m_S}{2}
\end{array}\right.
\end{align}
\end{small}
and
\begin{small}\label{eqn:expansion:ISLR:mfmmu}
\begin{align}
&I_{LR}^S=\nonumber\\
&\left\{\begin{array}{ll}
\frac{m_S^2}{4m_{\mu}^2}\Big[2+2(1-\frac{m_S^2}{m_{\mu}^2})\log\frac{m_S}{m_{\mu}}-\frac{2m_S}{\sqrt{m_S^2-4m_{\mu}^2}}(3-\frac{m_S^2}{m_{\mu}^2})\log\frac{m_S+\sqrt{m_S^2-4m_{\mu}^2}}{2m_{\mu}}\Big],&\mathrm{for}~m_\mu<\frac{m_S}{2}\\
\frac{m_S^2}{4m_{\mu}^2}\Big[2+2(1-\frac{m_S^2}{m_{\mu}^2})\log\frac{m_S}{m_{\mu}}-\frac{2m_S}{\sqrt{4m_{\mu}^2-m_S^2}}(3-\frac{m_S^2}{m_{\mu}^2})\arctan\sqrt{\frac{4m_{\mu}^2}{m_S^2}-1}\Big],&\mathrm{for}~m_\mu>\frac{m_S}{2}\\
4-6\log 2,&\mathrm{for}~m_\mu=\frac{m_S}{2}
\end{array}\right..
\end{align}
\end{small}
Then, they can be expanded as
\begin{align}\label{eqn:expansion:ISLL:mfmmu:esp}
I_{LL}^S\approx\left\{\begin{array}{ll}
-\frac{1}{12}+\frac{m_{\mu}^2}{8m_S^2}+\frac{m_{\mu}^4}{120m_S^4}(137-120\log\frac{m_S}{m_{\mu}}),& \hspace{15ex}\mathrm{for}~m_\mu\ll m_S\vspace{2ex}\\
\frac{m_S^2}{4m_{\mu}^2}(3-2\log\frac{m_{\mu}}{m_S})-\frac{5\pi m_S^3}{8m_{\mu}^3}+\frac{m_S^4}{8m_{\mu}^4}(12\log\frac{m_{\mu}}{m_S}+1)\vspace{1ex}\\
+\frac{35\pi m_S^5}{64m_{\mu}^5}-\frac{m_S^6}{48m_{\mu}^6}(24\log\frac{m_{\mu}}{m_S}+25),& \hspace{15ex}\mathrm{for}~m_S\ll m_\mu
\end{array}\right.
\end{align}
and
\begin{align}\label{eqn:expansion:ISLR:mfmmu:esp}
I_{LR}^S\approx\left\{\begin{array}{ll}
-\frac{1}{4}+\frac{m_{\mu}^2}{12m_S^2}(12\log\frac{m_S}{m_{\mu}}-11)+\frac{m_{\mu}^4}{24m_S^4}(120\log\frac{m_S}{m_{\mu}}-89),& \hspace{5ex}\mathrm{for}~m_\mu\ll m_S\vspace{1ex}\\
\frac{m_S^2}{2m_{\mu}^2}(1-\log\frac{m_{\mu}}{m_S})-\frac{3\pi m_S^3}{8m_{\mu}^3}+\frac{m_S^4}{8m_{\mu}^4}(4\log\frac{m_{\mu}}{m_S}+3)+\frac{5\pi m_S^5}{64m_{\mu}^5}-\frac{m_S^6}{16m_{\mu}^6},& \hspace{5ex}\mathrm{for}~m_S\ll m_\mu
\end{array}\right..
\end{align}

\subsubsection{Scenario of $m_\mu=m_S$}\label{sec:expansion:ISdeg:mmumS}
For $m_\mu=m_S$, the integrals $I_{LL}^S$ and $I_{LR}^S$ can be calculated as
\begin{align}\label{eqn:expansion:ISLL:mmumS}
&I_{LL}^S=\nonumber\\
&\left\{\begin{array}{ll}
-\frac{1}{4}+\frac{m_f^2}{2m_{\mu}^2}+\frac{m_f^2}{2m_{\mu}^2}(2-\frac{m_f^2}{m_{\mu}^2})\log\frac{m_f}{m_{\mu}}+\frac{m_f}{2\sqrt{m_f^2-4m_{\mu}^2}}(2-\frac{4m_f^2}{m_{\mu}^2}+\frac{m_f^4}{m_{\mu}^4})\log\frac{m_f+\sqrt{m_f^2-4m_{\mu}^2}}{2m_{\mu}},& \mathrm{for}~m_f>2m_{\mu}\\
-\frac{1}{4}+\frac{m_f^2}{2m_{\mu}^2}+\frac{m_f^2}{2m_{\mu}^2}(2-\frac{m_f^2}{m_{\mu}^2})\log\frac{m_f}{m_{\mu}}+\frac{m_f}{2\sqrt{4m_{\mu}^2-m_f^2}}(2-\frac{4m_f^2}{m_{\mu}^2}+\frac{m_f^4}{m_{\mu}^4})\arctan\sqrt{\frac{4m_{\mu}^2}{m_f^2}-1},& \mathrm{for}~m_f<2m_\mu\\
\frac{11}{4}-4\log2,& \mathrm{for}~m_f=2m_\mu
\end{array}\right.
\end{align}
and
\begin{align}\label{eqn:expansion:ISLR:mmumS}
I_{LR}^S=\left\{\begin{array}{ll}
\frac{m_f}{2m_{\mu}}[1+(1-\frac{m_f^2}{m_{\mu}^2})\log\frac{m_f}{m_{\mu}}]+\frac{m_f^2}{2m_{\mu}\sqrt{m_f^2-4m_{\mu}^2}}(\frac{m_f^2}{m_{\mu}^2}-3)\log\frac{m_f+\sqrt{m_f^2-4m_{\mu}^2}}{2m_{\mu}},& \mathrm{for}~m_f>2m_{\mu}\\
\frac{m_f}{2m_{\mu}}[1+(1-\frac{m_f^2}{m_{\mu}^2})\log\frac{m_f}{m_{\mu}}]+\frac{m_f^2}{2m_{\mu}\sqrt{4m_{\mu}^2-m_f^2}}(\frac{m_f^2}{m_{\mu}^2}-3)\arctan\sqrt{\frac{4m_{\mu}^2}{m_f^2}-1},& \mathrm{for}~m_f<2m_\mu\\
2-3\log2,& \mathrm{for}~m_f=2m_\mu
\end{array}\right. .
\end{align}
Then, they can be expanded as
\begin{align}\label{eqn:expansion:ISLL:mmumS:esp}
I_{LL}^S\approx\left\{\begin{array}{ll}
-\frac{m_{\mu}^2}{6m_f^2}+\frac{m_{\mu}^4}{24m_f^4}(24\log\frac{m_f}{m_{\mu}}-25)+\frac{m_{\mu}^6}{20m_f^6}(120\log\frac{m_f}{m_{\mu}}-97),& \hspace{5ex}\mathrm{for}~m_\mu\ll m_f\vspace{1ex}\\
-\frac{1}{4}+\frac{\pi m_f}{4m_{\mu}}+\frac{m_f^2}{4m_{\mu}^2}(1-4\log\frac{m_{\mu}}{m_f})-\frac{15\pi m_f^3}{32m_{\mu}^3}+\frac{m_f^4}{24m_{\mu}^4}(12\log\frac{m_{\mu}}{m_f}+11),& \hspace{5ex}\mathrm{for}~m_f\ll m_\mu
\end{array}\right.
\end{align}
and
\begin{align}\label{eqn:expansion:ISLR:mmumS:esp}
I_{LR}^S\approx\left\{\begin{array}{ll}
-\frac{m_{\mu}}{4m_f}+\frac{m_{\mu}^3}{12m_f^3}(12\log\frac{m_f}{m_{\mu}}-11)+\frac{m_{\mu}^5}{24m_f^5}(120\log\frac{m_f}{m_{\mu}}-89),& \hspace{10ex}\mathrm{for}~m_\mu\ll m_f\vspace{1ex}\\
\frac{m_f}{2m_{\mu}}(1-\log\frac{m_{\mu}}{m_f})-\frac{3\pi m_f^2}{8m_{\mu}^2}+\frac{m_f^3}{8m_{\mu}^3}(4\log\frac{m_{\mu}}{m_f}+3)+\frac{5\pi m_f^4}{64m_{\mu}^4}-\frac{m_f^5}{16m_{\mu}^5},& \hspace{10ex}\mathrm{for}~m_f\ll m_\mu
\end{array}\right..
\end{align}

\subsubsection{Scenario of $m_f=m_S$}\label{sec:expansion:ISdeg:mfmS}
For $m_f=m_S$, the integrals $I_{LL}^S$ and $I_{LR}^S$ can be calculated as
\begin{align}\label{eqn:expansion:ISLL:mfmS}
I_{LL}^S=\left\{\begin{array}{ll}
\frac{m_S^2}{4m_{\mu}^2}(1-\frac{4m_S^2}{m_{\mu}\sqrt{4m_S^2-m_{\mu}^2}}\arctan\frac{m_{\mu}}{\sqrt{4m_S^2-m_{\mu}^2}}),&\mathrm{for}~m_\mu<2m_S\\
\frac{m_S^2}{4m_{\mu}^2}[1+\frac{4m_S^2}{m_{\mu}\sqrt{m_{\mu}^2-4m_S^2}}(\log\frac{m_{\mu}+\sqrt{m_{\mu}^2-4m_S^2}}{2m_S}-\frac{i\pi}{2})],&\mathrm{for}~m_\mu>2m_S\\
\frac{1}{8},&\mathrm{for}~m_\mu=2m_S
\end{array}\right.
\end{align}
and
\begin{small}\label{eqn:expansion:ISLR:mfmS}
\begin{align}
I_{LR}^S=\left\{\begin{array}{ll}
\frac{m_S^3}{2m_{\mu}^3}(1-\frac{4m_S^2}{m_{\mu}\sqrt{4m_S^2-m_{\mu}^2}}\arctan\frac{m_{\mu}}{\sqrt{4m_S^2-m_{\mu}^2}}),&\mathrm{for}~m_\mu<2m_S\\
\frac{m_S^3}{2m_{\mu}^3}[1+\frac{4m_S^2}{m_{\mu}\sqrt{m_{\mu}^2-4m_S^2}}(\log\frac{m_{\mu}+\sqrt{m_{\mu}^2-4m_S^2}}{2m_S}-\frac{i\pi}{2})],&\mathrm{for}~m_\mu>2m_S\\
\frac{1}{8},&\mathrm{for}~m_\mu=2m_S
\end{array}\right..
\end{align}
\end{small}
Then, they can be expanded as
\begin{align}\label{eqn:expansion:ISLL:mfmS:esp}
I_{LL}^S\approx\left\{\begin{array}{ll}
-\frac{1}{24}(1+\frac{m_{\mu}^2}{5m_S^2}+\frac{3m_{\mu}^4}{70m_S^4}),& \hspace{5ex}\mathrm{for}~m_\mu\ll m_S\vspace{1ex}\\
\frac{m_S^2}{4m_{\mu}^2}\big[1+\frac{2m_S^2}{m_{\mu}^2}(2\log\frac{m_{\mu}}{m_S}-i\pi)+\frac{4m_S^4}{m_{\mu}^4}(2\log\frac{m_{\mu}}{m_S}-i\pi-1)\big],& \hspace{5ex}\mathrm{for}~m_S\ll m_\mu
\end{array}\right.
\end{align}
and
\begin{align}\label{eqn:expansion:ISLR:mfmS:esp}
I_{LR}^S\approx\left\{\begin{array}{ll}
-\frac{m_S}{12m_{\mu}}(1+\frac{m_{\mu}^2}{5m_S^2}+\frac{3m_{\mu}^4}{70m_S^4}),& \hspace{10ex}\mathrm{for}~m_\mu\ll m_S\vspace{1ex}\\
\frac{m_S^3}{2m_{\mu}^3}\big[1+\frac{2m_S^2}{m_{\mu}^2}(2\log\frac{m_{\mu}}{m_S}-i\pi)+\frac{4m_S^4}{m_{\mu}^4}(2\log\frac{m_{\mu}}{m_S}-i\pi-1)\big],& \hspace{10ex}\mathrm{for}~m_S\ll m_\mu
\end{array}\right..
\end{align}

\subsubsection{Scenario of $m_f=m_\mu=m_S$}
In this case, we have $I_{LL}^S=\frac{9-2\sqrt{3}\pi}{36}$ and
$I_{LR}^S=\frac{9-2\sqrt{3}\pi}{18}$.

\subsection{Short summary of the partial expansion results}\label{sec:Sexpansion:summary}
\subsubsection{Asymptotic behaviour of loop functions}
\begin{itemize}[leftmargin=10pt]
\item\textbf{Chiral limit of $m_{\mu}\rightarrow0$}\\
As $m_{\mu}$ goes to zero, the $\frac{m_\mu^2}{m_S^2}I_{LL}^f$, $\frac{m_\mu^2}{m_S^2}I_{LR}^f$, $\frac{m_\mu^2}{m_S^2}I_{LL}^S$, and $\frac{m_\mu^2}{m_S^2}I_{LR}^S$ vanish, which is expected from chiral symmetry.
\item\textbf{Chiral limit of $m_f\rightarrow0$}\\
As $m_f$ goes to zero, the $I_{LR}^S$ vanishes. As $m_f$ goes to zero, the $I_{LR}^f$ also vanishes except for the $m_\mu=m_S$ case.
\item\textbf{Limit of $m_S\rightarrow0$}\\
As $m_S$ goes to zero, there are non-zero contributions.
\item\textbf{Decoupling limit}\\
In the limit of $m_f\rightarrow\infty$ and $m_S\rightarrow\infty$, the $\frac{m_\mu^2}{m_S^2}I_{LL}^f$, $\frac{m_\mu^2}{m_S^2}I_{LR}^f$, $\frac{m_\mu^2}{m_S^2}I_{LL}^S$, and $\frac{m_\mu^2}{m_S^2}I_{LR}^S$ vanish, which is consistent with the spirit of decoupling theorem \cite{Appelquist:1974tg}. 
\end{itemize}

In Tab. \ref{tab:sum:asym:scalar}, we show the behaviour of $\frac{m_\mu^2}{m_S^2}I_{LL(LR)}^{f(S)}$ in the massless and infinite mass limits.
\begin{table}[!htb]
\begin{center}
\begin{tabular}{c|c|c|c}
\hline
\multicolumn{2}{c|}{} & \rule[-10pt]{0pt}{25pt}$\ds\frac{m_\mu^2}{m_S^2}I_{LL}^f$ & $\ds\frac{m_\mu^2}{m_S^2}I_{LR}^f$ \\
\hline
\multicolumn{2}{c|}{\rule[-9pt]{0pt}{20pt}Chiral limit $m_{\mu}\rightarrow0$} & 0 & 0 \\\hline
\multirow{2}{*}{\rule[-9pt]{0pt}{28pt}\makecell{Chiral limit \\$m_f\rightarrow0$}} & \rule[-10pt]{0pt}{25pt}$m_{\mu}\ne m_S$ & $\ds-\frac{1}{4}-\frac{m_S^2}{2m_{\mu}^2}-\frac{m_S^4}{2m_{\mu}^4}\log(1-\frac{m_{\mu}^2}{m_S^2})$ & 0 \\\cline{2-4}
 \rule[-10pt]{0pt}{25pt}& $m_{\mu}=m_S$ & $\ds-\frac{3}{4}+\frac{1}{2}\log\frac{m_{\mu}}{m_f}$ & $\ds\frac{\pi}{4}$ \\\hline
\multirow{2}{*}{\rule[-9pt]{0pt}{28pt}\makecell{Limit of \\$m_S\rightarrow0$}} & \rule[-11pt]{0pt}{26pt}$m_{\mu}\ne m_f$ & $\ds-\frac{1}{4}+\frac{m_f^2}{2m_\mu^2}+\frac{m_f^2(m_f^2-m_\mu^2)}{2m_\mu^4}\log(1-\frac{m_\mu^2}{m_f^2})$ & $\ds\frac{m_f}{2m_\mu}\big[1+(\frac{m_f^2}{m_\mu^2}-1)\log(1-\frac{m_\mu^2}{m_f^2})\big]$ \\\cline{2-4}
\rule[-10pt]{0pt}{25pt} & $m_{\mu}=m_f$ & $\ds\frac{1}{4}$ & $\ds\frac{1}{2}$ \\\hline
\multicolumn{2}{c|}{\rule[-9pt]{0pt}{22pt}\makecell{Decoupling limit\\ $m_f\rightarrow\infty$}} & 0 & 0 \\\hline
\multicolumn{2}{c|}{\rule[-9pt]{0pt}{22pt}\makecell{Decoupling limit\\ $m_S\rightarrow\infty$}} & 0 & 0 \\\hline\hline\hline
\multicolumn{2}{c|}{} & \rule[-10pt]{0pt}{25pt}$\ds\frac{m_\mu^2}{m_S^2}I_{LL}^S$ & $\ds\frac{m_\mu^2}{m_S^2}I_{LR}^S$ \\
\hline
\multicolumn{2}{c|}{\rule[-9pt]{0pt}{20pt}Chiral limit $m_{\mu}\rightarrow0$} & 0 & 0 \\\hline
\multirow{2}{*}{\rule[-9pt]{0pt}{28pt}\makecell{Chiral limit \\$m_f\rightarrow0$}} & \rule[-10pt]{0pt}{25pt}$m_{\mu}\ne m_S$ & $\ds\frac{1}{4}-\frac{m_S^2}{2m_{\mu}^2}+\frac{m_S^2(m_{\mu}^2-m_S^2)}{2m_{\mu}^4}\log(1-\frac{m_{\mu}^2}{m_S^2})$ & 0 \\\cline{2-4}
\rule[-9pt]{0pt}{22pt} & $m_{\mu}=m_S$ & $\ds-\frac{1}{4}$ & 0 \\\hline
\multirow{2}{*}{\rule[-9pt]{0pt}{28pt}\makecell{Limit of \\$m_S\rightarrow0$}} & \rule[-10pt]{0pt}{25pt}$m_{\mu}\ne m_f$ & $\ds\frac{1}{4}+\frac{m_f^2}{2m_\mu^2}+\frac{m_f^4}{2m_\mu^4}\log(1-\frac{m_\mu^2}{m_f^2})$ & $\ds\frac{m_f}{2m_\mu}\big[1+\frac{m_f^2}{m_\mu^2}\log(1-\frac{m_\mu^2}{m_f^2})\big]$ \\\cline{2-4}
\rule[-10pt]{0pt}{25pt} & $m_{\mu}=m_f$ & $\ds\frac{1}{4}(3-2\log\frac{m_{\mu}}{m_S})$ & $\ds\frac{1}{2}(1-\log\frac{m_{\mu}}{m_S})$ \\\hline
\multicolumn{2}{c|}{\rule[-9pt]{0pt}{22pt}\makecell{Decoupling limit\\ $m_f\rightarrow\infty$}} & 0 & 0 \\\hline
\multicolumn{2}{c|}{\rule[-9pt]{0pt}{22pt}\makecell{Decoupling limit\\ $m_S\rightarrow\infty$}} & 0 & 0 \\\hline
\end{tabular}
\caption{The behaviour of $\ds\frac{m_\mu^2}{m_S^2}I_{LL(LR)}^f$ (upper table) and $\ds\frac{m_\mu^2}{m_S^2}I_{LL(LR)}^S$ (lower table) for the scalar mediator case under different mass limits. In the above scenario of $m_{\mu}\ne m_i(i=S,f)$, the $\ds\log(1-\frac{m_\mu^2}{m_i^2})$ is valid for $m_{\mu}<m_i$, which should be taken as $\ds[\log(\frac{m_\mu^2}{m_i^2}-1)-i\pi]$ for $m_{\mu}>m_i$.} \label{tab:sum:asym:scalar}
\end{center}
\end{table}

\subsubsection{Expansion of contributions to $(g-2)_\mu$ under special scenarios}
In Tab. \ref{tab:sum:exp:scalar}, we list the leading order formulae of the $\Delta a_{\mu}$ for the scalar mediator cases in different scenarios. Here, we only collect the results for special scenarios. For complete expressions, please refer to Sec. \ref{sec:model:contri:ScalarRep}, Sec. \ref{sec:expansion:If}, Sec. \ref{sec:expansion:Ifdeg}, Sec. \ref{sec:expansion:IS}, and Sec. \ref{sec:expansion:ISdeg}.

\FloatBarrier
\begin{table}[!htb]
\begin{center}
\begin{tabular}{c|c}
\hline
\rule[-10pt]{0pt}{25pt} scenarios & $\ds\Delta a_{\mu}/(\frac{N_Cm_{\mu}^2}{8\pi^2m_S^2})$ \\
\hline
\rule[-10pt]{0pt}{25pt} $m_\mu\ll m_f\ll m_S$ & $\makecell{\ds(|y_L|^2+|y_R|^2)(-\frac{1}{6}Q_f+\frac{1}{12}Q_S)+\left[y_L(y_R)^\ast+y_R(y_L)^\ast\right]\cdot\frac{m_f}{m_{\mu}}\big[-(\log\frac{m_S}{m_f}-\frac{3}{4})Q_f+\frac{1}{4}Q_S\big]}$ \\
\hline
\rule[-10pt]{0pt}{25pt} $m_\mu\ll m_S\ll m_f$ & $\makecell{\ds(|y_L|^2+|y_R|^2)\cdot\frac{m_S^2}{m_f^2}(-\frac{1}{12}Q_f+\frac{1}{6}Q_S)+\left[y_L(y_R)^\ast+y_R(y_L)^\ast\right]\cdot\frac{m_S^2}{4m_fm_{\mu}}(-Q_f+Q_S)}$ \\
\hline
\rule[-20pt]{0pt}{45pt} $m_S\ll m_f\ll m_\mu$ & $\makecell{\ds(|y_L|^2+|y_R|^2)\cdot\frac{m_S^2}{4m_{\mu}^2}(Q_f-Q_S)\\\ds+\left[y_L(y_R)^\ast+y_R(y_L)^\ast\right]\cdot\frac{m_S^2m_f}{2m_{\mu}^3}[(2\log\frac{m_{\mu}}{m_f}-i\pi-1)Q_f-Q_S]}$ \\
\hline
\rule[-10pt]{0pt}{25pt} $m_S\ll m_\mu\ll m_f$ &  $\makecell{\ds(|y_L|^2+|y_R|^2)\cdot\frac{m_S^2}{m_f^2}(-\frac{1}{12}Q_f+\frac{1}{6}Q_S)+\left[y_L(y_R)^\ast+y_R(y_L)^\ast\right]\cdot\frac{m_S^2}{4m_{\mu}m_f}(-Q_f+Q_S)}$ \\
\hline
\rule[-10pt]{0pt}{25pt} $m_f\ll m_\mu\ll m_S$ & $\makecell{\ds(|y_L|^2+|y_R|^2)(-\frac{1}{6}Q_f+\frac{1}{12}Q_S)+\left[y_L(y_R)^\ast+y_R(y_L)^\ast\right]\cdot\frac{m_f}{m_{\mu}}[-(\log\frac{m_S}{m_f}-\frac{3}{4})Q_f+\frac{1}{4}Q_S]}$ \\
\hline
\rule[-20pt]{0pt}{45pt} $m_f\ll m_S\ll m_\mu$ & $\makecell{\ds(|y_L|^2+|y_R|^2)\cdot\frac{m_S^2}{4m_{\mu}^2}(Q_f-Q_S)\\\ds+\left[y_L(y_R)^\ast+y_R(y_L)^\ast\right]\cdot\frac{m_S^2m_f}{2m_{\mu}^3}[(2\log\frac{m_{\mu}}{m_f}-i\pi-1)Q_f-Q_S]}$ \\\hline\hline\hline
\rule[-20pt]{0pt}{45pt} $m_S\ll m_f=m_\mu$ & $\makecell{\ds(|y_L|^2+|y_R|^2)\cdot\frac{m_S^2}{4m_{\mu}^2}[-Q_f+(2\log\frac{m_{\mu}}{m_S}-3)Q_S]\\\ds+\left[y_L(y_R)^\ast+y_R(y_L)^\ast\right]\cdot\frac{m_S^2}{2m_{\mu}^2}[-Q_f+(\log\frac{m_{\mu}}{m_S}-1)Q_S]}$ \\
\hline
\rule[-10pt]{0pt}{25pt} $m_f=m_\mu\ll m_S$ & $\makecell{\ds(|y_L|^2+|y_R|^2)(-\frac{1}{6}Q_f+\frac{1}{12}Q_S)+\left[y_L(y_R)^\ast+y_R(y_L)^\ast\right]\cdot[-(\log\frac{m_S}{m_{\mu}}-\frac{3}{4})Q_f+\frac{1}{4}Q_S]}$ \\
\hline
\rule[-20pt]{0pt}{45pt} $m_f\ll m_\mu=m_S$ & $\makecell{\ds(|y_L|^2+|y_R|^2)\cdot[(\frac{3}{4}-\frac{1}{2}\log\frac{m_{\mu}}{m_f})Q_f+\frac{1}{4}Q_S]\vspace{1ex}\\\ds+\left[y_L(y_R)^\ast+y_R(y_L)^\ast\right]\cdot[-\frac{\pi}{4}Q_f+\frac{m_f}{2m_{\mu}}(\log\frac{m_{\mu}}{m_f}-1)Q_S]}$  \\
\hline
\rule[-10pt]{0pt}{25pt} $m_\mu=m_S\ll m_f$ & $\makecell{\ds(|y_L|^2+|y_R|^2)\cdot\frac{m_{\mu}^2}{m_f^2}(-\frac{1}{12}Q_f+\frac{1}{6}Q_S)+\left[y_L(y_R)^\ast+y_R(y_L)^\ast\right]\cdot\frac{m_{\mu}}{4m_f}(-Q_f+Q_S)}$ \\
\hline
\rule[-10pt]{0pt}{25pt} $m_\mu\ll m_f=m_S$ & $\makecell{\ds(|y_L|^2+|y_R|^2)\cdot\frac{1}{24}(-Q_f+Q_S)+\left[y_L(y_R)^\ast+y_R(y_L)^\ast\right]\cdot\frac{m_S}{m_{\mu}}(-\frac{1}{6}Q_f+\frac{1}{12}Q_S)}$ \\
\hline
\rule[-20pt]{0pt}{45pt} $m_f=m_S\ll m_\mu$ & $\makecell{\ds(|y_L|^2+|y_R|^2)\cdot\frac{m_S^2}{4m_{\mu}^2}(Q_f-Q_S)\\\ds+\left[y_L(y_R)^\ast+y_R(y_L)^\ast\right]\cdot\frac{m_S^3}{2m_{\mu}^3}[(2\log\frac{m_{\mu}}{m_S}-1-i\pi)Q_f-Q_S]}$ \\
\hline\hline
\rule[-20pt]{0pt}{45pt} $m_f=m_\mu=m_S$ & $\makecell{\ds(|y_L|^2+|y_R|^2)\cdot\frac{2\sqrt{3}\pi-9}{36}(-Q_f+Q_S)\\\ds+\left[y_L(y_R)^\ast+y_R(y_L)^\ast\right]\cdot(-\frac{9-\sqrt{3}\pi}{18}Q_f+\frac{2\sqrt{3}\pi-9}{18}Q_S)}$ \\
\hline
\end{tabular}
\caption{Leading order formulae of the $\Delta a_{\mu}$ for the scalar mediator case in different scenarios. These results are consistent with those in Ref. \cite{Queiroz:2014zfa}. Note that we have extracted the common factor $\ds\frac{N_Cm_{\mu}^2}{8\pi^2m_S^2}$. As aforementioned, the charge conservation requires $Q_f+Q_S=-1$.} \label{tab:sum:exp:scalar}
\end{center}
\end{table}
\FloatBarrier

Generally, the left-handed or right-handed Yukawa couplings are independent. The $(|y_L|^2+|y_R|^2)$ is always non-negative, while the sign of $\left[y_L(y_R)^\ast+y_R(y_L)^\ast\right]$ is indefinite. If both the left-handed and right-handed Yukawa couplings are present, it is possible to obtain the positive or negative contributions through adjusting the sign of $y_L$ and $y_R$. If either the left-handed or the right-handed Yukawa coupling is absent, the loop functions must be positive (negative) to obtain the positive (negative) contributions. In Tab. \ref{tab:sum:exp:scalarLL}, we list the leading order formulae of the $\Delta a_{\mu}$ with pure left-handed or right-handed Yukawa couplings for the scalar mediator cases in different scenarios. Furthermore, we give the electric charge bounds required by positive $\Delta a_{\mu}$.
\begin{table}[!htb]
\begin{center}
\begin{tabular}{c|c|c}
\hline
\rule[-10pt]{0pt}{25pt} scenarios & $\ds\Delta a_{\mu}/(\frac{N_Cm_{\mu}^2}{8\pi^2m_S^2})$ & condition of $\Delta a_{\mu}>0$ \\ 
\hline
\rule[-10pt]{0pt}{25pt} $m_\mu\ll m_f\ll m_S$ & $\ds|y_{L(R)}|^2\cdot(-\frac{1}{6}Q_f+\frac{1}{12}Q_S)$ & $\ds Q_f<-\frac{1}{3}$ or $\ds Q_S>-\frac{2}{3}$\\
\hline
\rule[-10pt]{0pt}{25pt} $m_\mu\ll m_S\ll m_f$ & $\ds|y_{L(R)}|^2\cdot\frac{m_S^2}{m_f^2}(-\frac{1}{12}Q_f+\frac{1}{6}Q_S)$ & $\ds Q_f<-\frac{2}{3}$ or $\ds Q_S>-\frac{1}{3}$ \\
\hline
\rule[-10pt]{0pt}{25pt} $m_S\ll m_f\ll m_\mu$ & $\ds|y_{L(R)}|^2\cdot\frac{m_S^2}{4m_{\mu}^2}(Q_f-Q_S)$ &  $\ds Q_f>-\frac{1}{2}$ or $\ds Q_S<-\frac{1}{2}$\\
\hline
\rule[-10pt]{0pt}{25pt} $m_S\ll m_\mu\ll m_f$ &  $\ds|y_{L(R)}|^2\cdot\frac{m_S^2}{m_f^2}(-\frac{1}{12}Q_f+\frac{1}{6}Q_S)$ & $\ds Q_f<-\frac{2}{3}$ or $\ds Q_S>-\frac{1}{3}$ \\
\hline
\rule[-10pt]{0pt}{25pt} $m_f\ll m_\mu\ll m_S$ & $\ds|y_{L(R)}|^2\cdot(-\frac{1}{6}Q_f+\frac{1}{12}Q_S)$ & $\ds Q_f<-\frac{1}{3}$ or $\ds Q_S>-\frac{2}{3}$ \\
\hline
\rule[-10pt]{0pt}{25pt} $m_f\ll m_S\ll m_\mu$ & $\ds|y_{L(R)}|^2\cdot\frac{m_S^2}{4m_{\mu}^2}(Q_f-Q_S)$ & $\ds Q_f>-\frac{1}{2}$ or $\ds Q_S<-\frac{1}{2}$ \\\hline\hline\hline
\rule[-10pt]{0pt}{25pt} $m_S\ll m_f=m_\mu$ & $\ds|y_{L(R)}|^2\cdot\frac{m_S^2}{4m_{\mu}^2}[-Q_f+(2\log\frac{m_{\mu}}{m_S}-3)Q_S]$ & $\ds Q_f\lesssim-1$ or $\ds Q_S\gtrsim0$\\
\hline
\rule[-10pt]{0pt}{25pt} $m_f=m_\mu\ll m_S$ & $\ds|y_{L(R)}|^2\cdot(-\frac{1}{6}Q_f+\frac{1}{12}Q_S)$ & $\ds Q_f<-\frac{1}{3}$ or $\ds Q_S>-\frac{2}{3}$ \\
\hline
\rule[-10pt]{0pt}{25pt} $m_f\ll m_\mu=m_S$ & $\ds|y_{L(R)}|^2\cdot[(\frac{3}{4}-\frac{1}{2}\log\frac{m_{\mu}}{m_f})Q_f+\frac{1}{4}Q_S]$ & $\ds Q_f\lesssim0$ or $\ds Q_S\gtrsim-1$\\
\hline
\rule[-10pt]{0pt}{25pt} $m_\mu=m_S\ll m_f$ & $\ds|y_{L(R)}|^2\cdot\frac{m_{\mu}^2}{m_f^2}(-\frac{1}{12}Q_f+\frac{1}{6}Q_S)$ & $\ds Q_f<-\frac{2}{3}$ or $\ds Q_S>-\frac{1}{3}$ \\
\hline
\rule[-10pt]{0pt}{25pt} $m_\mu\ll m_f=m_S$ & $\ds|y_{L(R)}|^2\cdot\frac{1}{24}(-Q_f+Q_S)$ & $\ds Q_f<-\frac{1}{2}$ or $\ds Q_S>-\frac{1}{2}$\\
\hline
\rule[-10pt]{0pt}{25pt} $m_f=m_S\ll m_\mu$ & $\ds|y_{L(R)}|^2\cdot\frac{m_S^2}{4m_{\mu}^2}(Q_f-Q_S)$ & $\ds Q_f>-\frac{1}{2}$ or $\ds Q_S<-\frac{1}{2}$ \\
\hline\hline
\rule[-10pt]{0pt}{25pt} $m_f=m_\mu=m_S$ & $\ds|y_{L(R)}|^2\cdot\frac{2\sqrt{3}\pi-9}{36}(-Q_f+Q_S)$ & $\ds Q_f<-\frac{1}{2}$ or $\ds Q_S>-\frac{1}{2}$\\
\hline
\end{tabular}
\caption{Leading order formulae of the $\Delta a_{\mu}$ with pure left-handed (right-handed) couplings for the scalar mediator case in different scenarios. Note that we have extracted the common factor $\ds\frac{N_Cm_{\mu}^2}{8\pi^2m_S^2}$. As aforementioned, the charge conservation requires $Q_f+Q_S=-1$.} \label{tab:sum:exp:scalarLL}
\end{center}
\end{table}


\section{Expansion of the loop functions: vector mediator case}\label{sec:expansionV}

In Sec. \ref{sec:model:contri:VectorRep}, we have exhibited the analytic forms of loop functions for the vector mediator case, including the PaVe representation, integral representation, and special function representation. Similarly, this section is dedicated to the expansion of loop functions. Again, we consider six hierarchical scenarios first: 
\begin{align}
&m_\mu\ll m_f,m_V;\qquad m_V\ll m_f,m_\mu;\qquad m_f\ll m_\mu,m_V; \nonumber\\
&m_f,m_V\ll m_\mu;\qquad m_f,m_\mu\ll m_V;\qquad m_\mu,m_V\ll m_f. 
\end{align}
Second, we also study the four degenerate scenarios:
\begin{align}
m_f=m_\mu,\qquad m_\mu=m_V,\qquad m_f=m_V,\qquad m_f=m_\mu=m_V.
\end{align}
Similar to the scalar mediator case, we retain the whole scenarios for completeness without delving into the possible UV realization for each scenario.

\subsection{Hierarchical mass expansion for the $L_{LL}^f$ and $L_{LR}^f$}\label{sec:expansion:Lf}
\subsubsection{Scenario of $m_\mu\ll m_f,m_V$}\label{sec:expansion:Lfhie:mmullmfmV}

Up to $m_{\mu}^4$, the integral $L_{LL}^f$ can be expanded as
\begin{align}\label{eqn:expansion:LfLL:mmullmfmV}
&L_{LL}^f=\frac{1}{12(m_V^2-m_f^2)^4}\big[(m_V^2-m_f^2)(5m_f^6-9m_f^4m_V^2+30m_f^2m_V^4-8m_V^6)+18m_f^4m_V^4\log\frac{m_f^2}{m_V^2}\big]\nonumber\\
&+\frac{m_\mu^2}{24(m_V^2-m_f^2)^6}\big[(m_V^2-m_f^2)(m_f^8+3m_f^6m_V^2+159m_f^4m_V^4+23m_f^2m_V^6-6m_V^8)\nonumber\\
	&+12m_f^2m_V^4(6m_f^4+10m_f^2m_V^2-m_V^4)\log\frac{m_f^2}{m_V^2}\big]\nonumber\\
	&+\frac{m_\mu^4}{120(m_V^2-m_f^2)^8}\big[(m_V^2-m_f^2)(2m_f^{10}+16m_f^8m_V^2+1551m_f^6m_V^4+2311m_f^4m_V^6-79m_f^2m_V^8-21m_V^{10})\nonumber\\
	&+60m_f^2m_V^4(9m_f^6+40m_f^4m_V^2+16m_f^2m_V^4-2m_V^6)\log\frac{m_f^2}{m_V^2}\big]+\mathcal{O}(m_\mu^6).
\end{align}
Up to $m_{\mu}^3$, the integral $L_{LR}^f$ can be expanded as
\begin{align}\label{eqn:expansion:LfLR:mmullmfmV}
&L_{LR}^f=\frac{m_f}{m_\mu}\Big\{\frac{1}{4(m_f^2-m_V^2)^3}\big[m_f^6+3m_f^2m_V^4-4m_V^6-6m_f^2m_V^4\log\frac{m_f^2}{m_V^2}\big]+\frac{m_\mu^2}{12(m_f^2-m_V^2)^5}\cdot\nonumber\\
	&\big[(m_f^2-m_V^2)(2m_f^6-3m_f^4m_V^2+78m_f^2m_V^4-5m_V^6)+6m_V^4(m_V^4-7m_f^2m_V^2-6m_f^4)\log\frac{m_f^2}{m_V^2}\big]\nonumber\\
&+\frac{m_\mu^4}{24(m_f^2-m_V^2)^7}\big[(m_f^4-m_V^4)(m_f^6+297m_f^2m_V^4-28m_V^6)\nonumber\\
	&+12m_V^4(-9m_f^6-31m_f^4m_V^2-6m_f^2m_V^4+m_V^6)\log\frac{m_f^2}{m_V^2}\big]\Big\}+\mathcal{O}(m_\mu^5).
\end{align}

Especially, we have the following approximations:
\begin{align}\label{eqn:expansion:LfLL:mmullmfmV:esp}
L_{LL}^f\approx\left\{\begin{array}{ll}
-\frac{2}{3}(1-\frac{3m_f^2}{4m_V^2}),& \mathrm{for}~m_\mu\ll m_f\ll m_V\vspace{1ex}\\
-\frac{5}{12}(1+\frac{6m_V^2}{5m_f^2}),& \mathrm{for}~m_\mu\ll m_V\ll m_f
\end{array}\right.
\end{align}
and
\begin{align}\label{eqn:expansion:LfLR:mmullmfmV:esp}
L_{LR}^f\approx\left\{\begin{array}{ll}
\frac{m_f}{m_\mu}\Big[1+\frac{3m_f^2}{4m_V^2}(3-4\log\frac{m_V}{m_f})\Big],& \mathrm{for}~m_\mu\ll m_f\ll m_V\vspace{1ex}\\
\frac{m_f}{4m_\mu}(1+\frac{3m_V^2}{m_f^2}),& \mathrm{for}~m_\mu\ll m_V\ll m_f
\end{array}\right. .
\end{align}

For the scenario of $m_\mu\ll m_f=m_V$, please refer to Eqs. \eqref{eqn:expansion:LfLL:mfmV:esp} and \eqref{eqn:expansion:LfLR:mfmV:esp}.
\subsubsection{Scenario of $m_V\ll m_f,m_\mu$}\label{sec:expansion:Lfhie:mVllmfmmu}

Up to $m_V^4$, the integrals $L_{LL}^f$ and $L_{LR}^f$ can be expanded as
\begin{align}\label{eqn:expansion:LfLL:mVllmfmmu}
&L_{LL}^f=\left\{\begin{array}{ll}
\frac{1}{4m_\mu^6}\big[m_\mu^2(2m_f^4-3m_f^2m_\mu^2-m_\mu^4)+2m_f^2(m_f^2-m_\mu^2)^2\log(1-\frac{m_\mu^2}{m_f^2})\big]+\frac{m_V^2}{m_\mu^4}\big[m_\mu^2+m_f^2\log(1-\frac{m_\mu^2}{m_f^2})\big]+&\vspace{1ex}\\
	\frac{m_V^4}{4m_\mu^6(m_f^2-m_\mu^2)^2}\big[m_\mu^2(3m_f^2m_\mu^2-6m_f^4-4m_\mu^4)+(6m_f^4m_\mu^2+6m_f^2m_\mu^4-6m_f^6+6m_\mu^6)\log(1-\frac{m_\mu^2}{m_f^2})&\vspace{1ex}\\
	+6m_\mu^6\log\frac{m_f^2}{m_V^2}\big]+\mathcal{O}(m_V^6), \hspace{45ex}\mathrm{for}~m_\mu<m_f&\vspace{2ex}\\
\frac{1}{4m_\mu^6}\big[m_\mu^2(2m_f^4-3m_f^2m_\mu^2-m_\mu^4)+2m_f^2(m_f^2-m_\mu^2)^2(\log\frac{m_\mu^2-m_f^2}{m_f^2}-i\pi)\big]&\vspace{1ex}\\
	+\frac{m_V^2}{m_\mu^4}\big[m_\mu^2+m_f^2(\log\frac{m_\mu^2-m_f^2}{m_f^2}-i\pi)\big]+\frac{m_V^4}{4m_\mu^6(m_f^2-m_\mu^2)^2}\big[m_\mu^2(3m_f^2m_\mu^2-6m_f^4-4m_\mu^4)&\vspace{1ex}\\
	+(6m_f^4m_\mu^2+6m_f^2m_\mu^4-6m_f^6+6m_\mu^6)(\log\frac{m_\mu^2-m_f^2}{m_f^2}-i\pi)+6m_\mu^6\log\frac{m_f^2}{m_V^2}\big]&\vspace{1ex}\\
	+\mathcal{O}(m_V^6), \hspace{60ex}\mathrm{for}~m_\mu>m_f&
\end{array}\right.
\end{align}
and
\begin{align}\label{eqn:expansion:LfLR:mVllmfmmu}
&L_{LR}^f=\left\{\begin{array}{ll}\frac{m_f}{m_\mu}\Big\{
\frac{1}{2m_\mu^4}\big[-m_f^2m_\mu^2+2m_\mu^4-(m_f^2-m_\mu^2)^2\log(1-\frac{m_\mu^2}{m_f^2})\big]+\frac{m_V^2}{2m_\mu^4}\big[-m_\mu^2-(m_f^2+m_\mu^2)\log(1-\frac{m_\mu^2}{m_f^2})\big]&\vspace{1ex}\\
	+\frac{m_V^4}{4m_\mu^4(m_f^2-m_\mu^2)^2}\big[4m_f^2m_\mu^2+3m_\mu^4+(4m_f^4-8m_f^2m_\mu^2-8m_\mu^4)\log(1-\frac{m_\mu^2}{m_f^2})-6m_\mu^4\log\frac{m_f^2}{m_V^2}\big]\Big\}&\vspace{1ex}\\
	+\mathcal{O}(m_V^6), \hspace{60ex}\mathrm{for}~m_\mu<m_f&\vspace{2ex}\\
\frac{m_f}{m_\mu}\Big\{\frac{1}{2m_\mu^4}\big[-m_f^2m_\mu^2+2m_\mu^4-(m_f^2-m_\mu^2)^2(\log\frac{m_\mu^2-m_f^2}{m_f^2}-i\pi)\big]&\vspace{1ex}\\
	+\frac{m_V^2}{2m_\mu^4}\big[-m_\mu^2-(m_f^2+m_\mu^2)(\log\frac{m_\mu^2-m_f^2}{m_f^2}-i\pi)\big]+\frac{m_V^4}{4m_\mu^4(m_f^2-m_\mu^2)^2}\big[4m_f^2m_\mu^2+3m_\mu^4&\vspace{1ex}\\
	+(4m_f^4-8m_f^2m_\mu^2-8m_\mu^4)(\log\frac{m_\mu^2-m_f^2}{m_f^2}-i\pi)-6m_\mu^4\log\frac{m_f^2}{m_V^2}\big]\Big\}&\vspace{1ex}\\
	+\mathcal{O}(m_V^6), \hspace{60ex}\mathrm{for}~m_\mu>m_f&
\end{array}\right..
\end{align}
Especially, we have the following approximations:
\begin{align}\label{eqn:expansion:LfLL:mVllmfmmu:esp}
L_{LL}^f\approx\left\{\begin{array}{ll}
-\frac{1}{4}\big[1+\frac{m_f^2}{m_\mu^2}(3-2\log\frac{m_\mu^2}{m_f^2}+2i\pi)\big],& \mathrm{for}~m_V\ll m_f\ll m_\mu\vspace{1ex}\\
-\frac{5}{12}(1+\frac{m_\mu^2}{10m_f^2}),& \mathrm{for}~m_V\ll m_\mu\ll m_f
\end{array}\right.
\end{align}
and
\begin{align}\label{eqn:expansion:LfLR:mVllmfmmu:esp}
L_{LR}^f\approx\left\{\begin{array}{ll}
\frac{m_f}{m_\mu}\big[1+\frac{1}{2}(-\log\frac{m_\mu^2}{m_f^2}+i\pi)+\frac{m_f^2}{m_\mu^2}(\log\frac{m_\mu^2}{m_f^2}-i\pi)\big],& \mathrm{for}~m_V\ll m_f\ll m_\mu\vspace{1ex}\\
\frac{m_f}{4m_\mu}(1+\frac{2m_\mu^2}{3m_f^2}),& \mathrm{for}~m_V\ll m_\mu\ll m_f
\end{array}\right. .
\end{align}

For the scenario of $m_V\ll m_f=m_\mu$, please refer to Eqs. \eqref{eqn:expansion:LfLL:mfmmu:esp} and \eqref{eqn:expansion:LfLR:mfmmu:esp}.
\subsubsection{Scenario of $m_f\ll m_\mu,m_V$}\label{sec:expansion:Lfhie:mfllmmumV}
Up to $m_f^4$, the integral $L_{LL}^f$ can be expanded as
\begin{align}\label{eqn:expansion:LfLL:mfllmmumV}
&L_{LL}^f=\left\{\begin{array}{ll}
\frac{1}{4m_\mu^6}\big[-m_\mu^2(m_\mu^2-2m_V^2)^2+2m_V^4(3m_\mu^2-2m_V^2)\log(1-\frac{m_\mu^2}{m_V^2})\big]+\frac{m_f^2}{4m_\mu^6(m_V^2-m_\mu^2)^2}\cdot\big[2m_\mu^8\log\frac{m_V^2}{m_f^2}&\vspace{1ex}\\
	+(6m_V^8-16m_\mu^2m_V^6+12m_\mu^4m_V^4+2m_\mu^8)\log(1-\frac{m_\mu^2}{m_V^2})+m_\mu^2(6m_V^6-13m_\mu^2m_V^4+8m_\mu^4m_V^2-3m_\mu^6)\big]&\vspace{1ex}\\
	+\frac{m_f^4}{4m_\mu^4(m_V^2-m_\mu^2)^4}\big[m_\mu^2m_V^2(4m_V^4-3m_\mu^2m_V^2-8m_\mu^4)+2m_\mu^4(8m_\mu^2m_V^2-3m_V^4-2m_\mu^4)\log\frac{m_V^2}{m_f^2}&\vspace{1ex}\\
	+(4m_V^8-16m_\mu^2m_V^6+12m_\mu^4m_V^4+16m_\mu^6m_V^2-4m_\mu^8)\log(1-\frac{m_\mu^2}{m_V^2})\big]+\mathcal{O}(m_f^6), \qquad\mathrm{for}~m_\mu<m_V&\vspace{4ex}\\
\frac{1}{4m_\mu^6}\big[-m_\mu^2(m_\mu^2-2m_V^2)^2+2m_V^4(3m_\mu^2-2m_V^2)(\log\frac{m_\mu^2-m_V^2}{m_V^2}-i\pi)\big]+\frac{m_f^2}{4m_\mu^6(m_V^2-m_\mu^2)^2}\cdot\big[2m_\mu^8\log\frac{m_V^2}{m_f^2}&\vspace{1ex}\\
	+(6m_V^8-16m_\mu^2m_V^6+12m_\mu^4m_V^4+2m_\mu^8)(\log\frac{m_\mu^2-m_V^2}{m_V^2}-i\pi)+m_\mu^2(6m_V^6-13m_\mu^2m_V^4+8m_\mu^4m_V^2-3m_\mu^6)\big]&\vspace{1ex}\\
	+\frac{m_f^4}{4m_\mu^4(m_V^2-m_\mu^2)^4}\big[m_\mu^2m_V^2(4m_V^4-3m_\mu^2m_V^2-8m_\mu^4)+2m_\mu^4(8m_\mu^2m_V^2-3m_V^4-2m_\mu^4)\log\frac{m_V^2}{m_f^2}&\vspace{1ex}\\
	+(4m_V^8-16m_\mu^2m_V^6+12m_\mu^4m_V^4+16m_\mu^6m_V^2-4m_\mu^8)(\log\frac{m_\mu^2-m_V^2}{m_V^2}-i\pi)\big]+\mathcal{O}(m_f^6), \quad\mathrm{for}~m_\mu>m_V&
\end{array}\right..
\end{align}

Up to $m_f^5$, the integral $L_{LR}^f$ can be expanded as
\begin{align}\label{eqn:expansion:LfLR:mfllmmumV}
&L_{LR}^f=\left\{\begin{array}{ll}
\frac{m_f}{m_\mu}\Big\{\frac{1}{4m_\mu^4(m_\mu^2-m_V^2)}\big[4m_\mu^2(m_\mu^2-m_V^2)^2+(4m_V^6-6m_\mu^2m_V^4-2m_\mu^6)\log(1-\frac{m_\mu^2}{m_V^2})-2m_\mu^6\log\frac{m_V^2}{m_f^2}\big]&\vspace{1ex}\\
	+\frac{m_f^2}{4m_\mu^4(m_\mu^2-m_V^2)^3}\big[2m_\mu^2m_V^2(3m_\mu^4-m_V^4)+m_\mu^4(6m_V^4-14m_\mu^2m_V^2+4m_\mu^4)\log\frac{m_V^2}{m_f^2}&\vspace{1ex}\\
	+(4m_\mu^8-14m_\mu^6m_V^2-6m_\mu^4m_V^4+10m_\mu^2m_V^6-2m_V^8)\log(1-\frac{m_\mu^2}{m_V^2})\big]&\vspace{1ex}\\
	+\frac{m_f^4}{4m_\mu^4(m_\mu^2-m_V^2)^5}\big[(-2m_\mu^{10}+10m_\mu^8m_V^2-56m_\mu^6m_V^4+16m_\mu^4m_V^6+10m_\mu^2m_V^8-2m_V^{10})\log(1-\frac{m_\mu^2}{m_V^2})&\vspace{1ex}\\
	+2m_\mu^4(9m_V^6-19m_\mu^2m_V^4+5m_\mu^4m_V^2-m_\mu^6)\log\frac{m_V^2}{m_f^2}&\vspace{1ex}\\
	+m_\mu^2(-3m_\mu^8+16m_\mu^6m_V^2+9m_\mu^4m_V^4-6m_\mu^2m_V^6-2m_V^8)\big]\Big\}+\mathcal{O}(m_f^7), \qquad\mathrm{for}~m_\mu<m_V&\vspace{4ex}\\
\frac{m_f}{m_\mu}\Big\{\frac{1}{4m_\mu^4(m_\mu^2-m_V^2)}\big[4m_\mu^2(m_\mu^2-m_V^2)^2+(4m_V^6-6m_\mu^2m_V^4-2m_\mu^6)(\log\frac{m_\mu^2-m_V^2}{m_V^2}-i\pi)-2m_\mu^6\log\frac{m_V^2}{m_f^2}\big]&\vspace{1ex}\\
	+\frac{m_f^2}{4m_\mu^4(m_\mu^2-m_V^2)^3}\big[2m_\mu^2m_V^2(3m_\mu^4-m_V^4)+m_\mu^4(6m_V^4-14m_\mu^2m_V^2+4m_\mu^4)\log\frac{m_V^2}{m_f^2}&\vspace{1ex}\\
	+(4m_\mu^8-14m_\mu^6m_V^2-6m_\mu^4m_V^4+10m_\mu^2m_V^6-2m_V^8)(\log\frac{m_\mu^2-m_V^2}{m_V^2}-i\pi)\big]+&\vspace{1ex}\\
	\frac{m_f^4}{4m_\mu^4(m_\mu^2-m_V^2)^5}\big[(-2m_\mu^{10}+10m_\mu^8m_V^2-56m_\mu^6m_V^4+16m_\mu^4m_V^6+10m_\mu^2m_V^8-2m_V^{10})(\log\frac{m_\mu^2-m_V^2}{m_V^2}-i\pi)&\vspace{1ex}\\
	+2m_\mu^4(9m_V^6-19m_\mu^2m_V^4+5m_\mu^4m_V^2-m_\mu^6)\log\frac{m_V^2}{m_f^2}&\vspace{1ex}\\
	+m_\mu^2(-3m_\mu^8+16m_\mu^6m_V^2+9m_\mu^4m_V^4-6m_\mu^2m_V^6-2m_V^8)\big]\Big\}+\mathcal{O}(m_f^7), \qquad\mathrm{for}~m_\mu>m_V
\end{array}\right..
\end{align}
Especially, we have the following approximations:
\begin{align}\label{eqn:expansion:LfLL:mfllmmumV:esp}
L_{LL}^f\approx\left\{\begin{array}{ll}
-\frac{2}{3}(1+\frac{3m_{\mu}^2}{8m_V^2}),& \mathrm{for}~m_f\ll m_\mu\ll m_V\vspace{1ex}\\
-\frac{1}{4}(1-\frac{4m_V^2}{m_{\mu}^2}),& \mathrm{for}~m_f\ll m_V\ll m_\mu
\end{array}\right.
\end{align}
and
\begin{align}\label{eqn:expansion:LfLR:mfllmmumV:esp}
L_{LR}^f\approx\left\{\begin{array}{ll}
\frac{m_f}{m_{\mu}}\big[1-\frac{m_{\mu}^2}{12m_V^2}(5-12\log\frac{m_V}{m_f})\big],& \mathrm{for}~m_f\ll m_\mu\ll m_V\vspace{1ex}\\
\frac{m_f}{m_{\mu}}\big[1-\frac{1}{2}(\log\frac{m_{\mu}^2}{m_f^2}-i\pi)-\frac{m_V^2}{2m_{\mu}^2}(1+\log\frac{m_{\mu}^2}{m_f^2}-i\pi)\big],& \mathrm{for}~m_f\ll m_V\ll m_\mu
\end{array}\right..
\end{align}

For the scenario of $m_f\ll m_\mu=m_V$, please refer to Eqs. \eqref{eqn:expansion:LfLL:mmumV:esp} and \eqref{eqn:expansion:LfLR:mmumV:esp}.
\subsubsection{Scenario of $m_f,m_V\ll m_\mu$}
Up to $1/m_{\mu}^4$, the integral $L_{LL}^f$ can be expanded as
\begin{align}
&L_{LL}^f=-\frac{1}{4}-\frac{1}{4m_\mu^2}\big[3m_f^2-4m_V^2-2m_f^2(\log\frac{m_\mu^2}{m_f^2}-i\pi)\big]\nonumber\\
&-\frac{1}{2m_\mu^4}\big[2m_V^4+2m_f^2(m_f^2-m_V^2)(\log\frac{m_\mu^2}{m_f^2}-i\pi)-3m_V^4(\log\frac{m_\mu^2}{m_f^2}-i\pi)\big]+\mathcal{O}(\frac{1}{m_\mu^6}).
\end{align}
Up to $1/m_{\mu}^5$, the integral $L_{LR}^f$ can be expanded as
\begin{align}
&L_{LR}^f=\frac{m_f}{m_\mu}\Big\{1-\frac{1}{2}(\log\frac{m_\mu^2}{m_f^2}-i\pi)+\frac{1}{2m_\mu^2}\big[-m_V^2+(2m_f^2-m_V^2)(\log\frac{m_\mu^2}{m_f^2}-i\pi)\big]+\frac{1}{4m_\mu^4}\big[-3m_f^4+2m_f^2m_V^2\nonumber\\
	&+3m_V^4-(2m_f^4+2m_f^2m_V^2+2m_V^4)(\log\frac{m_\mu^2}{m_f^2}-i\pi)-6m_V^4(\log\frac{m_\mu^2}{m_V^2}-i\pi)\big]\Big\}+\mathcal{O}(\frac{1}{m_\mu^7}).
\end{align}

For the scenario of $m_f\ll m_V\ll m_\mu$, the approximations have been given in Eqs. \eqref{eqn:expansion:LfLL:mfllmmumV:esp} and \eqref{eqn:expansion:LfLR:mfllmmumV:esp}. For the scenario of $m_V\ll m_f\ll m_\mu$, the approximations have been given in Eqs. \eqref{eqn:expansion:LfLL:mVllmfmmu:esp} and \eqref{eqn:expansion:LfLR:mVllmfmmu:esp}. For the scenario of $m_f=m_V\ll m_\mu$, please refer to Eqs. \eqref{eqn:expansion:LfLL:mfmV:esp} and \eqref{eqn:expansion:LfLR:mfmV:esp}.

\subsubsection{Scenario of $m_f,m_\mu\ll m_V$}

Up to $1/m_V^4$, the integrals $L_{LL}^f$ and $L_{LR}^f$ can be expanded as
\begin{align}
&L_{LL}^f=-\frac{2}{3}+\frac{2m_f^2-m_\mu^2}{4m_V^2}+\frac{330m_f^4-35m_f^2m_\mu^2-21m_\mu^4+60m_f^2(m_\mu^2-3m_f^2)\log\frac{m_V^2}{m_f^2}}{120m_V^4}+\mathcal{O}(\frac{1}{m_V^6})
\end{align}
and
\begin{align}
&L_{LR}^f=\frac{m_f}{m_\mu}\big[1+\frac{27m_f^2-5m_\mu^2+(6m_\mu^2-18m_f^2)\log\frac{m_V^2}{m_f^2}}{12m_V^2}\nonumber\\
	&+\frac{45m_f^4+58m_f^2m_\mu^2-14m_\mu^4+(6m_\mu^4-12m_f^2m_\mu^2-54m_f^4)\log\frac{m_V^2}{m_f^2}}{12m_V^4}\big]+\mathcal{O}(\frac{1}{m_V^6}).
\end{align}

For the scenario of $m_f\ll m_\mu\ll m_V$, the approximations have been given in Eqs. \eqref{eqn:expansion:LfLL:mfllmmumV:esp} and \eqref{eqn:expansion:LfLR:mfllmmumV:esp}. For the scenario of $m_\mu\ll m_f\ll m_V$, the approximations have been given in Eqs. \eqref{eqn:expansion:LfLL:mmullmfmV:esp} and \eqref{eqn:expansion:LfLR:mmullmfmV:esp}. For the scenario of $m_f=m_\mu\ll m_V$, please refer to Eqs. \eqref{eqn:expansion:LfLL:mfmmu:esp} and \eqref{eqn:expansion:LfLR:mfmmu:esp}.

\subsubsection{Scenario of $m_\mu,m_V\ll m_f$}

Up to $1/m_f^4$, the integral $L_{LL}^f$ can be expanded as
\begin{align}
&L_{LL}^f=-\frac{5}{12}-\frac{m_\mu^2+12m_V^2}{24m_f^2}+\frac{-m_\mu^4-20m_\mu^2m_V^2-165m_V^4+90m_V^4\log\frac{m_f^2}{m_V^2}}{60m_f^4}+\mathcal{O}(\frac{1}{m_f^6}).
\end{align}
Up to $1/m_f^3$, the integral $L_{LR}^f$ can be expanded as
\begin{align}
&L_{LR}^f=\frac{m_f}{m_\mu}\big[\frac{1}{4}+\frac{2m_\mu^2+9m_V^2}{12m_f^2}+\frac{m_\mu^4+10m_\mu^2m_V^2+54m_V^4-36m_V^4\log\frac{m_f^2}{m_V^2}}{24m_f^4}\big]+\mathcal{O}(\frac{1}{m_f^5}).
\end{align}

For the scenario of $m_\mu\ll m_V\ll m_f$, the approximations have been given in Eqs. \eqref{eqn:expansion:LfLL:mmullmfmV:esp} and \eqref{eqn:expansion:LfLR:mmullmfmV:esp}. For the scenario of $m_V\ll m_\mu\ll m_f$, the approximations have been given in Eqs. \eqref{eqn:expansion:LfLL:mVllmfmmu:esp} and \eqref{eqn:expansion:LfLR:mVllmfmmu:esp}. For the scenario of $m_\mu=m_V\ll m_f$, please refer to Eqs. \eqref{eqn:expansion:LfLL:mmumV:esp} and \eqref{eqn:expansion:LfLR:mmumV:esp}.

\subsection{Degenerate mass case for the $L_{LL}^f$ and $L_{LR}^f$}\label{sec:expansion:Lfdeg}
\subsubsection{Scenario of $m_f=m_\mu$}\label{sec:expansion:Lfdeg:mfmmu}
For $m_f=m_\mu$, the integrals $L_{LL}^f$ and $L_{LR}^f$ can be calculated as
\begin{align}\label{eqn:expansion:LfLL:mfmmu}
&L_{LL}^f=\left\{\begin{array}{ll}
-\frac{1}{2m_{\mu}^6}\big[m_{\mu}^6-3m_{\mu}^4m_V^2+2m_{\mu}^2m_V^4+m_V^2(m_{\mu}^4-3m_{\mu}^2m_V^2+m_V^4)\log\frac{m_{\mu}^2}{m_V^2}\big]\vspace{1ex}\\
	-\frac{m_V^3}{m_{\mu}^6\sqrt{m_V^2-4m_{\mu}^2}}(5m_{\mu}^4-5m_V^2m_{\mu}^2+m_V^4)\log\frac{m_V+\sqrt{m_V^2-4m_{\mu}^2}}{2m_{\mu}},&\mathrm{for}~m_\mu<\frac{m_V}{2}\vspace{3ex}\\
-\frac{1}{2m_{\mu}^6}\big[m_{\mu}^6-3m_{\mu}^4m_V^2+2m_{\mu}^2m_V^4+m_V^2(m_{\mu}^4-3m_{\mu}^2m_V^2+m_V^4)\log\frac{m_{\mu}^2}{m_V^2}\big]\vspace{1ex}\\
	-\frac{m_V^3}{m_{\mu}^6\sqrt{4m_{\mu}^2-m_V^2}}(5m_{\mu}^4-5m_{\mu}^2m_V^2+m_V^4)\arctan\sqrt{\frac{4m_{\mu}^2}{m_V^2}-1},&\mathrm{for}~m_\mu>\frac{m_V}{2}\vspace{3ex}\\
20\log 2-\frac{29}{2},&\mathrm{for}~m_\mu=\frac{m_V}{2}
\end{array}\right.
\end{align}
and
\begin{align}\label{eqn:expansion:LfLR:mfmmu}
&L_{LR}^f=\left\{\begin{array}{ll}
\frac{1}{2m_{\mu}^4}\big[m_{\mu}^4-2m_{\mu}^2m_V^2+m_V^2(m_{\mu}^2-m_V^2)\log\frac{m_{\mu}^2}{m_V^2}\big]\vspace{1ex}\\
	+\frac{m_V^3}{m_{\mu}^4\sqrt{m_V^2-4m_{\mu}^2}}(3m_{\mu}^2-m_V^2)\log\frac{m_V+\sqrt{m_V^2-4m_{\mu}^2}}{2m_{\mu}},&\mathrm{for}~m_\mu<\frac{m_V}{2}\vspace{3ex}\\
\frac{1}{2m_{\mu}^4}\big[m_{\mu}^4-2m_{\mu}^2m_V^2+m_V^2(m_{\mu}^2-m_V^2)\log\frac{m_{\mu}^2}{m_V^2}\big]\vspace{1ex}\\
	+\frac{m_V^3}{m_{\mu}^4\sqrt{4m_{\mu}^2-m_V^2}}(3m_{\mu}^2-m_V^2)\arctan\sqrt{\frac{4m_{\mu}^2}{m_V^2}-1},\quad\quad\quad&\mathrm{for}~m_\mu>\frac{m_V}{2}\vspace{3ex}\\
12\log 2-\frac{15}{2},&\mathrm{for}~m_\mu=\frac{m_V}{2}
\end{array}\right..
\end{align}

Then, they can be expanded as
\begin{align}\label{eqn:expansion:LfLL:mfmmu:esp}
L_{LL}^f\approx\left\{\begin{array}{ll}
-\frac{2}{3}+\frac{m_{\mu}^2}{4m_V^2}+\frac{m_{\mu}^4}{60m_V^4}(137-120\log\frac{m_V}{m_{\mu}}),\quad& \mathrm{for}~m_\mu\ll m_V\vspace{1ex}\\
-\frac{1}{2}+\frac{m_V^2}{2m_{\mu}^2}(3-2\log\frac{m_{\mu}}{m_V})-\frac{5\pi m_V^3}{4m_{\mu}^3}+\frac{m_V^4}{4m_{\mu}^4}(1+12\log\frac{m_{\mu}}{m_V}),\quad\quad\quad& \mathrm{for}~m_V\ll m_\mu
\end{array}\right.
\end{align}
and
\begin{align}\label{eqn:expansion:LfLR:mfmmu:esp}
L_{LR}^f\approx\left\{\begin{array}{ll}
1+\frac{m_{\mu}^2}{6m_V^2}(11-12\log\frac{m_V}{m_{\mu}})+\frac{m_{\mu}^4}{12m_V^4}(89-120\log\frac{m_V}{m_{\mu}}),\quad& \mathrm{for}~m_\mu\ll m_V\vspace{1ex}\\
\frac{1}{2}+\frac{m_V^2}{m_{\mu}^2}(\log\frac{m_{\mu}}{m_V}-1)+\frac{3\pi m_V^3}{4m_{\mu}^3}-\frac{m_V^4}{4m_{\mu}^4}(3+4\log\frac{m_{\mu}}{m_V}),\quad\quad\quad& \mathrm{for}~m_V\ll m_\mu
\end{array}\right..
\end{align}

\subsubsection{Scenario of $m_\mu=m_V$}\label{sec:expansion:Lfdeg:mmumV}
For $m_\mu=m_V$, the integrals $L_{LL}^f$ and $L_{LR}^f$ can be calculated as
\begin{align}\label{eqn:expansion:LfLL:mmumV}
L_{LL}^f=\left\{\begin{array}{ll}
\frac{1}{4m_{\mu}^6}\big[m_{\mu}^2(2m_f^4-m_f^2m_{\mu}^2-m_{\mu}^4)+(m_{\mu}^6+2m_f^4m_{\mu}^2-m_f^6)\log\frac{m_f^2}{m_{\mu}^2}\big]\vspace{1ex}\\
	+\frac{m_f}{2m_{\mu}^6\sqrt{m_f^2-4m_{\mu}^2}}(m_f^6-4m_f^4m_{\mu}^2+2m_f^2m_{\mu}^4-m_{\mu}^6)\log\frac{m_f+\sqrt{m_f^2-4m_{\mu}^2}}{2m_{\mu}},\quad\quad\quad& \mathrm{for}~m_{\mu}<\frac{m_f}{2}\vspace{3ex}\\
\frac{1}{4m_{\mu}^6}\big[m_{\mu}^2(2m_f^4-m_f^2m_{\mu}^2-m_{\mu}^4)+(m_{\mu}^6+2m_f^4m_{\mu}^2-m_f^6)\log\frac{m_f^2}{m_{\mu}^2}\big]\vspace{1ex}\\
	+\frac{m_f}{2m_{\mu}^6\sqrt{4m_{\mu}^2-m_f^2}}(m_f^6-4m_f^4m_{\mu}^2+2m_f^2m_{\mu}^4-m_{\mu}^6)\arctan\sqrt{\frac{4m_{\mu}^2}{m_f^2}-1},& \mathrm{for}~m_{\mu}>\frac{m_f}{2}\vspace{3ex}\\
\frac{41}{4}-\frac{31}{2}\log2,& \mathrm{for}~m_{\mu}=\frac{m_f}{2}
\end{array}\right.
\end{align}
and
\begin{align}\label{eqn:expansion:LfLR:mmumV}
L_{LR}^f=\left\{\begin{array}{ll}
\frac{m_f^3}{4m_{\mu}^5}\big[-2m_{\mu}^2+(m_f^2-m_{\mu}^2)\log\frac{m_f^2}{m_{\mu}^2}\big]\vspace{1ex}\\
	+\frac{1}{2m_{\mu}^5\sqrt{m_f^2-4m_{\mu}^2}}(-m_f^6+3m_f^4m_{\mu}^2+2m_{\mu}^6)\log\frac{m_f+\sqrt{m_f^2-4m_{\mu}^2}}{2m_{\mu}},\quad\quad\quad& \mathrm{for}~m_{\mu}<\frac{m_f}{2}\vspace{3ex}\\
\frac{m_f^3}{4m_{\mu}^5}\big[-2m_{\mu}^2+(m_f^2-m_{\mu}^2)\log\frac{m_f^2}{m_{\mu}^2}\big]\vspace{1ex}\\
	+\frac{1}{2m_{\mu}^5\sqrt{4m_{\mu}^2-m_f^2}}(-m_f^6+3m_f^4m_{\mu}^2+2m_{\mu}^6)\arctan\sqrt{\frac{4m_{\mu}^2}{m_f^2}-1},& \mathrm{for}~m_{\mu}>\frac{m_f}{2}\vspace{3ex}\\
-\frac{15}{2}+12\log2,& \mathrm{for}~m_{\mu}=\frac{m_f}{2}
\end{array}\right. .
\end{align}
Then, they can be expanded as
\begin{align}\label{eqn:expansion:LfLL:mmumV:esp}
L_{LL}^f\approx\left\{\begin{array}{ll}
-\frac{5}{12}-\frac{13m_{\mu}^2}{24m_f^2}+\frac{m_{\mu}^4}{10m_f^4}(-31+30\log\frac{m_f}{m_{\mu}}),\quad\quad\quad& \mathrm{for}~m_\mu\ll m_f\vspace{1ex}\\
-\frac{1}{4}-\frac{1}{2}\log\frac{m_{\mu}}{m_f}-\frac{\pi m_f}{8m_{\mu}}-\frac{m_f^2}{8m_{\mu}^2}+\frac{15\pi m_f^3}{64m_{\mu}^3}+\frac{m_f^4}{48m_{\mu}^4}(13-48\log\frac{m_{\mu}}{m_f}),\quad\quad& \mathrm{for}~m_f\ll m_\mu
\end{array}\right.
\end{align}
and
\begin{align}\label{eqn:expansion:LfLR:mmumV:esp}
L_{LR}^f\approx\left\{\begin{array}{ll}
\frac{m_f}{4m_{\mu}}+\frac{11m_{\mu}}{12m_f}+\frac{m_{\mu}^3}{24m_f^3}(65-72\log\frac{m_f}{m_{\mu}}),\quad\quad\quad& \mathrm{for}~m_\mu\ll m_f\vspace{1ex}\\
\frac{\pi}{4}-\frac{m_f}{4m_{\mu}}+\frac{\pi m_f^2}{32m_{\mu}^2}+\frac{m_f^3}{24m_{\mu}^3}(12\log\frac{m_{\mu}}{m_f}-13)+\frac{195\pi m_f^4}{512m_{\mu}^4}-\frac{m_f^5}{60m_{\mu}^5}(30\log\frac{m_{\mu}}{m_f}+23),\quad& \mathrm{for}~m_f\ll m_\mu
\end{array}\right..
\end{align}

\subsubsection{Scenario of $m_f=m_V$}\label{sec:expansion:Lfdeg:mfmV}
For $m_f=m_V$, the integrals $L_{LL}^f$ and $L_{LR}^f$ can be calculated as
\begin{align}\label{eqn:expansion:LfLL:mfmV}
L_{LL}^f=\left\{\begin{array}{ll}
\frac{m_V^2-m_{\mu}^2}{4m_{\mu}^2}+\frac{m_V^2(m_V^2+m_{\mu}^2)}{2m_{\mu}^3\sqrt{4m_V^2-m_{\mu}^2}}\arctan\frac{m_{\mu}\sqrt{4m_V^2-m_{\mu}^2}}{m_{\mu}^2-2m_V^2},&\mathrm{for}~m_\mu<2m_V\vspace{1ex}\\
\frac{m_V^2-m_{\mu}^2}{4m_{\mu}^2}+\frac{m_V^2(m_V^2+m_{\mu}^2)}{2m_{\mu}^3\sqrt{m_{\mu}^2-4m_V^2}}(2\log\frac{m_{\mu}+\sqrt{m_{\mu}^2-4m_V^2}}{2m_V}-i\pi),\quad\quad\quad&\mathrm{for}~m_\mu>2m_V\vspace{1ex}\\
\frac{1}{8},&\mathrm{for}~m_\mu=2m_V
\end{array}\right.
\end{align}
and
\begin{small}\label{eqn:expansion:LfLR:mfmV}
\begin{align}
L_{LR}^f=\left\{\begin{array}{ll}
\frac{m_V(2m_{\mu}^2-3m_V^2)}{2m_{\mu}^3}-\frac{m_V(m_{\mu}^4-3m_{\mu}^2m_V^2+6m_V^4)}{2m_{\mu}^4\sqrt{4m_V^2-m_{\mu}^2}}\arctan\frac{m_{\mu}\sqrt{4m_V^2-m_{\mu}^2}}{m_{\mu}^2-2m_V^2},&\mathrm{for}~m_\mu<2m_V\vspace{1ex}\\
\frac{m_V(2m_{\mu}^2-3m_V^2)}{2m_{\mu}^3}+\frac{m_V(m_{\mu}^4-3m_{\mu}^2m_V^2+6m_V^4)}{2m_{\mu}^4\sqrt{m_{\mu}^2-4m_V^2}}(-2\log\frac{m_{\mu}+\sqrt{m_{\mu}^2-4m_V^2}}{2m_V}+i\pi),\quad\quad\quad&\mathrm{for}~m_\mu>2m_V\vspace{1ex}\\
0,&\mathrm{for}~m_\mu=2m_V
\end{array}\right..
\end{align}
\end{small}
Then, they can be expanded as
\begin{align}\label{eqn:expansion:LfLL:mfmV:esp}
L_{LL}^f\approx\left\{\begin{array}{ll}
-\frac{13}{24}-\frac{m_{\mu}^2}{20m_V^2}-\frac{17m_{\mu}^4}{1680m_V^4},& \mathrm{for}~m_\mu\ll m_V\vspace{1ex}\\
-\frac{1}{4}+\frac{m_V^2}{4m_{\mu}^2}(1+4\log\frac{m_{\mu}}{m_V}-2i\pi)+\frac{m_V^4}{2m_{\mu}^4}(-2+6\log\frac{m_{\mu}}{m_V}-3i\pi),\quad\quad& \mathrm{for}~m_V\ll m_\mu
\end{array}\right.
\end{align}
and
\begin{align}\label{eqn:expansion:LfLR:mfmV:esp}
L_{LR}^f\approx\left\{\begin{array}{ll}
\frac{m_V}{2m_{\mu}}+\frac{7m_{\mu}}{40m_V}+\frac{23m_{\mu}^3}{840m_V^3},& \mathrm{for}~m_\mu\ll m_V\vspace{2ex}\\
\frac{m_V}{2m_{\mu}}(2-2\log\frac{m_{\mu}}{m_V}+i\pi)+\frac{m_V^3}{2m_{\mu}^3}(-1+2\log\frac{m_{\mu}}{m_V}-i\pi)\quad\quad\quad&\\
	+\frac{m_V^5}{2m_{\mu}^5}(1-12\log\frac{m_{\mu}}{m_V}+6i\pi), &\mathrm{for}~m_V\ll m_\mu
\end{array}\right..
\end{align}

\subsubsection{Scenario of $m_f=m_\mu=m_V$}
In this case, we have $L_{LL}^f=-\frac{\sqrt{3}\pi}{9}$ and $L_{LR}^f=\frac{4\sqrt{3}\pi-9}{18}$.

\subsection{Hierarchical mass expansion for the $L_{LL}^V$ and $L_{LR}^V$}\label{sec:expansion:LV}
\subsubsection{Scenario of $m_\mu\ll m_f,m_V$}\label{sec:expansion:LVhie:mmullmfmV}
Up to $m_{\mu}^4$, the integral $L_{LL}^V$ can be expanded as
\begin{align}\label{eqn:expansion:LVLL:mmullmfmV}
&L_{LL}^V=\frac{1}{12(m_V^2-m_f^2)^4}\big[(m_f^2-m_V^2)(4m_f^6-45m_f^4m_V^2+33m_f^2m_V^4-10m_V^6)+18m_f^6m_V^2\log\frac{m_f^2}{m_V^2}\big]\nonumber\\
&+\frac{m_\mu^2}{24(m_V^2-m_f^2)^6}\big[(m_f^4-m_V^4)(m_f^6-120m_f^4m_V^2+33m_f^2m_V^4-4m_V^6)\nonumber\\
	&+12m_f^4m_V^2(3m_f^4+13m_f^2m_V^2-m_V^4)\log\frac{m_f^2}{m_V^2}\big]\nonumber\\
	&+\frac{m_\mu^4}{40(m_V^2-m_f^2)^8}\big[(m_f^2-m_V^2)(m_f^{10}-242m_f^8m_V^2-957m_f^6m_V^4-97m_f^4m_V^6+38m_f^2m_V^8-3m_V^{10})\nonumber\\
	&+60m_f^4m_V^2(m_f^6+11m_f^4m_V^2+10m_f^2m_V^4-m_V^6)\log\frac{m_f^2}{m_V^2}\big]+\mathcal{O}(m_\mu^6).
\end{align}
Up to $m_{\mu}^3$, the integral $L_{LR}^V$ can be expanded as
\begin{align}\label{eqn:expansion:LVLR:mmullmfmV}
&L_{LR}^V=\frac{m_f}{m_\mu}\Big\{\frac{1}{4(m_f^2-m_V^2)^3}\big[(m_V^2-m_f^2)(m_f^4-11m_f^2m_V^2+4m_V^4)-6m_f^4m_V^2\log\frac{m_f^2}{m_V^2}\big]+\frac{m_\mu^2}{12(m_f^2-m_V^2)^5}\cdot\nonumber\\
	&\big[(m_V^2-m_f^2)(m_f^6-57m_f^4m_V^2-21m_f^2m_V^4+5m_V^6)+6m_f^2m_V^2(m_V^4-10m_f^2m_V^2-3m_f^4)\log\frac{m_f^2}{m_V^2}\big]\nonumber\\
&+\frac{m_\mu^4}{24(m_f^2-m_V^2)^7}\big[(m_V^2-m_f^2)(m_f^8-141m_f^6m_V^2-405m_f^4m_V^4-m_f^2m_V^6+6m_V^8)\nonumber\\
	&+12m_f^2m_V^2(2m_V^6-17m_f^2m_V^4-27m_f^4m_V^2-3m_f^6)\log\frac{m_f^2}{m_V^2}\big]\Big\}+\mathcal{O}(m_\mu^5).
\end{align}

Especially, we have the following approximations:
\begin{align}\label{eqn:expansion:LVLL:mmullmfmV:esp}
L_{LL}^V\approx\left\{\begin{array}{ll}
\frac{5}{6}(1-\frac{3m_f^2}{10m_V^2}),& \mathrm{for}~m_\mu\ll m_f\ll m_V\vspace{1ex}\\
\frac{1}{3}\big[1+\frac{3m_V^2}{4m_f^2}(12\log\frac{m_f}{m_V}-11)\big],\qquad\qquad& \mathrm{for}~m_\mu\ll m_V\ll m_f
\end{array}\right.
\end{align}
and
\begin{align}\label{eqn:expansion:LVLR:mmullmfmV:esp}
L_{LR}^V\approx\left\{\begin{array}{ll}
-\frac{m_f}{m_\mu}(1-\frac{3m_f^2}{4m_V^2}),& \mathrm{for}~m_\mu\ll m_f\ll m_V\vspace{1ex}\\
-\frac{m_f}{4m_\mu}\big[1+\frac{3m_V^2}{m_f^2}(4\log\frac{m_f}{m_V}-3)\big],\qquad\qquad& \mathrm{for}~m_\mu\ll m_V\ll m_f
\end{array}\right. .
\end{align}

For the scenario of $m_\mu\ll m_f=m_V$, please refer to Eqs. \eqref{eqn:expansion:LVLL:mfmV:esp} and \eqref{eqn:expansion:LVLR:mfmV:esp}.
\subsubsection{Scenario of $m_V\ll m_f,m_\mu$}\label{sec:expansion:LVhie:mVllmfmmu}
Up to $m_V^4$, the integrals $L_{LL}^V$ and $L_{LR}^V$ can be expanded as
\begin{align}\label{eqn:expansion:LVLL:mVllmfmmu}
&L_{LL}^V=\left\{\begin{array}{ll}
\frac{1}{4m_\mu^6}\big[m_\mu^2(2m_f^4-m_f^2m_\mu^2+m_\mu^4)+2m_f^4(m_f^2-m_\mu^2)\log\frac{m_f^2-m_\mu^2}{m_f^2}\big]+\frac{m_V^2}{2m_\mu^4(m_f^2-m_\mu^2)}\big[m_\mu^2(3m_f^2-4m_\mu^2)+&\vspace{1ex}\\
	(3m_f^4+3m_\mu^4)\log\frac{m_f^2-m_\mu^2}{m_f^2}+3m_\mu^4\log\frac{m_f^2}{m_V^2}\big]+\frac{m_V^4}{4m_\mu^6(m_f^2-m_\mu^2)^3}\big[m_\mu^2(-6m_f^6+5m_f^4m_\mu^2-12m_f^2m_\mu^4-2m_\mu^6)&\vspace{1ex}\\
	+(-6m_f^8+8m_f^6m_\mu^2+12m_f^4m_\mu^4+24m_f^2m_\mu^6-10m_\mu^8)\log\frac{m_f^2-m_\mu^2}{m_f^2}+2m_\mu^6(12m_f^2-5m_\mu^2)\log\frac{m_f^2}{m_V^2}\big]&\vspace{1ex}\\
	+\mathcal{O}(m_V^6), \hspace{60ex}\mathrm{for}~m_\mu<m_f&\vspace{2ex}\\
\frac{1}{4m_\mu^6}\big[m_\mu^2(2m_f^4-m_f^2m_\mu^2+m_\mu^4)+2m_f^4(m_f^2-m_\mu^2)(\log\frac{m_\mu^2-m_f^2}{m_f^2}-i\pi)\big]&\vspace{1ex}\\
	+\frac{m_V^2}{2m_\mu^4(m_f^2-m_\mu^2)}\big[m_\mu^2(3m_f^2-4m_\mu^2)+(3m_f^4+3m_\mu^4)(\log\frac{m_\mu^2-m_f^2}{m_f^2}-i\pi)+3m_\mu^4\log\frac{m_f^2}{m_V^2}\big]&\vspace{1ex}\\
	+\frac{m_V^4}{4m_\mu^6(m_f^2-m_\mu^2)^3}\big[m_\mu^2(-6m_f^6+5m_f^4m_\mu^2-12m_f^2m_\mu^4-2m_\mu^6)+2m_\mu^6(12m_f^2-5m_\mu^2)\log\frac{m_f^2}{m_V^2}&\vspace{1ex}\\
	+(-6m_f^8+8m_f^6m_\mu^2+12m_f^4m_\mu^4+24m_f^2m_\mu^6-10m_\mu^8)(\log\frac{m_\mu^2-m_f^2}{m_f^2}-i\pi)\big]&\vspace{1ex}\\
	+\mathcal{O}(m_V^6), \hspace{60ex}\mathrm{for}~m_\mu>m_f&
\end{array}\right.
\end{align}
and
\begin{align}\label{eqn:expansion:LVLR:mVllmfmmu}
&L_{LR}^V=\left\{\begin{array}{ll}\frac{m_f}{m_\mu}\Big\{
\frac{m_f^2}{2m_\mu^4}\big[-m_\mu^2+(m_\mu^2-m_f^2)\log\frac{m_f^2-m_\mu^2}{m_f^2}\big]+\frac{m_V^2}{2m_\mu^4(m_f^2-m_{\mu}^2)}\big[m_\mu^2(2m_{\mu}^2-m_f^2)&\vspace{1ex}\\
	-(3m_{\mu}^4+2m_{\mu}^2m_f^2+m_f^4)\log\frac{m_f^2-m_\mu^2}{m_f^2}-3m_{\mu}^4\log\frac{m_f^2}{m_V^2}\big]+\frac{m_V^4}{4m_\mu^4(m_f^2-m_\mu^2)^3}\cdot&\vspace{1ex}\\
	\big[m_\mu^2(6m_\mu^4+5m_\mu^2m_f^2+4m_f^4)+(4m_f^6-12m_f^4m_\mu^2-24m_f^2m_\mu^4+4m_\mu^6)\log\frac{m_f^2-m_\mu^2}{m_f^2}&\vspace{1ex}\\
	+2m_\mu^4(2m_\mu^2-9m_f^2)\log\frac{m_f^2}{m_V^2}\big]\Big\}+\mathcal{O}(m_V^6), &\hspace{-15ex}\mathrm{for}~m_\mu<m_f \vspace{2ex}\\
\frac{m_f}{m_\mu}\Big\{\frac{m_f^2}{2m_\mu^4}\big[-m_\mu^2+(m_\mu^2-m_f^2)(\log\frac{m_\mu^2-m_f^2}{m_f^2}-i\pi)\big]+\frac{m_V^2}{2m_\mu^4(m_f^2-m_{\mu}^2)}\big[m_\mu^2(2m_{\mu}^2-m_f^2)&\vspace{1ex}\\
	-(3m_{\mu}^4+2m_{\mu}^2m_f^2+m_f^4)(\log\frac{m_\mu^2-m_f^2}{m_f^2}-i\pi)-3m_{\mu}^4\log\frac{m_f^2}{m_V^2}\big]+\frac{m_V^4}{4m_\mu^4(m_f^2-m_\mu^2)^3}\cdot&\vspace{1ex}\\
	\big[m_\mu^2(6m_\mu^4+5m_\mu^2m_f^2+4m_f^4)+(4m_f^6-12m_f^4m_\mu^2-24m_f^2m_\mu^4+4m_\mu^6)(\log\frac{m_\mu^2-m_f^2}{m_f^2}-i\pi)&\vspace{1ex}\\
	+2m_\mu^4(2m_\mu^2-9m_f^2)\log\frac{m_f^2}{m_V^2}\big]\Big\}+\mathcal{O}(m_V^6), &\hspace{-15ex}\mathrm{for}~m_\mu>m_f
\end{array}\right..
\end{align}
Especially, we have the following approximations:
\begin{align}\label{eqn:expansion:LVLL:mVllmfmmu:esp}
L_{LL}^V\approx\left\{\begin{array}{ll}
\frac{1}{4}(1-\frac{m_f^2}{m_\mu^2}), &\hspace{20ex} \mathrm{for}~m_V\ll m_f\ll m_\mu\vspace{1ex}\\
\frac{1}{3}(1+\frac{m_\mu^2}{8m_f^2}),&\hspace{20ex} \mathrm{for}~m_V\ll m_\mu\ll m_f
\end{array}\right.
\end{align}
and
\begin{align}\label{eqn:expansion:LVLR:mVllmfmmu:esp}
L_{LR}^V\approx\left\{\begin{array}{ll}
\frac{m_f^3}{2m_\mu^3}(-1+\log\frac{m_\mu^2}{m_f^2}-i\pi),&\hspace{15ex} \mathrm{for}~m_V\ll m_f\ll m_\mu\vspace{1ex}\\
-\frac{m_f}{4m_\mu}(1+\frac{m_\mu^2}{3m_f^2}),&\hspace{15ex} \mathrm{for}~m_V\ll m_\mu\ll m_f
\end{array}\right. .
\end{align}

For the scenario of $m_V\ll m_f=m_\mu$, please refer to Eqs. \eqref{eqn:expansion:LVLL:mfmmu:esp} and \eqref{eqn:expansion:LVLR:mfmmu:esp}.
\subsubsection{Scenario of $m_f\ll m_\mu,m_V$}\label{sec:expansion:LVhie:mfllmmumV}
Up to $m_f^4$, the integral $L_{LL}^V$ can be expanded as
\begin{align}\label{eqn:expansion:LVLL:mfllmmumV}
&L_{LL}^V=\left\{\begin{array}{ll}
\frac{1}{4m_\mu^6}\big[m_\mu^6+8m_\mu^4m_V^2-4m_\mu^2m_V^4+2m_V^2(m_\mu^2-m_V^2)(2m_V^2-3m_\mu^2)\log\frac{m_V^2-m_\mu^2}{m_V^2}\big]&\vspace{1ex}\\
	+\frac{m_f^2}{4m_\mu^6(m_\mu^2-m_V^2)}\big[-m_\mu^6+9m_\mu^4m_V^2-6m_\mu^2m_V^4-6m_V^2(m_\mu^2-m_V^2)^2\log\frac{m_V^2-m_\mu^2}{m_V^2}\big]&\vspace{1ex}\\
	+\frac{m_f^4}{4m_\mu^4(m_\mu^2-m_V^2)^3}\big[(-2m_\mu^6-6m_\mu^4m_V^2+6m_\mu^2m_V^4-2m_V^6)\log\frac{m_V^2-m_\mu^2}{m_V^2}-2m_\mu^6\log\frac{m_V^2}{m_f^2}&\vspace{1ex}\\
	+m_\mu^2(2m_\mu^4+3m_\mu^2m_V^2-2m_V^4)\big]+\mathcal{O}(m_f^6), &\hspace{-30ex}\mathrm{for}~m_\mu<m_V\vspace{3ex}\\
\frac{1}{4m_\mu^6}\big[m_\mu^6+8m_\mu^4m_V^2-4m_\mu^2m_V^4+2m_V^2(m_\mu^2-m_V^2)(2m_V^2-3m_\mu^2)(\log\frac{m_\mu^2-m_V^2}{m_V^2}-i\pi)\big]&\vspace{1ex}\\
	+\frac{m_f^2}{4m_\mu^6(m_\mu^2-m_V^2)}\big[-m_\mu^6+9m_\mu^4m_V^2-6m_\mu^2m_V^4-6m_V^2(m_\mu^2-m_V^2)^2(\log\frac{m_\mu^2-m_V^2}{m_V^2}-i\pi)\big]&\vspace{1ex}\\
	+\frac{m_f^4}{4m_\mu^4(m_\mu^2-m_V^2)^3}\big[(-2m_\mu^6-6m_\mu^4m_V^2+6m_\mu^2m_V^4-2m_V^6)(\log\frac{m_\mu^2-m_V^2}{m_V^2}-i\pi)-2m_\mu^6\log\frac{m_V^2}{m_f^2}&\vspace{1ex}\\
	+m_\mu^2(2m_\mu^4+3m_\mu^2m_V^2-2m_V^4)\big]+\mathcal{O}(m_f^6),&\hspace{-30ex}\mathrm{for}~m_\mu>m_V
\end{array}\right..
\end{align}

Up to $m_f^5$, the integral $L_{LR}^V$ can be expanded as
\begin{align}\label{eqn:expansion:LVLR:mfllmmumV}
&L_{LR}^V=\left\{\begin{array}{ll}
\frac{m_f}{m_\mu}\Big\{\frac{m_V^2}{4m_\mu^4}\big[-4m_\mu^2+(6m_\mu^2-4m_V^2)\log\frac{m_V^2-m_\mu^2}{m_V^2}\big]+\frac{m_f^2}{4m_\mu^4(m_\mu^2-m_V^2)^2}\big[2m_\mu^2(m_V^4-m_\mu^2m_V^2-m_\mu^4)&\vspace{1ex}\\
	+(2m_\mu^6+6m_\mu^4m_V^2-6m_\mu^2m_V^4+2m_V^6)\log\frac{m_V^2-m_\mu^2}{m_V^2}+2m_\mu^6\log\frac{m_V^2}{m_f^2}\big]+\frac{m_f^4}{4m_\mu^4(m_\mu^2-m_V^2)^4}\cdot&\vspace{1ex}\\
	\big[(2m_V^8-8m_\mu^2m_V^6+20m_\mu^6m_V^2-2m_\mu^8)\log\frac{m_V^2-m_\mu^2}{m_V^2}+2m_\mu^4(7m_\mu^2m_V^2-m_\mu^4-3m_V^4)\log\frac{m_V^2}{m_f^2}&\vspace{1ex}\\
	+m_\mu^2(2m_V^6+2m_\mu^2m_V^4-9m_\mu^4m_V^2-2m_\mu^6)\big]\Big\}+\mathcal{O}(m_f^7), &\hspace{-25ex}\mathrm{for}~m_\mu<m_V\vspace{3ex}\\
\frac{m_f}{m_\mu}\Big\{\frac{m_V^2}{4m_\mu^4}\big[-4m_\mu^2+(6m_\mu^2-4m_V^2)(\log\frac{m_\mu^2-m_V^2}{m_V^2}-i\pi)\big]+\frac{m_f^2}{4m_\mu^4(m_\mu^2-m_V^2)^2}\big[2m_\mu^2(m_V^4-m_\mu^2m_V^2-m_\mu^4)&\vspace{1ex}\\
	+(2m_\mu^6+6m_\mu^4m_V^2-6m_\mu^2m_V^4+2m_V^6)(\log\frac{m_\mu^2-m_V^2}{m_V^2}-i\pi)+2m_\mu^6\log\frac{m_V^2}{m_f^2}\big]+\frac{m_f^4}{4m_\mu^4(m_\mu^2-m_V^2)^4}\cdot&\vspace{1ex}\\
	\big[(2m_V^8-8m_\mu^2m_V^6+20m_\mu^6m_V^2-2m_\mu^8)(\log\frac{m_\mu^2-m_V^2}{m_V^2}-i\pi)+2m_\mu^4(7m_\mu^2m_V^2-m_\mu^4-3m_V^4)\log\frac{m_V^2}{m_f^2}&\vspace{1ex}\\
	+m_\mu^2(2m_V^6+2m_\mu^2m_V^4-9m_\mu^4m_V^2-2m_\mu^6)\big]\Big\}+\mathcal{O}(m_f^7), &\hspace{-25ex}\mathrm{for}~m_\mu>m_V
\end{array}\right..
\end{align}
Especially, we have the following approximations:
\begin{align}\label{eqn:expansion:LVLL:mfllmmumV:esp}
L_{LL}^V\approx\left\{\begin{array}{ll}
\frac{5}{6}(1+\frac{m_{\mu}^2}{5m_V^2}),& \hspace{18ex}\mathrm{for}~m_f\ll m_\mu\ll m_V\vspace{1ex}\\
\frac{1}{4}\big[1+\frac{m_V^2}{m_{\mu}^2}(8+6i\pi-6\log\frac{m_{\mu}^2}{m_V^2})\big],& \hspace{18ex}\mathrm{for}~m_f\ll m_V\ll m_\mu
\end{array}\right.
\end{align}
and
\begin{align}\label{eqn:expansion:LVLR:mfllmmumV:esp}
L_{LR}^V\approx\left\{\begin{array}{ll}
-\frac{m_f}{m_{\mu}}(1+\frac{5m_{\mu}^2}{12m_V^2}),& \hspace{20ex}\mathrm{for}~m_f\ll m_\mu\ll m_V\vspace{1ex}\\
-\frac{m_fm_V^2}{m_{\mu}^3}\big[1-\frac{3}{2}(\log\frac{m_{\mu}^2}{m_V^2}-i\pi)\big],& \hspace{20ex}\mathrm{for}~m_f\ll m_V\ll m_\mu
\end{array}\right..
\end{align}

For the scenario of $m_f\ll m_\mu=m_V$, please refer to Eqs. \eqref{eqn:expansion:LVLL:mmumV:esp} and \eqref{eqn:expansion:LVLR:mmumV:esp}.
\subsubsection{Scenario of $m_f,m_V\ll m_\mu$}

Up to $1/m_{\mu}^4$, the integral $L_{LL}^V$ can be expanded as
\begin{align}
&L_{LL}^V=\frac{1}{4}+\frac{1}{4m_\mu^2}\big[8m_V^2-m_f^2-6m_V^2(\log\frac{m_\mu^2}{m_V^2}-i\pi)\big]\nonumber\\
&+\frac{1}{2m_\mu^4}\big[m_f^4+4m_f^2m_V^2+m_V^4-m_f^4(\log\frac{m_\mu^2}{m_f^2}-i\pi)+m_V^2(5m_V^2-3m_f^2)(\log\frac{m_\mu^2}{m_V^2}-i\pi)\big]+\mathcal{O}(\frac{1}{m_\mu^6}).
\end{align}
Up to $1/m_{\mu}^5$, the integral $L_{LR}^V$ can be expanded as
\begin{align}
&L_{LR}^V=\frac{m_f}{m_\mu}\Big\{\frac{1}{2m_\mu^2}\big[-m_f^2-2m_V^2+m_f^2(\log\frac{m_\mu^2}{m_f^2}-i\pi)+3m_V^2(\log\frac{m_\mu^2}{m_V^2}-i\pi)\big]+\frac{1}{2m_\mu^4}\big[-m_f^4-4m_f^2m_V^2\nonumber\\
	&-3m_V^4+m_f^2(2m_V^2-m_f^2)(\log\frac{m_\mu^2}{m_f^2}-i\pi)+m_V^2(3m_f^2-2m_V^2)(\log\frac{m_\mu^2}{m_V^2}-i\pi)\big]\Big\}+\mathcal{O}(\frac{1}{m_\mu^7}).
\end{align}

For the scenario of $m_f\ll m_V\ll m_\mu$, the approximations have been given in Eqs. \eqref{eqn:expansion:LVLL:mfllmmumV:esp} and \eqref{eqn:expansion:LVLR:mfllmmumV:esp}. For the scenario of $m_V\ll m_f\ll m_\mu$, the approximations have been given in Eqs. \eqref{eqn:expansion:LVLL:mVllmfmmu:esp} and \eqref{eqn:expansion:LVLR:mVllmfmmu:esp}. For the scenario of $m_f=m_V\ll m_\mu$, please refer to Eqs. \eqref{eqn:expansion:LVLL:mfmV:esp} and \eqref{eqn:expansion:LVLR:mfmV:esp}.

\subsubsection{Scenario of $m_f,m_\mu\ll m_V$}

Up to $1/m_V^4$, the integrals $L_{LL}^V$ and $L_{LR}^V$ can be expanded as
\begin{align}
&L_{LL}^V=\frac{5}{6}+\frac{2m_\mu^2-3m_f^2}{12m_V^2}+\frac{20m_f^4-15m_f^2m_\mu^2+3m_\mu^4}{40m_V^4}+\mathcal{O}(\frac{1}{m_V^6})
\end{align}
and
\begin{align}
&L_{LR}^V=\frac{m_f}{m_\mu}\big[-1+\frac{9m_f^2-5m_\mu^2}{12m_V^2}+\frac{27m_f^4+m_f^2m_\mu^2-3m_\mu^4+6m_f^2(m_\mu^2-3m_f^2)\log\frac{m_V^2}{m_f^2}}{12m_V^4}\big]+\mathcal{O}(\frac{1}{m_V^6}).
\end{align}

For the scenario of $m_f\ll m_\mu\ll m_V$, the approximations have been given in Eqs. \eqref{eqn:expansion:LVLL:mfllmmumV:esp} and \eqref{eqn:expansion:LVLR:mfllmmumV:esp}. For the scenario of $m_\mu\ll m_f\ll m_V$, the approximations have been given in Eqs. \eqref{eqn:expansion:LVLL:mmullmfmV:esp} and \eqref{eqn:expansion:LVLR:mmullmfmV:esp}. For the scenario of $m_f=m_\mu\ll m_V$, please refer to Eqs. \eqref{eqn:expansion:LVLL:mfmmu:esp} and \eqref{eqn:expansion:LVLR:mfmmu:esp}.

\subsubsection{Scenario of $m_\mu,m_V\ll m_f$}

Up to $1/m_f^4$, the integral $L_{LL}^V$ can be expanded as
\begin{align}
&L_{LL}^V=\frac{1}{3}+\frac{m_\mu^2-66m_V^2+36m_V^2\log\frac{m_f^2}{m_V^2}}{24m_f^2}\nonumber\\
&+\frac{m_\mu^4-190m_\mu^2m_V^2-260m_V^4+60m_V^2(m_\mu^2+4m_V^2)\log\frac{m_f^2}{m_V^2}}{40m_f^4}+\mathcal{O}(\frac{1}{m_f^6}).
\end{align}
Up to $1/m_f^3$, the integral $L_{LR}^f$ can be expanded as
\begin{align}
&L_{LR}^V=\frac{m_f}{m_\mu}\big[-\frac{1}{4}+\frac{27m_V^2-m_\mu^2-18m_V^2\log\frac{m_f^2}{m_V^2}}{12m_f^2}\nonumber\\
&+\frac{90m_V^4+106m_\mu^2m_V^2-m_\mu^4-36m_V^2(m_\mu^2+3m_V^2)\log\frac{m_f^2}{m_V^2}}{24m_f^4}\big]+\mathcal{O}(\frac{1}{m_f^5}).
\end{align}

For the scenario of $m_\mu\ll m_V\ll m_f$, the approximations have been given in Eqs. \eqref{eqn:expansion:LVLL:mmullmfmV:esp} and \eqref{eqn:expansion:LVLR:mmullmfmV:esp}. For the scenario of $m_V\ll m_\mu\ll m_f$, the approximations have been given in Eqs. \eqref{eqn:expansion:LVLL:mVllmfmmu:esp} and \eqref{eqn:expansion:LVLR:mVllmfmmu:esp}. For the scenario of $m_\mu=m_V\ll m_f$, please refer to Eqs. \eqref{eqn:expansion:LVLL:mmumV:esp} and \eqref{eqn:expansion:LVLR:mmumV:esp}.

\subsection{Degenerate mass case for the $L_{LL}^V$ and $L_{LR}^V$}\label{sec:expansion:LVdeg}

\subsubsection{Scenario of $m_f=m_\mu$}\label{sec:expansion:LVdeg:mfmu}

For $m_f=m_\mu$, the integrals $L_{LL}^V$ and $L_{LR}^V$ can be calculated as
\begin{align}\label{eqn:expansion:LVLL:mfmmu}
&L_{LL}^V=\left\{\begin{array}{ll}
\frac{1}{2m_{\mu}^6}\big[m_{\mu}^6+5m_{\mu}^4m_V^2-2m_{\mu}^2m_V^4-m_V^2(3m_{\mu}^4-4m_{\mu}^2m_V^2+m_V^4)\log\frac{m_{\mu}^2}{m_V^2}\big]\vspace{1ex}\\
	+\frac{m_V}{m_{\mu}^6\sqrt{m_V^2-4m_{\mu}^2}}(2m_{\mu}^6-9m_{\mu}^4m_V^2+6m_{\mu}^2m_V^4-m_V^6)\log\frac{m_V+\sqrt{m_V^2-4m_{\mu}^2}}{2m_{\mu}},&\mathrm{for}~m_\mu<\frac{m_V}{2}\vspace{3ex}\\
\frac{1}{2m_{\mu}^6}\big[m_{\mu}^6+5m_{\mu}^4m_V^2-2m_{\mu}^2m_V^4-m_V^2(3m_{\mu}^4-4m_{\mu}^2m_V^2+m_V^4)\log\frac{m_{\mu}^2}{m_V^2}\big]\vspace{1ex}\\
	+\frac{m_V}{m_{\mu}^6\sqrt{4m_{\mu}^2-m_V^2}}(2m_{\mu}^6-9m_{\mu}^4m_V^2+6m_{\mu}^2m_V^4-m_V^6)\arctan\sqrt{\frac{4m_{\mu}^2}{m_V^2}-1},&\mathrm{for}~m_\mu>\frac{m_V}{2}\vspace{3ex}\\
12\log 2-\frac{15}{2},&\mathrm{for}~m_\mu=\frac{m_V}{2}
\end{array}\right.
\end{align}
and
\begin{align}\label{eqn:expansion:LVLR:mfmmu}
&L_{LR}^V=\left\{\begin{array}{ll}
-\frac{1}{2m_{\mu}^4}\big[m_{\mu}^4+2m_{\mu}^2m_V^2+m_V^2(m_V^2-2m_{\mu}^2)\log\frac{m_{\mu}^2}{m_V^2}\big]\vspace{1ex}\\
	-\frac{m_V}{m_{\mu}^4\sqrt{m_V^2-4m_{\mu}^2}}(2m_{\mu}^4-4m_{\mu}^2m_V^2+m_V^4)\log\frac{m_V+\sqrt{m_V^2-4m_{\mu}^2}}{2m_{\mu}},&\mathrm{for}~m_\mu<\frac{m_V}{2}\vspace{3ex}\\
-\frac{1}{2m_{\mu}^4}\big[m_{\mu}^4+2m_{\mu}^2m_V^2+m_V^2(m_V^2-2m_{\mu}^2)\log\frac{m_{\mu}^2}{m_V^2}\big]\vspace{1ex}\\
	-\frac{m_V}{m_{\mu}^4\sqrt{4m_{\mu}^2-m_V^2}}(2m_{\mu}^4-4m_{\mu}^2m_V^2+m_V^4)\arctan\sqrt{\frac{4m_{\mu}^2}{m_V^2}-1},\quad\quad\quad&\mathrm{for}~m_\mu>\frac{m_V}{2}\vspace{3ex}\\
8\log 2-\frac{13}{2},&\mathrm{for}~m_\mu=\frac{m_V}{2}
\end{array}\right..
\end{align}

Then, they can be expanded as
\begin{align}\label{eqn:expansion:LVLL:mfmmu:esp}
L_{LL}^V\approx\left\{\begin{array}{ll}
\frac{5}{6}-\frac{m_{\mu}^2}{12m_V^2}+\frac{m_{\mu}^4}{5m_V^4},\quad& \mathrm{for}~m_\mu\ll m_V\vspace{1ex}\\
\frac{1}{2}+\frac{\pi m_V}{2m_{\mu}}+\frac{m_V^2}{m_{\mu}^2}(2-3\log\frac{m_{\mu}}{m_V})-\frac{35\pi m_V^3}{16m_{\mu}^3}+\frac{m_V^4}{6m_{\mu}^4}(7+24\log\frac{m_{\mu}}{m_V}),\quad\quad\quad& \mathrm{for}~m_V\ll m_\mu
\end{array}\right.
\end{align}
and
\begin{align}\label{eqn:expansion:LVLR:mfmmu:esp}
L_{LR}^V\approx\left\{\begin{array}{ll}
-1+\frac{m_{\mu}^2}{3m_V^2}+\frac{m_{\mu}^4}{12m_V^4}(25-24\log\frac{m_V}{m_{\mu}}),\quad& \mathrm{for}~m_\mu\ll m_V\vspace{1ex}\\
-\frac{1}{2}-\frac{\pi m_V}{2m_{\mu}}+\frac{m_V^2}{2m_{\mu}^2}(4\log\frac{m_{\mu}}{m_V}-1)+\frac{15\pi m_V^3}{16m_{\mu}^3}-\frac{m_V^4}{12m_{\mu}^4}(11+12\log\frac{m_{\mu}}{m_V}),\quad\quad\quad& \mathrm{for}~m_V\ll m_\mu
\end{array}\right..
\end{align}

\subsubsection{Scenario of $m_\mu=m_V$}\label{sec:expansion:LVdeg:mmumV}
For $m_\mu=m_V$, the integrals $L_{LL}^V$ and $L_{LR}^V$ can be calculated as
\begin{align}\label{eqn:expansion:LVLL:mmumV}
L_{LL}^V=\left\{\begin{array}{ll}
\frac{1}{4m_{\mu}^6}\big[m_{\mu}^2(2m_f^4+m_f^2m_{\mu}^2+5m_{\mu}^4)+m_f^4(m_{\mu}^2-m_f^2)\log\frac{m_f^2}{m_{\mu}^2}\big]\vspace{1ex}\\
	+\frac{m_f}{2m_{\mu}^6\sqrt{m_f^2-4m_{\mu}^2}}(m_f^6-3m_f^4m_{\mu}^2-2m_{\mu}^6)\log\frac{m_f+\sqrt{m_f^2-4m_{\mu}^2}}{2m_{\mu}},\quad\quad\quad& \mathrm{for}~m_{\mu}<\frac{m_f}{2}\vspace{3ex}\\
\frac{1}{4m_{\mu}^6}\big[m_{\mu}^2(2m_f^4+m_f^2m_{\mu}^2+5m_{\mu}^4)+m_f^4(m_{\mu}^2-m_f^2)\log\frac{m_f^2}{m_{\mu}^2}\big]\vspace{1ex}\\
	+\frac{m_f}{2m_{\mu}^6\sqrt{4m_{\mu}^2-m_f^2}}(m_f^6-3m_f^4m_{\mu}^2-2m_{\mu}^6)\arctan\sqrt{\frac{4m_{\mu}^2}{m_f^2}-1},& \mathrm{for}~m_{\mu}>\frac{m_f}{2}\vspace{3ex}\\
\frac{69}{4}-24\log2,& \mathrm{for}~m_{\mu}=\frac{m_f}{2}
\end{array}\right.
\end{align}
and
\begin{align}\label{eqn:expansion:LVLR:mmumV}
L_{LR}^V=\left\{\begin{array}{ll}
\frac{m_f}{4m_{\mu}^5}\big[-2m_{\mu}^2(m_f^2+2m_{\mu}^2)+(m_f^4+m_{\mu}^4)\log\frac{m_f^2}{m_{\mu}^2}\big]\vspace{1ex}\\
	+\frac{m_f^2}{2m_{\mu}^5\sqrt{m_f^2-4m_{\mu}^2}}(-m_f^4+2m_f^2m_{\mu}^2+m_{\mu}^4)\log\frac{m_f+\sqrt{m_f^2-4m_{\mu}^2}}{2m_{\mu}},\quad\quad\quad& \mathrm{for}~m_{\mu}<\frac{m_f}{2}\vspace{3ex}\\
\frac{m_f}{4m_{\mu}^5}\big[-2m_{\mu}^2(m_f^2+2m_{\mu}^2)+(m_f^4+m_{\mu}^4)\log\frac{m_f^2}{m_{\mu}^2}\big]\vspace{1ex}\\
	+\frac{m_f^2}{2m_{\mu}^5\sqrt{4m_{\mu}^2-m_f^2}}(-m_f^4+2m_f^2m_{\mu}^2+m_{\mu}^4)\arctan\sqrt{\frac{4m_{\mu}^2}{m_f^2}-1},& \mathrm{for}~m_{\mu}>\frac{m_f}{2}\vspace{3ex}\\
-13+17\log2,& \mathrm{for}~m_{\mu}=\frac{m_f}{2}
\end{array}\right. .
\end{align}
Then, they can be expanded as
\begin{align}\label{eqn:expansion:LVLL:mmumV:esp}
L_{LL}^V\approx\left\{\begin{array}{ll}
\frac{1}{3}+\frac{m_{\mu}^2}{24m_f^2}(-65+72\log\frac{m_f}{m_{\mu}})+\frac{m_{\mu}^4}{40m_f^4}(-449+600\log\frac{m_f}{m_{\mu}}),\quad\quad\quad& \mathrm{for}~m_\mu\ll m_f\vspace{1ex}\\
\frac{5}{4}-\frac{\pi m_f}{4m_{\mu}}+\frac{m_f^2}{2m_{\mu}^2}-\frac{\pi m_f^3}{32m_{\mu}^3}+\frac{m_f^4}{24m_{\mu}^4}(13-12\log\frac{m_{\mu}}{m_f}),\quad\quad& \mathrm{for}~m_f\ll m_\mu
\end{array}\right.
\end{align}
and
\begin{align}\label{eqn:expansion:LVLR:mmumV:esp}
L_{LR}^V\approx\left\{\begin{array}{ll}
-\frac{m_f}{4m_{\mu}}+\frac{m_{\mu}}{6m_f}(13-18\log\frac{m_f}{m_{\mu}})+\frac{m_{\mu}^3}{8m_f^3}(65-96\log\frac{m_f}{m_{\mu}}),\quad\quad\quad& \mathrm{for}~m_\mu\ll m_f\vspace{1ex}\\
-\frac{m_f}{2m_{\mu}}(\log\frac{m_{\mu}}{m_f}+2)+\frac{\pi m_f^2}{8m_{\mu}^2}-\frac{5m_f^3}{8m_{\mu}^3}+\frac{17\pi m_f^4}{64m_{\mu}^4}-\frac{m_f^5}{48m_{\mu}^5}(24\log\frac{m_{\mu}}{m_f}+13),\quad\quad& \mathrm{for}~m_f\ll m_\mu
\end{array}\right..
\end{align}

\subsubsection{Scenario of $m_f=m_V$}\label{sec:expansion:LVdeg:mfmV}
For $m_f=m_V$, the integrals $L_{LL}^V$ and $L_{LR}^V$ can be calculated as
\begin{align}\label{eqn:expansion:LVLL:mfmV}
L_{LL}^V=\left\{\begin{array}{ll}
\frac{7m_V^2+m_{\mu}^2}{4m_{\mu}^2}+\frac{m_V^2(7m_V^2-3m_{\mu}^2)}{2m_{\mu}^3\sqrt{4m_V^2-m_{\mu}^2}}\arctan\frac{m_{\mu}\sqrt{4m_V^2-m_{\mu}^2}}{m_{\mu}^2-2m_V^2},&\mathrm{for}~m_\mu<2m_V\vspace{1ex}\\
\frac{7m_V^2+m_{\mu}^2}{4m_{\mu}^2}+\frac{m_V^2(7m_V^2-3m_{\mu}^2)}{2m_{\mu}^3\sqrt{m_{\mu}^2-4m_V^2}}(2\log\frac{m_{\mu}+\sqrt{m_{\mu}^2-4m_V^2}}{2m_V}-i\pi),\quad\quad\quad&\mathrm{for}~m_\mu>2m_V\vspace{1ex}\\
\frac{3}{8},&\mathrm{for}~m_\mu=2m_V
\end{array}\right.
\end{align}
and
\begin{small}\label{eqn:expansion:LVLR:mfmV}
\begin{align}
L_{LR}^V=\left\{\begin{array}{ll}
-\frac{3m_V^3}{2m_{\mu}^3}+\frac{m_V^3(2m_{\mu}^2-3m_V^2)}{m_{\mu}^4\sqrt{4m_V^2-m_{\mu}^2}}\arctan\frac{m_{\mu}\sqrt{4m_V^2-m_{\mu}^2}}{m_{\mu}^2-2m_V^2},&\mathrm{for}~m_\mu<2m_V\vspace{1ex}\\
-\frac{3m_V^3}{2m_{\mu}^3}+\frac{m_V^3(2m_{\mu}^2-3m_V^2)}{m_{\mu}^4\sqrt{m_{\mu}^2-4m_V^2}}(2\log\frac{m_{\mu}+\sqrt{m_{\mu}^2-4m_V^2}}{2m_V}-i\pi),\quad\quad\quad&\mathrm{for}~m_\mu>2m_V\vspace{1ex}\\
\frac{1}{8},&\mathrm{for}~m_\mu=2m_V
\end{array}\right..
\end{align}
\end{small}
Then, they can be expanded as
\begin{align}\label{eqn:expansion:LVLL:mfmV:esp}
L_{LL}^V\approx\left\{\begin{array}{ll}
\frac{17}{24}+\frac{m_{\mu}^2}{15m_V^2}+\frac{m_{\mu}^4}{80m_V^4},& \mathrm{for}~m_\mu\ll m_V\vspace{1ex}\\
\frac{1}{4}+\frac{m_V^2}{4m_{\mu}^2}(7-12\log\frac{m_{\mu}}{m_V}+6i\pi)+\frac{m_V^4}{2m_{\mu}^4}(6+2\log\frac{m_{\mu}}{m_V}-i\pi),\quad\quad& \mathrm{for}~m_V\ll m_\mu
\end{array}\right.
\end{align}
and
\begin{align}\label{eqn:expansion:LVLR:mfmV:esp}
L_{LR}^V\approx\left\{\begin{array}{ll}
-\frac{3m_V}{4m_{\mu}}-\frac{7m_{\mu}}{60m_V}-\frac{19m_{\mu}^3}{840m_V^3},& \mathrm{for}~m_\mu\ll m_V\vspace{2ex}\\
\frac{m_V^3}{2m_{\mu}^3}(-3+8\log\frac{m_{\mu}}{m_V}-4i\pi)+\frac{m_V^5}{m_{\mu}^5}(-4+2\log\frac{m_{\mu}}{m_V}-i\pi), \quad\quad&\mathrm{for}~m_V\ll m_\mu
\end{array}\right..
\end{align}

\subsubsection{Scenario of $m_f=m_\mu=m_V$}
In this case, we have $L_{LL}^V=\frac{18-2\sqrt{3}\pi}{9}$ and
$L_{LR}^V=\frac{2\sqrt{3}\pi-27}{18}$.

\subsection{Short summary of the partial expansion results}\label{sec:Vexpansion:summary}
\subsubsection{Asymptotic behaviour of loop functions}
\begin{itemize}[leftmargin=10pt]
\item\textbf{Chiral limit of $m_{\mu}\rightarrow0$}\\
As $m_{\mu}$ goes to zero, the $\frac{m_\mu^2}{m_V^2}L_{LL}^f$, $\frac{m_\mu^2}{m_V^2}L_{LR}^f$, $\frac{m_\mu^2}{m_V^2}L_{LL}^V$, and $\frac{m_\mu^2}{m_V^2}L_{LR}^V$ vanish, which is expected from chiral symmetry.
\item\textbf{Chiral limit of $m_f\rightarrow0$}\\
As $m_f$ goes to zero, the $L_{LR}^V$ vanishes. As $m_f$ goes to zero, the $L_{LR}^f$ also vanishes except for the $m_\mu=m_V$ case.
\item\textbf{Limit of $m_V\rightarrow0$}\\
As $m_V$ goes to zero, we find that $\frac{m_\mu^2}{m_V^2}L_{LL}^{f(V)}$ and $\frac{m_\mu^2}{m_V^2}L_{LR}^{f(V)}$ diverges. In the scenario of $m_{\mu}=m_f$, the $\frac{m_\mu^2}{m_V^2}(L_{LL}^f+L_{LR}^f)$ is finite because the divergent terms $\frac{m_{\mu}^2}{m_V^2}$ and $\log\frac{m_{\mu}}{m_V}$ are exactly cancelled. This corresponds to the photon contribution to $(g-2)_{\mu}$ in QED. For vectorial couplings $g_L=g_R=g$, the contribution $(|g_L|^2+|g_R|^2)L_{LL}^f+[g_L(g_R)^\ast+g_R(g_L)^\ast]L_{LR}^f$ only depend on the sum $(L_{LL}^f+L_{LR}^f)$.
\item\textbf{Decoupling limit of $m_f\rightarrow\infty$}\\
As $m_V$ goes to infinity, we find that $L_{LL}^{f(V)}$ is a constant and $L_{LR}^{f(V)}$ is proportional to $m_f$. However, the key point is that the couplings $g_L$ and $g_R$ are suppressed by the heavy fermion mass, for example $1/m_f$. Then, both the $(|g_L|^2+|g_R|^2)L_{LL}^{f(V)}$ and $[g_L(g_R)^\ast+g_R(g_L)^\ast]L_{LR}^{f(V)}$ vanish, which is consistent with the spirit of decoupling theorem \cite{Appelquist:1974tg}. 
\item\textbf{Decoupling limit of $m_V\rightarrow\infty$}\\
As $m_V$ goes to infinity, the $\frac{m_\mu^2}{m_V^2}L_{LL}^f$, $\frac{m_\mu^2}{m_V^2}L_{LR}^f$, $\frac{m_\mu^2}{m_V^2}L_{LL}^V$, and $\frac{m_\mu^2}{m_V^2}L_{LR}^V$ vanish, which is consistent with the spirit of decoupling theorem \cite{Appelquist:1974tg}. 
\end{itemize}

In Tab. \ref{tab:sum:asym:vector}, we show the behaviour of $\frac{m_\mu^2}{m_V^2}L_{LL(LR)}^{f(V)}$ in the massless and infinite mass limits.
\begin{table}[!htb]
\begin{center}
\begin{tabular}{c|c|c|c}
\hline
\multicolumn{2}{c|}{} & \rule[-9pt]{0pt}{22pt}$\ds\frac{m_\mu^2}{m_V^2}L_{LL}^f$ & $\ds\frac{m_\mu^2}{m_V^2}L_{LR}^f$ \\
\hline
\multicolumn{2}{c|}{\rule[-9pt]{0pt}{20pt}Chiral limit $m_{\mu}\rightarrow0$} & 0 & 0 \\\hline
\multirow{2}{*}{\rule[-9pt]{0pt}{28pt}\makecell{Chiral limit \\$m_f\rightarrow0$}} & $m_{\mu}\ne m_V$ & \rule[-24pt]{0pt}{52pt}$\makecell{\ds-\frac{(m_\mu^2-2m_V^2)^2}{4m_\mu^2m_V^2}+\frac{m_V^2(3m_\mu^2-2m_V^2)}{2m_\mu^4}\log(1-\frac{m_\mu^2}{m_V^2})}$ & 0\\\cline{2-4}
 \rule[-10pt]{0pt}{24pt}& $m_{\mu}=m_V$ & $\ds-\frac{1}{4}-\frac{1}{2}\log\frac{m_{\mu}}{m_f}$ & $\ds\frac{\pi}{4}$ \\\hline
\multirow{2}{*}{\rule[-9pt]{0pt}{28pt}\makecell{Limit of \\$m_V\rightarrow0$}} & \rule[-10pt]{0pt}{25pt}$m_{\mu}\ne m_f$ & $\rule[-24pt]{0pt}{52pt}\makecell{\ds\frac{1}{4m_\mu^4m_V^2}\Big[m_\mu^2(2m_f^4-3m_f^2m_\mu^2-m_\mu^4)+\vspace{1ex}\\ \ds2m_f^2(m_f^2-m_\mu^2)^2\log(1-\frac{m_\mu^2}{m_f^2})\Big]+1+\frac{m_f^2}{m_\mu^2}\log(1-\frac{m_\mu^2}{m_f^2})}$ & $\makecell{\ds\frac{m_f}{m_\mu}\Big\{\frac{1}{2m_\mu^2m_V^2}\Big[-m_f^2m_\mu^2+2m_\mu^4-(m_f^2-m_\mu^2)^2\cdot\vspace{1ex}\\  \ds\log(1-\frac{m_\mu^2}{m_f^2})\Big]-\frac{1}{2}-\frac{m_f^2+m_\mu^2}{2m_\mu^2}\log(1-\frac{m_\mu^2}{m_f^2})\Big\}}$ \\\cline{2-4}
\rule[-11pt]{0pt}{26pt} & $m_{\mu}=m_f$ & $\ds-\frac{m_{\mu}^2}{2m_V^2}+\frac{3}{2}-\log\frac{m_{\mu}}{m_V}$ & $\ds\frac{m_{\mu}^2}{2m_V^2}+\log\frac{m_{\mu}}{m_V}-1$ \\\hline
\multicolumn{2}{c|}{\rule[-10pt]{0pt}{24pt}\makecell{Decoupling limit\\ $m_f\rightarrow\infty$}} & $\ds-\frac{5m_\mu^2}{12m_V^2}$ & $\ds\frac{m_fm_\mu}{4m_V^2}$ \\\hline
\multicolumn{2}{c|}{\makecell{Decoupling limit\\ $m_V\rightarrow\infty$}} & 0 & 0 \\\hline\hline\hline
\multicolumn{2}{c|}{} & \rule[-9pt]{0pt}{22pt}$\ds\frac{m_\mu^2}{m_V^2}L_{LL}^V$ & $\ds\frac{m_\mu^2}{m_V^2}L_{LR}^V$ \\
\hline
\multicolumn{2}{c|}{\rule[-9pt]{0pt}{20pt}Chiral limit $m_{\mu}\rightarrow0$} & 0 & 0 \\\hline
\multirow{2}{*}{\rule[-9pt]{0pt}{28pt}\makecell{Chiral limit \\$m_f\rightarrow0$}} & $m_{\mu}\ne m_V$ & \rule[-24pt]{0pt}{52pt}$\makecell{\ds\frac{1}{4m_\mu^4m_V^2}\Big[m_\mu^6+8m_\mu^4m_V^2-4m_\mu^2m_V^4\vspace{1ex}\\ \ds+2m_V^2(m_\mu^2-m_V^2)(2m_V^2-3m_\mu^2)\log\frac{m_V^2-m_\mu^2}{m_V^2}\Big]}
$ & 0 \\\cline{2-4}
\rule[-10pt]{0pt}{24pt} & $m_{\mu}=m_V$ & $\ds\frac{5}{4}$ & 0 \\\hline
\multirow{2}{*}{\rule[-9pt]{0pt}{28pt}\makecell{Limit of \\$m_V\rightarrow0$}} & \rule[-30pt]{0pt}{65pt}$m_{\mu}\ne m_f$ & $\makecell{\frac{1}{4m_\mu^4m_V^2}\Big[m_\mu^2(2m_f^4-m_f^2m_\mu^2+m_\mu^4)+2m_f^4(m_f^2-m_\mu^2)\vspace{1ex}\\ \cdot\log\frac{m_f^2-m_\mu^2}{m_f^2}\Big]+\frac{1}{2m_\mu^2(m_f^2-m_\mu^2)}\Big[3m_\mu^4\log\frac{m_f^2}{m_V^2}\vspace{1ex}\\ +(3m_f^4+3m_\mu^4)\log\frac{m_f^2-m_\mu^2}{m_f^2}+m_\mu^2(3m_f^2-4m_\mu^2)\Big]}$ & $\makecell{\frac{m_f}{m_\mu}\Big\{\frac{m_f^2}{2m_\mu^2m_V^2}\Big[-m_\mu^2+(m_\mu^2-m_f^2)\log\frac{m_f^2-m_\mu^2}{m_f^2}\Big]\vspace{1ex}\\ +\frac{1}{2m_\mu^2(m_f^2-m_{\mu}^2)}\Big[m_\mu^2(2m_{\mu}^2-m_f^2)-3m_{\mu}^4\log\frac{m_f^2}{m_V^2}\vspace{1ex}\\ -(3m_{\mu}^4+2m_{\mu}^2m_f^2+m_f^4)\log\frac{m_f^2-m_\mu^2}{m_f^2}\Big]\Big\}}$ \\\cline{2-4}
\rule[-11pt]{0pt}{26pt} & $m_{\mu}=m_f$ & $\ds\frac{m_{\mu}^2}{2m_V^2}+\frac{\pi m_{\mu}}{2m_V}+2-3\log\frac{m_{\mu}}{m_V}$ & $\ds-\frac{m_{\mu}^2}{2m_V^2}-\frac{\pi m_{\mu}}{2m_V}+2\log\frac{m_{\mu}}{m_V}-\frac{1}{2}$ \\\hline
\multicolumn{2}{c|}{\rule[-10pt]{0pt}{24pt}\makecell{Decoupling limit\\ $m_f\rightarrow\infty$}} & $\ds\frac{m_\mu^2}{3m_V^2}$ & $\ds-\frac{m_fm_\mu}{4m_V^2}$ \\\hline
\multicolumn{2}{c|}{\makecell{Decoupling limit\\ $m_V\rightarrow\infty$}} & 0 & 0 \\\hline
\end{tabular}
\caption{The behaviour of $\ds\frac{m_\mu^2}{m_V^2}L_{LL(LR)}^f$ (upper table) and $\ds\frac{m_\mu^2}{m_V^2}L_{LL(LR)}^V$ (lower table) for the vector mediator case under different mass limits. In the above scenario of $m_{\mu}\ne m_i(i=V,f)$, the $\ds\log(1-\frac{m_\mu^2}{m_i^2})$ is valid for $m_{\mu}<m_i$, which should be taken as $\ds[\log(\frac{m_\mu^2}{m_i^2}-1)-i\pi]$ for $m_{\mu}>m_i$.} \label{tab:sum:asym:vector}
\end{center}
\end{table}

\subsubsection{Expansion of contributions to $(g-2)_\mu$ under special scenarios}
In Tab. \ref{tab:sum:exp:vector}, we list the leading order formulae of the $\Delta a_{\mu}$ for the scalar mediator cases in different scenarios. Here, we only collect the results for special scenarios. For complete expressions, please refer to Sec. \ref{sec:model:contri:VectorRep}, Sec. \ref{sec:expansion:Lf}, Sec. \ref{sec:expansion:Lfdeg}, Sec. \ref{sec:expansion:LV}, and Sec. \ref{sec:expansion:LVdeg}.

\FloatBarrier
\begin{table}[!htb]
\begin{center}
\begin{tabular}{c|c}
\hline
\rule[-10pt]{0pt}{25pt} scenarios & $\ds\Delta a_{\mu}/(\frac{N_Cm_{\mu}^2}{8\pi^2m_V^2})$ \\
\hline
\rule[-10pt]{0pt}{25pt} $m_\mu\ll m_f\ll m_V$ & $\makecell{\ds(|g_L|^2+|g_R|^2)\cdot(\frac{2}{3}Q_f-\frac{5}{6}Q_V)+\left[g_L(g_R)^\ast+g_R(g_L)^\ast\right]\cdot\frac{m_f}{m_{\mu}}(-Q_f+Q_V)}$ \\
\hline
\rule[-10pt]{0pt}{25pt} $m_\mu\ll m_V\ll m_f$ & $\makecell{\ds(|g_L|^2+|g_R|^2)(\frac{5}{12}Q_f-\frac{1}{3}Q_V)+\left[g_L(g_R)^\ast+g_R(g_L)^\ast\right]\cdot\frac{m_f}{4m_{\mu}}(-Q_f+Q_V)}$ \\
\hline
\rule[-19pt]{0pt}{43pt} $m_V\ll m_f\ll m_\mu$ & $\makecell{\ds(|g_L|^2+|g_R|^2)(\frac{1}{4}Q_f-\frac{1}{4}Q_V)\\\ds+\left[g_L(g_R)^\ast+g_R(g_L)^\ast\right]\cdot\frac{m_f}{m_{\mu}}[(\log\frac{m_{\mu}}{m_f}-\frac{1}{2}i\pi-1)Q_f+\frac{m_f^2}{m_{\mu}^2}(-\log\frac{m_{\mu}}{m_f}+\frac{1}{2}i\pi+\frac{1}{2})Q_V]}$ \\
\hline
\rule[-10pt]{0pt}{25pt} $m_V\ll m_\mu\ll m_f$ &  $\makecell{\ds(|g_L|^2+|g_R|^2)(\frac{5}{12}Q_f-\frac{1}{3}Q_V)+\left[g_L(g_R)^\ast+g_R(g_L)^\ast\right]\cdot\frac{m_f}{4m_{\mu}}(-Q_f+Q_V)}$ \\
\hline
\rule[-10pt]{0pt}{25pt} $m_f\ll m_\mu\ll m_V$ & $\makecell{\ds(|g_L|^2+|g_R|^2)(\frac{2}{3}Q_f-\frac{5}{6}Q_V)+\left[g_L(g_R)^\ast+g_R(g_L)^\ast\right]\cdot\frac{m_f}{m_{\mu}}(-Q_f+Q_V)}$ \\
\hline
\rule[-19pt]{0pt}{43pt} $m_f\ll m_V\ll m_\mu$ & $\makecell{\ds(|g_L|^2+|g_R|^2)(\frac{1}{4}Q_f-\frac{1}{4}Q_V)\\\ds+\left[g_L(g_R)^\ast+g_R(g_L)^\ast\right]\cdot\frac{m_f}{m_{\mu}}[(\log\frac{m_{\mu}}{m_f}-\frac{1}{2}i\pi-1)Q_f+\frac{m_V^2}{m_{\mu}^2}(-3\log\frac{m_{\mu}}{m_V}+\frac{3}{2}i\pi+1)Q_V]}$ \\\hline\hline\hline
\rule[-10pt]{0pt}{25pt} $m_V\ll m_f=m_\mu$ & $\makecell{\ds(|g_L|^2+|g_R|^2)(\frac{1}{2}Q_f-\frac{1}{2}Q_V)+\left[g_L(g_R)^\ast+g_R(g_L)^\ast\right](-\frac{1}{2}Q_f+\frac{1}{2}Q_V)}$ \\
\hline
\rule[-10pt]{0pt}{25pt} $m_f=m_\mu\ll m_V$ & $\makecell{\ds(|g_L|^2+|g_R|^2)(\frac{2}{3}Q_f-\frac{5}{6}Q_V)+\left[g_L(g_R)^\ast+g_R(g_L)^\ast\right](-Q_f+Q_V)}$ \\
\hline
\rule[-20pt]{0pt}{45pt} $m_f\ll m_\mu=m_V$ & $\makecell{\ds(|g_L|^2+|g_R|^2)\cdot[Q_f(\frac{1}{4}+\frac{1}{2}\log\frac{m_{\mu}}{m_f})-\frac{5}{4}Q_V]\vspace{1ex}\\\ds+\left[g_L(g_R)^\ast+g_R(g_L)^\ast\right]\cdot[-\frac{\pi}{4}Q_f+\frac{m_f}{2m_{\mu}}(\log\frac{m_{\mu}}{m_f}+2)Q_V]}$  \\
\hline
\rule[-10pt]{0pt}{25pt} $m_\mu=m_V\ll m_f$ & $\makecell{\ds(|g_L|^2+|g_R|^2)(\frac{5}{12}Q_f-\frac{1}{3}Q_V)+\left[g_L(g_R)^\ast+g_R(g_L)^\ast\right]\cdot\frac{m_f}{4m_{\mu}}(-Q_f+Q_V)}$ \\
\hline
\rule[-10pt]{0pt}{25pt} $m_\mu\ll m_f=m_V$ & $\makecell{\ds(|g_L|^2+|g_R|^2)(\frac{13}{24}Q_f-\frac{17}{24}Q_V)+\left[g_L(g_R)^\ast+g_R(g_L)^\ast\right]\cdot\frac{m_V}{m_{\mu}}(-\frac{1}{2}Q_f+\frac{3}{4}Q_V)}$ \\
\hline
\rule[-20pt]{0pt}{45pt} $m_f=m_V\ll m_\mu$ & $\makecell{\ds(|g_L|^2+|g_R|^2)(\frac{1}{4}Q_f-\frac{1}{4}Q_V)\vspace{1ex}\\\ds+\left[g_L(g_R)^\ast+g_R(g_L)^\ast\right]\cdot\frac{m_V}{m_{\mu}}[(\log\frac{m_{\mu}}{m_V}-\frac{1}{2}i\pi-1)Q_f+\frac{m_V^2}{m_{\mu}^2}(-4\log\frac{m_{\mu}}{m_V}+2i\pi+\frac{3}{2})Q_V]}$ \\\hline\hline\hline
\rule[-10pt]{0pt}{25pt} $m_f=m_\mu=m_V$ & $\makecell{\ds(|g_L|^2+|g_R|^2)(\frac{\sqrt{3}\pi}{9}Q_f+\frac{2\sqrt{3}\pi-18}{9}Q_V)+\left[g_L(g_R)^\ast+g_R(g_L)^\ast\right](\frac{9-4\sqrt{3}\pi}{18}Q_f+\frac{27-2\sqrt{3}\pi}{18}Q_V)}$ \\
\hline
\end{tabular}
\caption{Leading order formulae of the $\Delta a_{\mu}$ for the vector mediator case in different scenarios. These results are consistent with those in Ref. \cite{Queiroz:2014zfa}. Note that we have extracted the common factor $\ds\frac{N_Cm_{\mu}^2}{8\pi^2m_V^2}$. As aforementioned, the charge conservation requires $Q_f+Q_V=-1$.} \label{tab:sum:exp:vector}
\end{center}
\end{table}
\FloatBarrier

Similarly, the left-handed or right-handed gauge couplings are independent. The $(|g_L|^2+|g_R|^2)$ is always non-negative, while the sign of $\left[g_L(g_R)^\ast+g_R(g_L)^\ast\right]$ is indefinite. If both the left-handed and right-handed gauge couplings are present, it is possible to obtain the positive or negative contributions through adjusting the sign of $g_L$ and $g_R$. If either the left-handed or the right-handed gauge coupling is absent, the loop functions must be positive (negative) to obtain the positive (negative) contributions. In Tab. \ref{tab:sum:exp:vectorLL}, we list the leading order formulae of the $\Delta a_{\mu}$ with pure left-handed or right-handed gauge couplings for the vector mediator cases in different scenarios. Furthermore, we give the electric charge bounds required by positive $\Delta a_{\mu}$.
\begin{table}[!htb]
\begin{center}
\begin{tabular}{c|c|c}
\hline
\rule[-9pt]{0pt}{22pt} scenarios & $\ds\Delta a_{\mu}/(\frac{N_Cm_{\mu}^2}{8\pi^2m_V^2})$ & condition of $\Delta a_{\mu}>0$ \\ 
\hline
\rule[-9pt]{0pt}{22pt} $m_\mu\ll m_f\ll m_V$ & $\ds|g_{L(R)}|^2\cdot(\frac{2}{3}Q_f-\frac{5}{6}Q_V)$ & $\ds Q_f>-\frac{5}{9}$ or $\ds Q_V<-\frac{4}{9}$\\
\hline
\rule[-9pt]{0pt}{22pt} $m_\mu\ll m_V\ll m_f$ & $\ds|g_{L(R)}|^2\cdot(\frac{5}{12}Q_f-\frac{1}{3}Q_V)$ & $\ds Q_f>-\frac{4}{9}$ or $\ds Q_V<-\frac{5}{9}$ \\
\hline
\rule[-9pt]{0pt}{22pt} $m_V\ll m_f\ll m_\mu$ & $\ds|g_{L(R)}|^2\cdot(\frac{1}{4}Q_f-\frac{1}{4}Q_V)$ &  $\ds Q_f>-\frac{1}{2}$ or $\ds Q_V<-\frac{1}{2}$\\
\hline
\rule[-9pt]{0pt}{22pt} $m_V\ll m_\mu\ll m_f$ &  $\ds|g_{L(R)}|^2\cdot(\frac{5}{12}Q_f-\frac{1}{3}Q_V)$ & $\ds Q_f>-\frac{4}{9}$ or $\ds Q_V<-\frac{5}{9}$ \\
\hline
\rule[-9pt]{0pt}{22pt} $m_f\ll m_\mu\ll m_V$ & $\ds|g_{L(R)}|^2\cdot(\frac{2}{3}Q_f-\frac{5}{6}Q_V)$ & $\ds Q_f>-\frac{5}{9}$ or $\ds Q_V<-\frac{4}{9}$ \\
\hline
\rule[-9pt]{0pt}{22pt} $m_f\ll m_V\ll m_\mu$ & $\ds|g_{L(R)}|^2\cdot(\frac{1}{4}Q_f-\frac{1}{4}Q_V)$ & $\ds Q_f>-\frac{1}{2}$ or $\ds Q_V<-\frac{1}{2}$ \\\hline\hline\hline
\rule[-9pt]{0pt}{22pt} $m_V\ll m_f=m_\mu$ & $\ds|g_{L(R)}|^2\cdot(\frac{1}{2}Q_f-\frac{1}{2}Q_V)$ & $\ds Q_f>-\frac{1}{2}$ or $\ds Q_V<-\frac{1}{2}$\\
\hline
\rule[-9pt]{0pt}{22pt} $m_f=m_\mu\ll m_V$ & $\ds|g_{L(R)}|^2\cdot(\frac{2}{3}Q_f-\frac{5}{6}Q_V)$ & $\ds Q_f>-\frac{5}{9}$ or $\ds Q_V<-\frac{4}{9}$ \\
\hline
\rule[-10pt]{0pt}{24pt} $m_f\ll m_\mu=m_V$ & $\ds|g_{L(R)}|^2\cdot[Q_f(\frac{1}{4}+\frac{1}{2}\log\frac{m_{\mu}}{m_f})-\frac{5}{4}Q_V]$ & $\ds Q_f\gtrsim0$ or $\ds Q_V\lesssim-1$\\
\hline
\rule[-9pt]{0pt}{22pt} $m_\mu=m_V\ll m_f$ & $\ds|g_{L(R)}|^2\cdot(\frac{5}{12}Q_f-\frac{1}{3}Q_V)$ & $\ds Q_f>-\frac{4}{9}$ or $\ds Q_V<-\frac{5}{9}$ \\
\hline
\rule[-9pt]{0pt}{22pt} $m_\mu\ll m_f=m_V$ & $\ds|g_{L(R)}|^2\cdot(\frac{13}{24}Q_f-\frac{17}{24}Q_V)$ & $\ds Q_f>-\frac{17}{30}$ or $\ds Q_V<-\frac{13}{30}$\\
\hline
\rule[-9pt]{0pt}{22pt} $m_f=m_V\ll m_\mu$ & $\ds|g_{L(R)}|^2\cdot(\frac{1}{4}Q_f-\frac{1}{4}Q_V)$ & $\ds Q_f>-\frac{1}{2}$ or $\ds Q_V<-\frac{1}{2}$ \\
\hline\hline\hline
\rule[-24pt]{0pt}{52pt} $m_f=m_\mu=m_V$ & $\ds|g_{L(R)}|^2\cdot(\frac{\sqrt{3}\pi}{9}Q_f+\frac{2\sqrt{3}\pi-18}{9}Q_V)$ & \makecell{$\ds Q_f>-\frac{6\sqrt{3}-2\pi}{6\sqrt{3}-\pi}\approx-0.57$ \vspace{1ex}\\ or $\ds Q_V<-\frac{\pi}{6\sqrt{3}-\pi}\approx-0.43$} \\
\hline
\end{tabular}
\caption{Leading order formulae of the $\Delta a_{\mu}$ with pure left-handed (right-handed) couplings for the vector mediator case in different scenarios. Note that we have extracted the common factor $\ds\frac{N_Cm_{\mu}^2}{8\pi^2m_V^2}$. As aforementioned, the charge conservation requires $Q_f+Q_V=-1$.} \label{tab:sum:exp:vectorLL}
\end{center}
\end{table}


\section{Applications and examples in specific models}\label{sec:summary:examples}

\subsection{Compute the muon MDM in SM}
$\bullet$ For the photon contributions, the interactions can be written as $-e\bar{\mu}\gamma^{\mu}\mu A_{\mu}$. Hence, we should take $g_L=g_R=-e$. According to the parameterizations in Eq. \eqref{eqn:contri:structure:chiral}, the muon MDM here is calculated as
\begin{align}
&\Delta a_{\mu}^{\gamma}=\lim\limits_{m_{\gamma}\rightarrow0}\frac{m_{\mu}^2}{8\pi^2m_{\gamma}^2}\big\{(|g_L|^2+|g_R|^2)L_{LL}^f(m_{\mu},m_{\mu},m_{\gamma})+[g_L(g_R)^\ast+g_R(g_L)^\ast]L_{LR}^f(m_{\mu},m_{\mu},m_{\gamma})\big\}.
\end{align}
Assuming $m_V\ll m_f=m_{\mu}$ and inserting the expansion forms of $L_{LL}^f$ in Eq. \eqref{eqn:expansion:LfLL:mfmmu} and $L_{LR}^f$ in Eq. \eqref{eqn:expansion:LfLR:mfmmu}, we can obtain the following expressions of muon MDM: 
\begin{align}
&\Delta a_{\mu}^{\gamma}=\lim\limits_{m_{\gamma}\rightarrow0}\frac{e^2m_{\mu}^2}{4\pi^2m_{\gamma}^2}[-\frac{1}{2m_{\mu}^6}(m_{\mu}^6-3m_{\mu}^4m_{\gamma}^2+m_{\gamma}^2m_{\mu}^4\log\frac{m_{\mu}^2}{m_{\gamma}^2})\nonumber\\
	&+\frac{1}{2m_{\mu}^4}(m_{\mu}^4-2m_{\mu}^2m_{\gamma}^2+m_{\gamma}^2m_{\mu}^2\log\frac{m_{\mu}^2}{m_{\gamma}^2})]=\frac{e^2}{8\pi^2}.
\end{align}

$\bullet$ For the $W$ boson contributions, the interactions can be written as $(g/\sqrt{2})\overline{\mu_L}\gamma^{\mu}\nu_LW_{\mu}^-+\mathrm{h.c.}$. Hence, we should take $g_L=g/\sqrt{2}$ and $g_R=0$. According to the parameterizations in Eq. \eqref{eqn:contri:structure:chiral}, the muon MDM here is calculated as
\begin{align}
&\Delta a_{\mu}^W=\frac{m_{\mu}^2}{8\pi^2m_W^2}|g_L|^2L_{LL}^V(m_{\mu},m_{\nu},m_W).
\end{align}
Assuming $m_f\ll m_{\mu}\ll m_V$ and inserting the expansion forms of $L_{LL}^V$ in Eq. \eqref{eqn:expansion:LVLL:mfllmmumV}, we can obtain the following expressions of muon MDM: 
\begin{align}
&\Delta a_{\mu}^W\approx\frac{g^2m_{\mu}^2}{16\pi^2m_W^2}[\frac{1}{4}+\frac{2m_W^2}{m_{\mu}^2}-\frac{m_W^4}{m_{\mu}^4}+\frac{m_W^2}{2m_{\mu}^2}(1-\frac{m_W^2}{m_{\mu}^2})(\frac{2m_W^2}{m_{\mu}^2}-3)\log(1-\frac{m_{\mu}^2}{m_W^2})]\nonumber\\
&\approx\frac{5g^2m_{\mu}^2}{96\pi^2m_W^2}(1+\frac{m_{\mu}^2}{5m_W^2}+\frac{9m_{\mu}^4}{100m_W^4}).
\end{align}
Note the subleading terms are suppressed by $m_{\mu}^2/m_W^2\sim\mc{O}(10^{-6})$, which is far below the current experimental precision.\\[-2ex]

$\bullet$ For the $Z$ boson contributions, the interactions can be written as $(g/c_W)[(-1/2+s_W^2)\overline{\mu_L}\gamma^{\mu}\mu_L+s_W^2\overline{\mu_R}\gamma^{\mu}\mu_R]Z_{\mu}$. Hence, we should take $g_L=g(2s_W^2-1)/(2c_W)$ and $g_R=gs_W^2/c_W$. According to the parameterizations in Eq. \eqref{eqn:contri:structure:chiral}, the muon MDM here is calculated as
\begin{align}
&\Delta a_{\mu}^Z=\frac{m_{\mu}^2}{8\pi^2m_Z^2}\big\{(|g_L|^2+|g_R|^2)L_{LL}^f(m_{\mu},m_{\mu},m_Z)+[g_L(g_R)^\ast+g_R(g_L)^\ast]L_{LR}^f(m_{\mu},m_{\mu},m_Z)\big\}.
\end{align}
Assuming $m_f=m_{\mu}\ll m_V$ and inserting the expansion forms of $L_{LL}^f$ in Eq. \eqref{eqn:expansion:LfLL:mfmmu:esp} and $L_{LR}^f$ in Eq. \eqref{eqn:expansion:LfLR:mfmmu:esp}, we can obtain the following expressions of muon MDM: 
\begin{align}
&\Delta a_{\mu}^Z\approx\frac{g^2m_{\mu}^2}{8\pi^2m_Z^2c_W^2}\big\{[(s_W^2-\frac{1}{2})^2+s_W^4][-\frac{2}{3}+\frac{m_{\mu}^2}{4m_Z^2}+\frac{m_{\mu}^4}{60m_Z^4}(137-120\log\frac{m_Z}{m_{\mu}})]\nonumber\\
	&+s_W^2(2s_W^2-1)[1+\frac{m_{\mu}^2}{6m_Z^2}(11-12\log\frac{m_Z}{m_{\mu}})+\frac{m_{\mu}^4}{12m_Z^4}(89-120\log\frac{m_Z}{m_{\mu}})]\big\}\nonumber\\
&\approx\frac{g^2m_{\mu}^2}{48\pi^2m_Z^2c_W^2}(4s_W^4-2s_W^2-1).
\end{align}

$\bullet$ For the Higgs contributions, the interactions can be written as $(-m_{\mu}/v)\bar{\mu}\mu h$. Hence, we should take $y_L=y_R=-m_{\mu}/v$. According to the parameterizations in Eq. \eqref{eqn:contri:structure:chiral}, the muon MDM here is calculated as
\begin{align}
&\Delta a_{\mu}^h=\frac{m_{\mu}^2}{8\pi^2m_h^2}\big\{(|y_L|^2+|y_R|^2)I_{LL}^f(m_{\mu},m_{\mu},m_h)+[y_L(y_R)^\ast+y_R(y_L)^\ast]I_{LR}^f(m_{\mu},m_{\mu},m_h)\big\}.
\end{align}
Assuming $m_f=m_{\mu}\ll m_S$ and inserting the expansion forms of $I_{LL}^f$ in Eq. \eqref{eqn:expansion:IfLL:mfmmu:esp} and $I_{LR}^f$ in Eq. \eqref{eqn:expansion:IfLR:mfmmu:esp}, we can obtain the following expressions of muon MDM: 
\begin{align}
&\Delta a_{\mu}^h\approx\frac{m_{\mu}^4}{4\pi^2m_h^2v^2}[\log\frac{m_h}{m_{\mu}}-\frac{7}{12}+\frac{m_{\mu}^2}{m_h^2}(3\log\frac{m_h}{m_{\mu}}-\frac{13}{8})+\frac{m_{\mu}^4}{m_h^4}(9\log\frac{m_h}{m_{\mu}}-\frac{201}{40})].
\end{align}

These results agree exactly with those in Refs. \cite{Jackiw:1972jz, Bardeen:1972vi, Leveille:1977rc}.

\subsection{Compute the muon MDM induced by doubly charged mediators}\label{sec:summary:examples:doublycharged}
For the interactions $\bar{\mu}(y_L\omega_-+y_R\omega_+)\mu^CS+\bar{\mu}\gamma^{\mu}(g_L\omega_-+g_R\omega_+)\mu^CV_{\mu}+\mathrm{h.c.}$, they can be transformed into the canonical interactions through field redefinition of $f\equiv \mu^C$. Here, $Q_f=Q_{\mu^C}=-Q_{\mu}=1$ and $Q_S=Q_V=-2$. One tricky thing is that we should compensate the symmetry factor due to identical muon fields as pointed in Ref. \cite{Queiroz:2014zfa}. According to the parameterizations in Eq. \eqref{eqn:contri:structure:chiral}, the contributions here are calculated as
\begin{align}
&\Delta a_{\mu}=\frac{m_{\mu}^2}{2\pi^2}\Big\{\frac{(|y_L|^2+|y_R|^2)(-I_{LL}^f+2I_{LL}^S)+\left[y_L(y_R)^\ast+y_R(y_L)^\ast\right](-I_{LR}^f+2I_{LR}^S)}{m_S^2}\nonumber\\
	&+\frac{(|g_L|^2+|g_R|^2)(-L_{LL}^f+2L_{LL}^V)+\left[g_L(g_R)^\ast+g_R(g_L)^\ast\right](-L_{LR}^f+2L_{LR}^V)}{m_V^2}\Big\}.
\end{align}
In the above, the variables of scalar and vector loop functions are $(m_{\mu},m_{\mu},m_S)$ and $(m_{\mu},m_{\mu},m_V)$. The total extra factor 4 is the multiplication of symmetry factor 2 in each $\bar{\mu}\mu^CS(V)$ vertex. As a matter of fact, this can also be checked by the traditional contraction method of the canonical quantization in quantum field theory.

Because the doubly charged mediators are always as heavy as TeV, we can assume $m_f=m_{\mu}\ll m_{S(V)}$ here. For the scalar case, we should adopt the expansion forms of $I_{LL}^f$ in Eq. \eqref{eqn:expansion:IfLL:mfmmu:esp}, $I_{LR}^f$ in Eq. \eqref{eqn:expansion:IfLR:mfmmu:esp}, $I_{LL}^S$ in Eq. \eqref{eqn:expansion:ISLL:mfmmu:esp}, $I_{LR}^S$ in Eq. \eqref{eqn:expansion:ISLR:mfmmu:esp}. For the vector case, we should adopt the expansion forms of $L_{LL}^f$ in Eq. \eqref{eqn:expansion:LfLL:mfmmu:esp}, $L_{LR}^f$ in Eq. \eqref{eqn:expansion:LfLR:mfmmu:esp}, $L_{LL}^V$ in Eq. \eqref{eqn:expansion:LVLL:mfmmu:esp}, $L_{LR}^V$ in Eq. \eqref{eqn:expansion:LVLR:mfmmu:esp}. After inserting these expansion formulae, we can obtain the following expressions of muon MDM: 
\begin{align}
&\Delta a_{\mu}\approx\frac{m_{\mu}^2}{2\pi^2m_S^2}\Big\{(|y_L|^2+|y_R|^2)\cdot[-\frac{1}{3}+\frac{m_{\mu}^2}{m_S^2}(\log\frac{m_S}{m_{\mu}}-\frac{19}{24})+\frac{m_{\mu}^4}{m_S^4}(4\log\frac{m_S}{m_{\mu}}-\frac{77}{30})]\nonumber\\
&+\left[y_L(y_R)^\ast+y_R(y_L)^\ast\right]\cdot[-\log\frac{m_S}{m_{\mu}}+\frac{1}{4}+\frac{m_{\mu}^2}{m_S^2}(-2\log\frac{m_S}{m_{\mu}}+\frac{5}{6})+\frac{m_{\mu}^4}{m_S^4}(-5\log\frac{m_S}{m_{\mu}}+\frac{59}{24})]\Big\}\nonumber\\
&+\frac{m_{\mu}^2}{2\pi^2m_V^2}\Big\{(|g_L|^2+|g_R|^2)\cdot[\frac{7}{3}-\frac{5m_{\mu}^2}{12m_V^2}+\frac{m_{\mu}^4}{m_V^4}(2\log\frac{m_V}{m_{\mu}}-\frac{113}{60})]\nonumber\\
&+\left[g_L(g_R)^\ast+g_R(g_L)^\ast\right]\cdot[-3+\frac{m_{\mu}^2}{m_V^2}(2\log\frac{m_V}{m_{\mu}}-\frac{7}{6})+\frac{m_{\mu}^4}{m_V^4}(6\log\frac{m_V}{m_{\mu}}-\frac{13}{4})]\Big\}\nonumber\\
&\approx\frac{m_{\mu}^2}{2\pi^2m_S^2}\Big\{-\frac{1}{3}(|y_L|^2+|y_R|^2)+\left[y_L(y_R)^\ast+y_R(y_L)^\ast\right]\cdot(-\log\frac{m_S}{m_{\mu}}+\frac{1}{4})\Big\}\nonumber\\
&+\frac{m_{\mu}^2}{2\pi^2m_V^2}\Big\{\frac{7}{3}(|g_L|^2+|g_R|^2)-3\left[g_L(g_R)^\ast+g_R(g_L)^\ast\right]\Big\}.
\end{align}
This exactly matches the contributions derived in Ref. \cite{Queiroz:2014zfa}.
\subsection{Compute the dark matter MDM}
For the interactions $\bar{\chi}(y_L\omega_-+y_R\omega_+)fS^{\ast}+\mathrm{h.c.}$, the electric charge conservation condition is taken as $Q_f+Q_{S^{\ast}}=Q_f-Q_S=Q_{\chi}=0$. According to the parameterizations in Eq. \eqref{eqn:contri:structure:chiral}, the MDM of dark matter $\chi$ here is calculated as
\begin{align}
&\Delta a_{\mu}=\frac{m_{\chi}^2Q_f}{8\pi^2m_S^2}\big\{(|y_L|^2+|y_R|^2)(-I_{LL}^f+I_{LL}^S)+\left[y_L(y_R)^\ast+y_R(y_L)^\ast\right](-I_{LR}^f+I_{LR}^S)\big\}.
\end{align}
In the above, the variables of scalar loop functions are $(m_{\chi},m_f,m_S)$. Inserting the analytic forms of $I_{LL}^f$ in Eq. \eqref{eqn:con:ILLf},  $I_{LR}^f$ in Eq. \eqref{eqn:con:ILRf},  $I_{LL}^S$ in Eq. \eqref{eqn:con:ILLS}, and $I_{LR}^S$ in Eq. \eqref{eqn:con:ILRS}, we can obtain the following contributions:
\begin{align}
&\Delta a_{\mu}=-\frac{Q_f}{16\pi^2}\Big\{(|y_L|^2+|y_R|^2)\cdot\big[\frac{-m_f^4+(2m_S^2+m_{\chi}^2)m_f^2+m_S^2(m_{\chi}^2-m_S^2)}{2m_{\chi}^2}f(m_{\chi}^2,m_f^2,m_S^2)\nonumber\\
&+\frac{m_f^2-m_S^2}{m_{\chi}^2}\log\frac{m_f}{m_S}-1\big]+\frac{m_f}{m_{\chi}}\left[y_L(y_R)^\ast+y_R(y_L)^\ast\right]\cdot\big[\frac{-m_f^2+m_S^2+m_{\chi}^2}{2}f(m_{\chi}^2,m_f^2,m_S^2)+\log\frac{m_f}{m_S}\big]\Big\}.
\end{align}
These results reproduce the expressions of dark matter MDM in Ref. \cite{Ibarra:2024mpq} exactly.

\subsection{Compute the muon MDM from axion-like particle interactions}
$\bullet$ For the diagonal interactions $y_Pa\bar{\mu}(i\gamma^5)\mu$, we have $y_L=-iy_P$ and $y_R=iy_P$. Then, the muon MDM is calculated as
\begin{align}
&\Delta a_{\mu}=\frac{m_{\mu}^2y_P^2}{4\pi^2m_a^2}[I_{LL}^f(m_{\mu},m_{\mu},m_a)-I_{LR}^f(m_{\mu},m_{\mu},m_a)].
\end{align}
Inserting the analytic forms of $I_{LL}^f$ in Eq. \eqref{eqn:expansion:IfLL:mfmmu} and $I_{LR}^f$ in Eq. \eqref{eqn:expansion:IfLR:mfmmu}, we can obtain the following contributions:
\begin{align}
&\Delta a_{\mu}=\frac{m_{\mu}^2y_P^2}{4\pi^2m_a^2}[I_{LL}^f(m_{\mu},m_{\mu},m_a)-I_{LR}^f(m_{\mu},m_{\mu},m_a)]\nonumber\\
&=\frac{y_P^2}{16\pi^2}\cdot
\left\{\begin{array}{ll}
\Big[-\frac{m_{\mu}^2+2m_a^2}{m_{\mu}^2}+\frac{2m_a^2(m_a^2-m_{\mu}^2)}{m_{\mu}^4}\log\frac{m_a}{m_{\mu}}+\frac{2m_a^3(3m_{\mu}^2-m_a^2)}{m_{\mu}^4\sqrt{m_a^2-4m_{\mu}^2}}\log\frac{m_a+\sqrt{m_a^2-4m_{\mu}^2}}{2m_{\mu}}\Big],&\mathrm{for}~2m_\mu<m_a\\[2ex]
\Big[-\frac{m_{\mu}^2+2m_a^2}{m_{\mu}^2}+\frac{2m_a^2(m_a^2-m_{\mu}^2)}{m_{\mu}^4}\log\frac{m_a}{m_{\mu}}+\frac{2m_a^3(3m_{\mu}^2-m_a^2)}{m_{\mu}^4\sqrt{4m_{\mu}^2-m_a^2}}\arctan\sqrt{\frac{4m_{\mu}^2}{m_a^2}-1}\Big],&\mathrm{for}~2m_\mu>m_a
\end{array}\right..
\end{align}

$\bullet$ For the off-diagonal interactions $a\bar{\mu}(y_L\omega_-+y_R\omega_+)f+\mathrm{h.c.}$, the muon MDM is calculated as
\begin{align}
&\Delta a_{\mu}=\frac{m_{\mu}^2}{8\pi^2m_a^2}\big\{(|y_L|^2+|y_R|^2)I_{LL}^f(m_{\mu},m_f,m_a)+[y_L(y_R)^\ast+y_R(y_L)^\ast]I_{LR}^f(m_{\mu},m_f,m_a)\big\}.
\end{align}

When $f$ is the tau lepton, we have $m_{\mu}\ll m_{\tau}$. Typically, we can neglect the $I_{LL}^f$ term suppressed by $m_{\mu}/m_{\tau}$. Inserting the expansion forms of $I_{LR}^f$ in Eqs. \eqref{eqn:expansion:IfLR:mmullmfmS} and \eqref{eqn:expansion:IfLR:mmumSllmf}, we can obtain the following contributions:
\begin{align}
&\Delta a_{\mu}\approx\frac{m_{\mu}^2}{8\pi^2m_a^2}[y_L(y_R)^\ast+y_R(y_L)^\ast]I_{LR}^f(m_{\mu},m_{\tau},m_a)\nonumber\\
&\approx\frac{m_{\mu}m_{\tau}}{32\pi^2}[y_L(y_R)^\ast+y_R(y_L)^\ast]\cdot\nonumber\\
&\left\{\begin{array}{ll}
\ds\frac{1}{(m_a^2-m_{\tau}^2)^3}(-3m_a^4+4m_a^2m_{\tau}^2-m_{\tau}^4-2m_a^4\log\frac{m_{\tau}^2}{m_a^2}),&\quad\mathrm{for}~m_\mu\ll m_{\tau}, m_a\\[2ex]
\ds\frac{1}{m_{\tau}^2}\big[1+\frac{1}{3m_{\tau}^2}(m_\mu^2-3m_a^2)\big],&\quad\mathrm{for}~m_\mu,m_a\ll m_{\tau}
\end{array}\right..
\end{align}

When $f$ is the electron, we have $m_e\ll m_{\mu}$. Typically, we can neglect the $I_{LR}^f$ term suppressed by $m_e/m_{\mu}$. Inserting the expansion forms of $I_{LL}^f$ in Eqs. \eqref{eqn:expansion:IfLL:mfllmmumS} and \eqref{eqn:expansion:IfLL:mfmSllmmu}, we can obtain the following contributions:
\begin{align}
&\Delta a_{\mu}\approx\frac{m_{\mu}^2}{8\pi^2m_a^2}(|y_L|^2+|y_R|^2)I_{LL}^f(m_{\mu},m_e,m_a)\nonumber\\
&\approx-\frac{1}{32\pi^2}(|y_L|^2+|y_R|^2)\cdot\nonumber\\
&\left\{\begin{array}{ll}
\ds\frac{1}{m_{\mu}^4}\big[m_{\mu}^4+2m_a^2m_{\mu}^2+2m_a^4\log\frac{m_a^2-m_{\mu}^2}{m_a^2}\big],&\quad\mathrm{for}~m_e\ll m_\mu,m_a~\mathrm{and}~m_\mu<m_a\\[2ex]
\ds\big[1+\frac{2}{m_\mu^2}(m_a^2-m_e^2-i\pi m_e^2+m_e^2\log\frac{m_\mu^2}{m_e^2})\big],&\quad\mathrm{for}~m_e,m_a\ll m_\mu
\end{array}\right..
\end{align}

Then, all these results reproduce the expressions of muon MDM in Ref. \cite{Galda:2023qjx} exactly \footnote{The relation between $\mr{arccos}$ and $\mr{arctan}$ is given in App. \ref{app:functions}.}.\\[-1ex]

\section{Summary and conclusions}\label{sec:summary}

Many new physics models are motivated to explain the $(g-2)_{\mu}$ anomaly, and numerous approximate formulae are employed to compute the new physics contributions. Therefore, it is necessary and essential to perform a systematic investigation of the relevant formulae including their scope of validity. In this manuscript, we focus on particles with spin not greater than one. Respecting the Lorentz invariance and electromagnetic gauge invariance within local quantum field theory, we study the most general renormalizable interactions termed as the canonical interactions. Based on the canonical interactions, we provide the comprehensive analytic and approximate expressions of muon MDM at one-loop level. The key features of this paper are summarized below:

\begin{itemize}[itemindent=0em, leftmargin=10pt, listparindent=1em]
\item \textit{Universality of the formulae for spin-half fermions}

Starting from the canonical interactions, we investigate the contributions to muon MDM at one-loop level comprehensively. Our formulae are not only valid for the canonical interactions, but also applicable to the quasi-canonical interactions through field redefinition, total derivative, and equation of motion. When generalizing these formulae of the muon MDM, we should pay more attention to the proper substitutions of masses, couplings, and electric charge conservation conditions. Especially, we give some examples to demonstrate the computation of particle MDM.

\item \textit{Equivalent representations of the contributions}

Although the analytic results for scalar and vector mediators have been established for a long time \cite{Leveille:1977rc}, we re-derive the loop results in the integral representation. Moreover, we additionally present the forms in the PaVe representation and special function representation. The three representations are equivalent, and each offers distinct advantages. The PaVe representation is expressed in terms of the Passarino-Veltman integrals; thus, it is helpful to the loop calculation check and can be generalized to arbitrary dimension. The integral representation is not unique due to terms like $(1-2x)$ in integrands, which is suitable to analyze the results in different mass limits and perform the expansion. The special function representation is useful to clarify the singularity and threshold behaviours, because all the loop functions can be expressed in terms of the $f$ function, whose analytic properties are rigorously studied in this work.

\item \textit{Analytical and approximate results of the contributions}

We study Landau singularities of the loop functions and identify the singular locations at $m_f=0$, $m_{S(V)}=0$, and $m_{\mu}=m_f+m_{S(V)}$. In Tab. \ref{tab:sum:asym:scalar} and Tab. \ref{tab:sum:asym:vector}, we summarize the loop contributions in the chiral limit of $m_f\rightarrow0$ and limit of $m_{S(V)}\rightarrow0$. The $m_{\mu}=m_f+m_{S(V)}$ is the branch cut. More importantly, we perform the complete expansions for the hierarchical and degenerate mass scales, which can be applied to various models. The validity of each formulae is clearly exhibited.

\item \textit{Physical implications for new physics}

Based on the analytic and expansion formulae, we make physical discussions in Sec. \ref{sec:Sexpansion:summary} and Sec. \ref{sec:Vexpansion:summary}. First, we consider the massless limit and decoupling behaviours. Keeping the left-handed and right-handed couplings simultaneously, we then summarize the leading order formulae of the muon MDM in different scenarios, which depends on the Yukawa coupling (or gauge coupling) and electric charge. Because the left-handed and right-handed couplings are independent generally, it is always possible to make the contributions of $\left[y_L(y_R)^\ast+y_R(y_L)^\ast\right]$ (or $\left[g_L(g_R)^\ast+g_R(g_L)^\ast\right]$) part positive or negative through adjusting the sign of the couplings. Keeping only one of the left-handed or the right-handed couplings, we also summarize the leading order formulae of the muon MDM in different scenarios. Because the sign of $|y_{L(R)}|^2$ (or $|g_{L(R)}|^2$) is definitely non-negative, the positivity of $(g-2)_{\mu}$ constrains the electric charge of fermion or mediator as presented in Tab. \ref{tab:sum:exp:scalarLL} and Tab. \ref{tab:sum:exp:vectorLL}. These insights in view of the simplified interactions remain relevant even in models with extended fields and interactions.

Following the theoretical and experimental updates in 2025, the total new physics contributions can accommodate both the positive and negative values with small magnitude such that $\Delta a_{\mu}=(38\pm63)\times10^{-11}$. The muon $g-2$ as a precision probe can put constraints on the new physics models. Based on the specific muon interactions, we can systematically estimate both the sign and magnitude of the contributions under certain mass scales. Let us take the doubly charged scalar appearing in the complex triplet scalar model as an illustrative example. Due to the electroweak gauge symmetry, there are only left-handed Yukawa interactions between the charged lepton and doubly charged scalar (namely $y_L\bar{\mu}\omega_-\mu^CS$). Then, we should apply the $m_f=m_\mu\ll m_S$ ($f=\mu^C$) scenario shown in Tab. \ref{tab:sum:exp:scalarLL}. Note that $Q_f=Q_{\mu^C}=1$ and $Q_S=-2$, which violates the condition of $\Delta a_{\mu}>0$ (namely $Q_f<-1/3$ or $Q_S>-2/3$). We can conclude that contribution from $y_L\bar{\mu}\omega_-\mu^CS$ interactions is negative, which is incompatible with the old results. However, it is permissible by the 2025 results. According to the expressions in Tab. \ref{tab:sum:exp:scalarLL} (note the extra symmetry factor 2 in each vertex as explained in Sec. \ref{sec:summary:examples:doublycharged}), the released new data lead to the constraints $|y_L|/m_S\lesssim10^{-3}\mr{GeV}^{-1}$ roughly. For fixed coupling, this sets a lower bound on the doubly charged scalar mass. Similar analysis can also extend generically to the muon interactions in other new physics models.

\end{itemize}

\textit{Note added:} The formulae shown in this manuscript are cross-checked using Package-X \cite{Patel:2015tea, Patel:2016fam}. Based on the analytic expressions and approximate behaviours, we can also exemplify the possible benchmark points for numerical estimation. The evaluations can reveal implications on the new physics mass scales and coupling strengths further, while such an analysis is beyond the scope of this work.

\begin{acknowledgments}
We would like to thank Jian-Bin Chen, Qing-Hong Cao, Jian Wang, Yefan Wang, Bin Yan, Chen Zhang, and Shou-hua Zhu for helpful discussions. This work was supported in part by the Basic Research Program of Shanxi Province (Grant No. 202403021222062) and the startup research fund of Taiyuan University of Technology (Grant No. RY2400001554). This research also received partial support through an appointment to the Young Scientist Training Program at the APCTP from 2021 to 2023, during which we completed the expansion of the scalar mediator part.
\end{acknowledgments}

\vspace{2ex}
\textbf{Data Availability Statement}: No Data associated in the manuscript.
\appendix
\section{Details on the one-loop integrals}\label{app:loopintegrals}
\subsection{One-loop scalar integrals}
The $B_0$ function is defined as \cite{Denner:1991kt}
\begin{align}
&B_0(k^2,m_0^2,m_1^2)\equiv\frac{(2\pi\mu)^{4-D}}{i\pi^2}\int d^Dq\frac{1}{(q^2-m_0^2)[(q+k)^2-m_1^2]}.
\end{align}
In integral representation, the $B_0$ function can be expressed as
\begin{align}\label{eqn:app:1L:definition:B0int}
B_0(k^2,m_0^2,m_1^2)\equiv\Delta_{\epsilon}-\int_0^1dx\log\frac{xm_0^2+(1-x)m_1^2-x(1-x)k^2}{\mu^2}+\mc{O}(\epsilon),
\end{align}
where $\Delta_{\epsilon}=1/\epsilon-\gamma_E+\log4\pi$ with $\gamma_E$ being the Euler's constant. Here, $D=4-2\epsilon$ is the spacetime dimension, and $\mu$ is the renormalization scale. From ultraviolet behaviour of the $B_0$ function, we find $\lim\limits_{D\rightarrow4}(D-4)B_0(k^2,m_0^2,m_1^2)=-2$. In the special function representation, it is  given as
\begin{align}\label{eqn:app:1L:definition:B0SF}
B_0(k^2,m_0^2,m_1^2)=\Delta_{\epsilon}+2-\log\frac{m_0m_1}{\mu^2}-\frac{m_0^2-m_1^2}{2k^2}\log\frac{m_0^2}{m_1^2}+\frac{\Delta}{2k^2}f(k^2,m_0^2,m_1^2).
\end{align}

The $C_0$ function is defined as \cite{Denner:1991kt}
\begin{align}
&C_0(k_1^2,k_{12}^2,k_2^2,m_0^2,m_1^2,m_2^2)\nonumber\\
\equiv&\frac{(2\pi\mu)^{4-D}}{i\pi^2}\int d^Dq\frac{1}{(q^2-m_0^2)[(q+k_1)^2-m_1^2][(q+k_2)^2-m_2^2]}~(k_{12}\equiv k_1-k_2).
\end{align}
In integral representation, the $C_0$ function can be expressed as
\begin{align}
&C_0(k_1^2,k_{12}^2,k_2^2,m_0^2,m_1^2,m_2^2)=-\int_0^1\int_0^1\int_0^1dxdydz\frac{\delta(x+y+z-1)}{xm_0^2+ym_1^2+zm_2^2-xyk_1^2-xzk_2^2-yzk_{12}^2}+\mc{O}(\epsilon).
\end{align}
Note that the $C_0$ function is ultraviolet finite. In particular, we have the following results:
\begin{align}\label{eqn:app:1L:definition:C0espInt}
&C_0(k^2,0,k^2,m_0^2,m_1^2,m_1^2)=-\int_0^1dx\frac{1-x}{xm_0^2+(1-x)m_1^2-x(1-x)k^2}+\mc{O}(\epsilon).
\end{align}
 In the special function representation, it is given as
\begin{align}\label{eqn:app:1L:definition:C0espSF}
&C_0(k^2,0,k^2,m_0^2,m_1^2,m_1^2)=\frac{1}{2k^2}\log\frac{m_0^2}{m_1^2}+\frac{m_1^2-m_0^2-k^2}{2k^2}f(k^2,m_0^2,m_1^2).
\end{align}
 
\subsection{Reduction of the one-loop tensor integrals}
The loop contributions are calculated as one-loop three-point tensor integrals, defined as follows \cite{Denner:1991kt}:
\begin{align}
&\frac{(2\pi\mu)^{4-D}}{i\pi^2}\int d^Dq\frac{q^{\mu}}{(q^2-m_0^2)[(q+k_1)^2-m_1^2][(q+k_2)^2-m_2^2]}\equiv k_1^{\mu}C_1+k_2^{\mu}C_2,\nonumber\\[1ex]
&\frac{(2\pi\mu)^{4-D}}{i\pi^2}\int d^Dq\frac{q^{\mu}q^{\nu}}{(q^2-m_0^2)[(q+k_1)^2-m_1^2][(q+k_2)^2-m_2^2]}\nonumber\\
	\equiv& g^{\mu\nu}C_{00}+k_1^{\mu}k_1^{\nu}C_{11}+(k_1^{\mu}k_2^{\nu}+k_2^{\mu}k_1^{\nu})C_{12}+k_2^{\mu}k_2^{\nu}C_{22},\nonumber\\[1ex]
&\frac{(2\pi\mu)^{4-D}}{i\pi^2}\int d^Dq\frac{q^{\mu}q^{\nu}q^{\rho}}{(q^2-m_0^2)[(q+k_1)^2-m_1^2][(q+k_2)^2-m_2^2]}\nonumber\\
	\equiv& (g^{\mu\nu}k_1^{\rho}+g^{\mu\rho}k_1^{\nu}+g^{\nu\rho}k_1^{\mu})C_{001}+(g^{\mu\nu}k_2^{\rho}+g^{\mu\rho}k_2^{\nu}+g^{\nu\rho}k_2^{\mu})C_{002}+k_1^{\mu}k_1^{\nu}k_1^{\rho}C_{111}\nonumber\\
	&+(k_1^{\mu}k_1^{\nu}k_2^{\rho}+k_1^{\mu}k_2^{\nu}k_1^{\rho}+k_2^{\mu}k_1^{\nu}k_1^{\rho})C_{112}+(k_1^{\mu}k_2^{\nu}k_2^{\rho}+k_2^{\mu}k_1^{\nu}k_2^{\rho}+k_2^{\mu}k_2^{\nu}k_1^{\rho})C_{122}+k_2^{\mu}k_2^{\nu}k_2^{\rho}C_{222}.
\end{align}
In the above, the $C$ functions with variables are written as $C_{...}\big((k_1^2,(k_1-k_2)^2,k_2^2,m_0^2,m_1^2,m_2^2\big)$. Because the $q^{\mu}q^{\nu}$ integrals produce the same results as $q^{\nu}q^{\mu}$, we have $C_{12}=C_{21}$. Thus, the terms $(k_1^{\mu}k_2^{\nu}C_{12}+k_2^{\mu}k_1^{\nu}C_{21})$ can be merged. For the $q^{\mu}q^{\nu}q^{\rho}$ integrals, the permutation invariance also leads to merging properties.\\
Note that all the three-point tensor integrals can be reduced to $B_0$ and $C_0$ scalar integrals. In the following, we will perform the tensor reduction by means of FeynCalc \cite{Mertig:1990an, Shtabovenko:2016sxi, Shtabovenko:2020gxv}.\\

$\bullet$ The $C_1(m_{\mu}^2,k^2,m_{\mu}^2,m_0^2,m_1^2,m_1^2)$ can be reduced as
\begin{align}
C_1(m_{\mu}^2,k^2,m_{\mu}^2,m_0^2,m_1^2,m_1^2)=\frac{-B_0(k^2,m_1^2,m_1^2)+B_0(m_{\mu}^2,m_0^2,m_1^2)+(m_1^2-m_{\mu}^2-m_0^2)C_0(m_{\mu}^2,k^2,m_{\mu}^2,m_0^2,m_1^2,m_1^2)}{4m_{\mu}^2-k^2}.
\end{align}

$\bullet$ Note that $\lim\limits_{k^2\rightarrow0}C_{11}(m_{\mu}^2,k^2,m_{\mu}^2,m_0^2,m_1^2,m_1^2)$ and $\lim\limits_{k^2\rightarrow0}C_{12}(m_{\mu}^2,k^2,m_{\mu}^2,m_0^2,m_1^2,m_1^2)$ are both divergent. The reason is that the following Gram determinant \cite{Denner:1991kt} becomes singular:
 \begin{align}
 \left| \begin{array}{cc}p_1^2 & -p_1\cdot p_2\\[1ex] -p_1\cdot p_2 & p_2^2\end{array}\right| = k^2(m_{\mu}^2-\frac{1}{4}k^2).
\end{align}
However, the sum $C_{11}+C_{12}$ is convergent and can be reduced as
\begin{align}
&\lim_{k^2\rightarrow0}[C_{11}(m_{\mu}^2,k^2,m_{\mu}^2,m_0^2,m_1^2,m_1^2)+C_{12}(m_{\mu}^2,k^2,m_{\mu}^2,m_0^2,m_1^2,m_1^2)]\nonumber\\
&=\frac{1}{8(D-2)m_{\mu}^4}\Big\{[(D-1)m_0^2+(3D-7)m_{\mu}^2-(D+1)m_1^2]B_0(0,m_1^2,m_1^2)+2m_0^2B_0(0,m_0^2,m_0^2)\nonumber\\
	&+(3-2D)(m_{\mu}^2+m_0^2-m_1^2)B_0(m_{\mu}^2,m_0^2,m_1^2)\nonumber\\
	&+[(D-1)\big(m_{\mu}^4-2m_{\mu}^2m_1^2+(m_0^2-m_1^2)^2\big)+2(D-3)m_{\mu}^2m_0^2]C_0(m_{\mu}^2,0,m_{\mu}^2,m_0^2,m_1^2,m_1^2)\Big\}\nonumber\\
&=\frac{1}{16m_{\mu}^4}\Big\{(3m_0^2-5m_1^2+5m_{\mu}^2)B_0(0,m_1^2,m_1^2)+2m_0^2B_0(0,m_0^2,m_0^2)+(5m_1^2-5m_0^2-5m_{\mu}^2)B_0(m_{\mu}^2,m_0^2,m_1^2)\nonumber\\
	&+[3m_{\mu}^4+m_{\mu}^2(2m_0^2-6m_1^2)+3(m_0^2-m_1^2)^2]C_0(m_{\mu}^2,0,m_{\mu}^2,m_0^2,m_1^2,m_1^2)+2m_0^2-2m_1^2-2m_{\mu}^2\Big\}+\mc{O}(D-4).
\end{align}

$\bullet$ The $\lim\limits_{k^2\rightarrow0}k^2C_{12}(m_{\mu}^2,k^2,m_{\mu}^2,m_0^2,m_1^2,m_1^2)$ can be reduced as
\begin{align}
&\lim_{k^2\rightarrow0}k^2C_{12}(m_{\mu}^2,k^2,m_{\mu}^2,m_0^2,m_1^2,m_1^2)\nonumber\\
&=\frac{1}{4(2-D)m_{\mu}^2}\Big\{(m_0^2-m_{\mu}^2+m_1^2)B_0(0,m_1^2,m_1^2)-2m_0^2B_0(0,m_0^2,m_0^2)+(D-3)(m_{\mu}^2+m_0^2-m_1^2)B_0(m_{\mu}^2,m_0^2,m_1^2)\nonumber\\
	&+[m_{\mu}^4-2m_{\mu}^2(m_0^2+m_1^2)+(m_0^2-m_1^2)^2]C_0(m_{\mu}^2,0,m_{\mu}^2,m_0^2,m_1^2,m_1^2)\Big\}\nonumber\\
&=-\frac{1}{8m_{\mu}^2}\Big\{(m_0^2+m_1^2-m_{\mu}^2)B_0(0,m_1^2,m_1^2)-2m_0^2B_0(0,m_0^2,m_0^2)+(m_0^2-m_1^2+m_{\mu}^2)B_0(m_{\mu}^2,m_0^2,m_1^2)\nonumber\\
	&+[m_{\mu}^4-2m_{\mu}^2(m_0^2+m_1^2)+(m_0^2-m_1^2)^2]C_0(m_{\mu}^2,0,m_{\mu}^2,m_0^2,m_1^2,m_1^2)+2m_1^2-2m_0^2-2m_{\mu}^2\Big\}+\mc{O}(D-4)\nonumber\\
&=0.
\end{align}
This can be checked by inserting the special function representation of $B_0$ in Eq. \eqref{eqn:app:1L:definition:B0SF} and $C_0$ in Eq. \eqref{eqn:app:1L:definition:C0espSF}. It can also be checked through integral representation of $B_0$ in Eq. \eqref{eqn:app:1L:definition:B0int} and $C_0$ in Eq. \eqref{eqn:app:1L:definition:C0espInt}.

$\bullet$ The $\lim\limits_{k^2\rightarrow0}k^2C_{112}(m_{\mu}^2,k^2,m_{\mu}^2,m_0^2,m_1^2,m_1^2)$ can be reduced as
\begin{align}
&\lim_{k^2\rightarrow0}k^2C_{112}(m_{\mu}^2,k^2,m_{\mu}^2,m_0^2,m_1^2,m_1^2)\nonumber\\
&=\frac{1}{16(D-2)m_{\mu}^4}\Big\{(-m_{\mu}^4+2m_1^2m_{\mu}^2+m_0^4-m_1^4)B_0(0,m_1^2,m_1^2)+2m_0^2(m_1^2-m_0^2-m_{\mu}^2)B_0(0,m_0^2,m_0^2)\nonumber\\
	&+(D-3)(m_{\mu}^2+m_0^2-m_1^2)^2B_0(m_{\mu}^2,m_0^2,m_1^2)\nonumber\\
	&+(m_{\mu}^2+m_0^2-m_1^2)[m_{\mu}^4-2m_{\mu}^2(m_0^2+m_1^2)+(m_0^2-m_1^2)^2]C_0(m_{\mu}^2,0,m_{\mu}^2,m_0^2,m_1^2,m_1^2)\Big\}\nonumber\\
&=\frac{1}{32m_{\mu}^4}\Big\{(-m_{\mu}^4+2m_1^2m_{\mu}^2+m_0^4-m_1^4)B_0(0,m_1^2,m_1^2)+2m_0^2(m_1^2-m_0^2-m_{\mu}^2)B_0(0,m_0^2,m_0^2)\nonumber\\
	&+(m_{\mu}^2+m_0^2-m_1^2)^2B_0(m_{\mu}^2,m_0^2,m_1^2)-2(m_{\mu}^2+m_0^2-m_1^2)^2\nonumber\\
	&+(m_{\mu}^2+m_0^2-m_1^2)[m_{\mu}^4-2m_{\mu}^2(m_0^2+m_1^2)+(m_0^2-m_1^2)^2]C_0(m_{\mu}^2,0,m_{\mu}^2,m_0^2,m_1^2,m_1^2)\Big\}+\mc{O}(D-4)\nonumber\\
&=0.
\end{align}

$\bullet$ The $\lim\limits_{k^2\rightarrow0}C_{00}(m_{\mu}^2,k^2,m_{\mu}^2,m_0^2,m_1^2,m_1^2)$ can be reduced as
\begin{align}
&\lim_{k^2\rightarrow0}C_{00}(m_{\mu}^2,k^2,m_{\mu}^2,m_0^2,m_1^2,m_1^2)\nonumber\\
&=\frac{1}{4(D-2)m_{\mu}^2}\Big\{(m_{\mu}^2-m_0^2+m_1^2)B_0(0,m_1^2,m_1^2)+(m_{\mu}^2+m_0^2-m_1^2)B_0(m_{\mu}^2,m_0^2,m_1^2)\nonumber\\
	&-[m_{\mu}^4-2m_{\mu}^2(m_0^2+m_1^2)+(m_0^2-m_1^2)^2]C_0(m_{\mu}^2,0,m_{\mu}^2,m_0^2,m_1^2,m_1^2)\Big\}\nonumber\\
&=\frac{1}{8m_{\mu}^2}\Big\{(m_{\mu}^2-m_0^2+m_1^2)B_0(0,m_1^2,m_1^2)+(m_{\mu}^2+m_0^2-m_1^2)B_0(m_{\mu}^2,m_0^2,m_1^2)+2m_{\mu}^2\nonumber\\
	&-[m_{\mu}^4-2m_{\mu}^2(m_0^2+m_1^2)+(m_0^2-m_1^2)^2]C_0(m_{\mu}^2,0,m_{\mu}^2,m_0^2,m_1^2,m_1^2)\Big\}+\mc{O}(D-4).
\end{align}

$\bullet$ The $\lim\limits_{k^2\rightarrow0}C_{001}(m_{\mu}^2,k^2,m_{\mu}^2,m_0^2,m_1^2,m_1^2)$ can be reduced as
\begin{align}
&\lim_{k^2\rightarrow0}C_{001}(m_{\mu}^2,k^2,m_{\mu}^2,m_0^2,m_1^2,m_1^2)\nonumber\\
&=\frac{1}{16(D-1)(D-2)m_{\mu}^4}\Big\{[(D-1)(m_0^4-m_{\mu}^4)+2(3-2D)m_1^2m_{\mu}^2-2Dm_0^2m_1^2+(D+1)m_1^4]B_0(0,m_1^2,m_1^2)\nonumber\\
	&+2m_0^2(m_{\mu}^2+m_0^2-m_1^2)B_0(0,m_0^2,m_0^2)+[(3-2D)\big(m_{\mu}^4-2m_{\mu}^2m_1^2+(m_0^2-m_1^2)^2\big)-2m_{\mu}^2m_0^2]B_0(m_{\mu}^2,m_0^2,m_1^2)\nonumber\\
	&+(D-1)(m_{\mu}^2+m_0^2-m_1^2)[m_{\mu}^4-2m_{\mu}^2(m_0^2+m_1^2)+(m_0^2-m_1^2)^2]C_0(m_{\mu}^2,0,m_{\mu}^2,m_0^2,m_1^2,m_1^2)\Big\}\nonumber\\
&=\frac{1}{288m_{\mu}^4}\Big\{3(-3m_{\mu}^4-10m_1^2m_{\mu}^2+3m_0^4-8m_0^2m_1^2+5m_1^4)B_0(0,m_1^2,m_1^2)+6m_0^2(m_{\mu}^2+m_0^2-m_1^2)B_0(0,m_0^2,m_0^2)\nonumber\\
	&+3[-5m_{\mu}^4+2m_{\mu}^2(5m_1^2-m_0^2)-5(m_0^2-m_1^2)^2]B_0(m_{\mu}^2,m_0^2,m_1^2)+6(m_0^2-m_1^2)^2-22m_{\mu}^4\nonumber\\
	&+9(m_{\mu}^2+m_0^2-m_1^2)[m_{\mu}^4-2m_{\mu}^2(m_0^2+m_1^2)+(m_0^2-m_1^2)^2]C_0(m_{\mu}^2,0,m_{\mu}^2,m_0^2,m_1^2,m_1^2)\Big\}+\mc{O}(D-4).
\end{align}

$\bullet$ Note that $\lim\limits_{k^2\rightarrow0}C_{111}(m_{\mu}^2,k^2,m_{\mu}^2,m_0^2,m_1^2,m_1^2)$ and $\lim\limits_{k^2\rightarrow0}C_{112}(m_{\mu}^2,k^2,m_{\mu}^2,m_0^2,m_1^2,m_1^2)$ are both divergent, while the sum $C_{111}+3C_{112}$ is convergent and can be reduced as
\begin{align}
&\lim_{k^2\rightarrow0}[C_{111}(m_{\mu}^2,k^2,m_{\mu}^2,m_0^2,m_1^2,m_1^2)+3C_{112}(m_{\mu}^2,k^2,m_{\mu}^2,m_0^2,m_1^2,m_1^2)]\nonumber\\
&=\frac{1}{16(D-1)(D-2)m_{\mu}^6}\Big\{[(-7D^2+24D-17)m_{\mu}^4-4(D-1)(D-2)m_0^2m_{\mu}^2+2(2D^2+2D-7)m_{\mu}^2m_1^2\nonumber\\
	&+(m_1^2-m_0^2)\big((D^2-1)m_0^2-(D^2+4D+1)m_1^2\big)]B_0(0,m_1^2,m_1^2)+2(2D+1)m_0^2(m_1^2-m_{\mu}^2-m_0^2)B_0(0,m_0^2,m_0^2)\nonumber\\
	&+[3(D^2-D-1)\big(m_{\mu}^4-2m_{\mu}^2m_1^2+(m_0^2-m_1^2)^2\big)+6(D^2-3D+3)m_{\mu}^2m_0^2]B_0(m_{\mu}^2,m_0^2,m_1^2)\nonumber\\
	&+(1-D)(m_{\mu}^2+m_0^2-m_1^2)[(D+1)\big(m_{\mu}^4-2m_{\mu}^2m_1^2+(m_0^2-m_1^2)^2\big)+2(D-5)m_{\mu}^2m_0^2]C_0(m_{\mu}^2,0,m_{\mu}^2,m_0^2,m_1^2,m_1^2)\Big\}\nonumber\\
&=\frac{1}{96m_{\mu}^6}\Big\{3[-11(m_{\mu}^2-m_1^2)^2-8m_{\mu}^2m_0^2-5m_0^4+16m_0^2m_1^2]B_0(0,m_1^2,m_1^2)+18m_0^2(m_1^2-m_{\mu}^2-m_0^2)B_0(0,m_0^2,m_0^2)\nonumber\\
	&+3[11m_{\mu}^4+2m_{\mu}^2(7m_0^2-11m_1^2)+11(m_0^2-m_1^2)^2]B_0(m_{\mu}^2,m_0^2,m_1^2)-18(m_0^2-m_1^2)^2+12m_{\mu}^2(m_1^2-m_0^2)+22m_{\mu}^4\nonumber\\
	&+3(m_1^2-m_{\mu}^2-m_0^2)[5m_{\mu}^4-2m_{\mu}^2(m_0^2+5m_1^2)+5(m_0^2-m_1^2)^2]C_0(m_{\mu}^2,0,m_{\mu}^2,m_0^2,m_1^2,m_1^2)\Big\}+\mc{O}(D-4).
\end{align}

$\bullet$ As given in Eq. \eqref{eqn:app:1L:definition:B0int}, the divergent part of $B_0(k^2,m_0^2,m_1^2)$ is $-2/(D-4)$. Based on the reduction formulae, we find the following non-vanishing 
results:
\begin{align}
&\lim_{D\rightarrow4}(D-4)C_{00}(m_{\mu}^2,k^2,m_{\mu}^2,m_0^2,m_1^2,m_1^2)=-\frac{1}{2},\quad \lim_{D\rightarrow4}(D-4)C_{001}(m_{\mu}^2,k^2,m_{\mu}^2,m_0^2,m_1^2,m_1^2)=\frac{1}{6}.
\end{align}
This verifies the ultraviolet divergent part shown in Denner's paper \cite{Denner:1991kt}.

\section{Technical details on the properties of $f(k^2,m_0^2,m_1^2)$ function}\label{app:ftech}
First, we note that the function $f(k^2,m_0^2,m_1^2)$ satisfies the following differential equations
\begin{align}
\frac{\partial f(k^2,m_0^2,m_1^2)}{\partial k^2}+\frac{k^2-m_0^2-m_1^2}{\Delta}f(k^2,m_0^2,m_1^2)+\frac{2}{\Delta}=0,
\end{align}
and
\begin{align}
\frac{\partial f(k^2,m_0^2,m_1^2)}{\partial m_0^2}+\frac{m_0^2-m_1^2-k^2}{\Delta}f(k^2,m_0^2,m_1^2)+\frac{m_1^2-m_0^2-k^2}{m_0^2\Delta}=0.
\end{align}
Here, we will not investigate the $f(k^2,m_0^2,m_1^2)$ from above differential equations. Instead, we study its properties based on the explicit expressions integrated out in Eq. \eqref{eqn:rep:f}.

As shown in Sec. \ref{sec:model:contri:ScalarRep} and Sec. \ref{sec:model:contri:VectorRep}, all the loop functions can be expressed in terms of the $f$ function in the special function representation. Hence, it is of great importance to study the analytic and asymptotic behaviours of the function $f$, which is the task of this section.
\begin{enumerate}[leftmargin=10pt,label=\textcircled{\arabic*}]

\item\textbf{Analytic properties}\\
In general, the function $f$ can be treated as a complex function with three complex variables. In the three-dimensional complex space, there is a branch cut for $k^2=(m_0+m_1)^2$. This is consistent with the fact that the imaginary (absorptive) parts of $I_{LL(LR)}^{f(S)}$ and $L_{LL(LR)}^{f(V)}$ can appear in the case of $m_{\mu}>m_f+m_{S(V)}$, namely, the loop particles are on-shell.

Furthermore, $m_0=0$ and $m_1=0$ are the singular points, as evidenced by the following identity:
\begin{align}
\frac{1}{\sqrt{\Delta}}\log\frac{m_0^2+m_1^2-k^2+\sqrt{\Delta}}{m_0^2+m_1^2-k^2-\sqrt{\Delta}}=\frac{1}{\sqrt{\Delta}}\log\frac{(m_0^2+m_1^2-k^2+\sqrt{\Delta})^2}{4m_0^2m_1^2}
\end{align}

\item\textbf{Symmetric properties}\\
By definition, the function $f$ satisfies the exchange relation $f(k^2,m_0^2,m_1^2)=f(k^2,m_1^2,m_0^2)$.
\item\textbf{Asymptotic properties}
\begin{itemize}[leftmargin=10pt]
\item Expansion at $k^2=0$:
\begin{align}
&f(k^2,m_0^2,m_1^2)=\left\{ \begin{array}{ll}
\frac{1}{m_0^2-m_1^2}\log\frac{m_0^2}{m_1^2}+\frac{(m_0^2+m_1^2)\log\frac{m_0^2}{m_1^2}-2(m_0^2-m_1^2)}{(m_0^2-m_1^2)^3}k^2\\
+\frac{(m_0^4+4m_0^2m_1^2+m_1^4)\log\frac{m_0^2}{m_1^2}-3(m_0^4-m_1^4)}{(m_0^2-m_1^2)^5}k^4+\mc{O}(k^6)& ,\quad\textrm{for}\quad m_0\ne m_1\\\\
\frac{1}{m_0^2}(1+\frac{k^2}{6m_0^2}+\frac{k^4}{30m_0^4})+\mc{O}(k^6)&,\quad\textrm{for}\quad m_0=m_1
\end{array} \right..
\end{align}
\item Expansion at $k^2\rightarrow\infty$:
\begin{align}
&f(k^2,m_0^2,m_1^2)=\frac{2}{k^2}\big[1+\frac{m_0^2+m_1^2}{k^2}+\frac{m_0^4+4m_0^2m_1^2+m_1^4}{k^4}\big](-\log\frac{k^2}{m_0m_1}+i\pi)\nonumber\\
&+\frac{2(m_0^2+m_1^2)}{k^4}+\frac{3m_0^4+8m_0^2m_1^2+3m_1^4}{k^6}+\mc{O}(\frac{1}{k^8}).
\end{align}

\item Expansion at $m_0=0$:
\begin{align}\label{eqn:contri:fprop:m0eq0}
&f(k^2,m_0^2,m_1^2)\nonumber\\
&=\left\{ \begin{array}{ll}
\frac{1}{m_1^2-k^2}\big[1+\frac{k^2+m_1^2}{(k^2-m_1^2)^2}m_0^2+\frac{k^4+4k^2m_1^2+m_1^4}{(k^2-m_1^2)^4}m_0^4\big]\cdot\big[\log\frac{(m_1^2-k^2)^2}{m_0^2m_1^2}-2i\pi\theta(k^2-m_1^2)\big]\\
+\frac{2k^2}{(k^2-m_1^2)^3}m_0^2+\frac{k^2(3k^2+4m_1^2)}{(k^2-m_1^2)^5}m_0^4+\mc{O}(m_0^6)\qquad\qquad\qquad\qquad\qquad\qquad, \textrm{for}\quad k^2\ne m_1^2\\ \\
\frac{\pi}{2\sqrt{m_0^2k^2}}(1+\frac{m_0^2}{8k^2}+\frac{3m_0^4}{128k^4}+\frac{5m_0^6}{1024k^6})-\frac{1}{2k^2}(1+\frac{m_0^2}{6k^2}+\frac{m_0^4}{30k^4})+\mc{O}(m_0^6)\quad,\textrm{for}\quad k^2=m_1^2
\end{array} \right..
\end{align}
In the above, $\theta(x)$ is the Heaviside step function, defined as $\theta(x)=1$ for $x>0$ and $\theta(x)=0$ for $x<0$. Note that $f(k^2,m_0^2,m_1^2)$ diverges as $m_0\rightarrow0$, which can lead to infrared divergence.

\item Expansion at $m_0\rightarrow\infty$:
\begin{align}\label{eqn:contri:fprop:m0eqinf}
&f(k^2,m_0^2,m_1^2)=\frac{1}{m_0^2}\log\frac{m_0^2}{m_1^2}+\big[(k^2+m_1^2)\log\frac{m_0^2}{m_1^2}-2k^2\big]\frac{1}{m_0^4}\nonumber\\
&+\big[(k^4+4k^2m_1^2+m_1^4)\log\frac{m_0^2}{m_1^2}-k^2(3k^2+4m_1^2)\big]\frac{1}{m_0^6}+\mc{O}(\frac{1}{m_0^8}).
\end{align}

\item Expansion at $m_1=0$ and $m_1\rightarrow\infty$:\\
The results for the two cases can be obtained through the replacement of $m_0\leftrightarrow m_1$ in Eqs. \eqref{eqn:contri:fprop:m0eq0} and \eqref{eqn:contri:fprop:m0eqinf}.

\end{itemize}

\item\textbf{Degenerate mass properties}\\
\begin{itemize}[leftmargin=10pt]
\item Case of $m_0^2=m_1^2\equiv m^2$:\\
The function $f(k^2,m^2,m^2)$ can be reduced as
\begin{align}
&f(k^2,m^2,m^2)=\left\{ \begin{array}{ll}
\frac{1}{\sqrt{k^4-4m^2k^2}}(\log\frac{2m^2-k^2+\sqrt{k^4-4m^2k^2}}{2m^2-k^2-\sqrt{k^4-4m^2k^2}}+2i\pi)& ,\quad\textrm{for}\quad k^2>4m^2\\
\frac{2}{\sqrt{4m^2k^2-k^4}}\arctan\frac{\sqrt{4m^2k^2-k^4}}{2m^2-k^2}& ,\quad\textrm{for}\quad 0<k^2<4m^2\\
\frac{1}{m^2} & ,\quad\textrm{for}\quad k^2=0\\
-\frac{1}{m^2} & ,\quad\textrm{for}\quad k^2=4m^2
\end{array} \right..
\end{align}
Then, it can be expanded as
\begin{align}
&f(k^2,m^2,m^2)=\left\{ \begin{array}{ll}
\frac{2}{k^2}\big[(1+\frac{2m^2}{k^2}+\frac{6m^4}{k^4})(-\log\frac{k^2}{m^2}+i\pi)+\frac{2m^2}{k^2}+\frac{7m^4}{k^4}+\mc{O}(\frac{m^6}{k^6})\big]& ,\;\textrm{for}\; k^2\gg m^2\\\\
\frac{1}{m^2}\big[1+\frac{k^2}{6m^2}+\frac{k^4}{30m^4}+\mc{O}(\frac{k^6}{m^6})\big]& ,\;\textrm{for}\; k^2\ll m^2
\end{array} \right..
\end{align}

\item Case of $k^2=m_0^2$:\\
The function $f(m_0^2,m_0^2,m_1^2)$ can be reduced as
\begin{align}
&f(m_0^2,m_0^2,m_1^2)=\left\{ \begin{array}{ll}
\frac{1}{m_1\sqrt{m_1^2-4m_0^2}}\log\frac{m_1+\sqrt{m_1^2-4m_0^2}}{m_1-\sqrt{m_1^2-4m_0^2}}& ,\quad\textrm{for}\quad m_1>2m_0\\
\frac{2}{m_1\sqrt{4m_0^2-m_1^2}}\arctan\frac{\sqrt{4m_0^2-m_1^2}}{m_1}& ,\quad\textrm{for}\quad m_1<2m_0\\
\frac{1}{m_0m_1} & ,\quad\textrm{for}\quad m_1=2m_0
\end{array} \right..
\end{align}
Then, it can be expanded as
\begin{align}
&f(m_0^2,m_0^2,m_1^2)\nonumber\\
&=\left\{ \begin{array}{ll}
\frac{2}{m_1^2}\big[(1+\frac{2m_0^2}{m_1^2}+\frac{6m_0^4}{m_1^4})\log\frac{m_1}{m_0}-\frac{m_0^2}{m_1^2}-\frac{7m_0^4}{2m_1^4}+\mc{O}(\frac{m_0^6}{m_1^6})\big]& ,\;\textrm{for}\; m_1\gg m_0\\\\
-\frac{1}{2m_0^2}\big[1+\frac{m_1^2}{6m_0^2}+\frac{m_1^4}{30m_0^4}+\mc{O}(\frac{m_1^6}{m_0^6})\big]+\frac{\pi}{2m_0m_1}\big[1+\frac{m_1^2}{8m_0^2}+\frac{3m_1^4}{128m_0^4}+\mc{O}(\frac{m_1^6}{m_0^6})\big]& ,\;\textrm{for}\; m_1\ll m_0
\end{array} \right..
\end{align}

\item Case of $k^2=m_1^2$:\\
This case can be reduced to the case of $k^2=m_0^2$ due to the symmetric properties, namely, $f(m_1^2,m_0^2,m_1^2)=f(m_1^2,m_1^2,m_0^2)$.

\item Case of $k^2=m_0^2=m_1^2$:\\
In this case, we have $f(k^2,k^2,k^2)=\frac{2\pi}{3\sqrt{3}k^2}$.
\end{itemize}
\end{enumerate}

\section{Properties of the elementary functions}\label{app:functions}
In this section, we collect the properties of some elementary functions, particularly the inverse trigonometric and inverse hyperbolic functions, which are relevant to this manuscript.\\[-2ex]

$\bullet$ Inverse trigonometric functions\\[1ex]
The inverse tangent is labelled as $\mr{tan}^{-1}(z)\equiv\mr{arctan}(z)$, and similar for the other inverse trigonometric functions. The domain of $z$ is all real numbers, while the range of usual principal value for $\mr{arctan}(z)$ is $(-\pi/2,\pi/2)$. Here are some of the properties:
\begin{align}
\mr{arctan}(-z)=-\mr{arctan}(z),\quad &\textrm{for all}~z\nonumber\\
\mr{arctan}(x)+\mr{arctan}(y)=\mr{arctan}(\frac{x+y}{1-xy}),\quad &\textrm{for}~ x y<1\nonumber\\
\mr{arctan}(z)=\mr{arcsin}\frac{z}{\sqrt{1+z^2}},\quad &\textrm{for all}~z\nonumber\\
\mr{arctan}(z)=\mr{arccos}\frac{1}{\sqrt{1+z^2}},\quad &\textrm{for}~z>0.
\end{align}
The Taylor series expansion of the inverse tangent is as follows:
\begin{align}
\mr{arctan}(\frac{1}{z})=\left\{ \begin{array}{cl}
\ds\frac{\pi}{2}-z+\frac{z^3}{3}-\frac{z^5}{5}+\frac{z^7}{7}+\mc{O}(z^9) & ,\quad\quad\quad\textrm{for}\quad z>0\vspace{1ex}\\
\ds-\frac{\pi}{2}-z+\frac{z^3}{3}-\frac{z^5}{5}+\frac{z^7}{7}+\mc{O}(z^9) & ,\quad\quad\quad\textrm{for}\quad z<0
\end{array} \right.
\end{align}

$\bullet$ Inverse hyperbolic functions\\[1ex] 
The inverse hyperbolic tangent function is defined as
\begin{align}
\mr{tanh}^{-1}(z)\equiv\mr{arctanh}(z)=\frac{1}{2}\log\frac{1+z}{1-z}.
\end{align}
 The domain of $z$ is $(-1,1)$, while the range of $\mr{arctanh}(z)$ is $(-\infty,\infty)$. It has the following properties:
\begin{align}
\mr{arctanh}(-z)=-\mr{arctanh}(z).\quad \mr{arctanh}(x)+\mr{arctanh}(y)=\mr{arctanh}(\frac{x+y}{1+xy}).
\end{align}

$\bullet$ Matching between inverse trigonometric tangent and inverse hyperbolic tangent functions\\[1ex] 
In general, variables of these functions can be complex. Then, the relationships between inverse trigonometric tangent and inverse hyperbolic tangent are given as \cite{Gradshteyn:2007cpj}
\begin{align}
&i~\mr{arctan}(z)=\frac{1}{2}\log\frac{1+iz}{1-iz}=\mr{arctanh}(iz),\nonumber\\
&\mr{arctan}(iz)=\frac{i}{2}\log\frac{1+z}{1-z}=i~\mr{arctanh}(z).
\end{align}

\section{Singularity of the scalar loop integrals}\label{app:singularity}
In perturbation quantum field theory, loop integrals always appear in the computation of Feynman diagram amplitudes. It is important to study their analyticity structure and singularities. Landau equations can provide a systematic way to identify locations of the singularities. For the one-loop scalar integral of
\begin{align}
&\int d^Dq\frac{1}{[(q+k_1)^2-m_1^2][(q+k_2)^2-m_2^2]\cdots[(q+k_n)^2-m_n^2]},
\end{align}
the equivalent form in the Feynman parameterization is
\begin{align}
&\int_0^1dx_1dx_2\cdots dx_n\int d^Dq\frac{\delta(x_1+x_2+\cdots+x_n-1)(n-1)!}{\big\{x_1[(q+k_1)^2-m_1^2]+x_2[(q+k_2)^2-m_2^2]+\cdots+x_n[(q+k_n)^2-m_n^2]\big\}^n}.
\end{align}
If the denominator encounters zero, the integral becomes singular. The Landau equations read as \cite{Landau:1959fi}
\begin{align}\label{eqn:singularity:Landaueq}
&x_i[(q+p_i)^2-m_i^2]=0,\qquad \mr{for~every~} i=1,2,\cdots,n,\nonumber\\
&\sum_{i=1}^nx_i(q+p_i)=0.
\end{align}
From Landau equations in Eq. \eqref{eqn:singularity:Landaueq}, we deduce $\sum_{i=1}^nx_iY_{ij}=0$ with the Cayley matrix $Y_{ij}$ defined as $Y_{ij}=(q+p_i)\cdot(q+p_j)$ ($i,j=1,2,\cdots,n$). The existence of non-zero $x_i$ implies $\mr{det}Y=0$. 

For $x_i\ne0$, Eq. \eqref{eqn:singularity:Landaueq} means $(q+p_i)^2=m_i^2$. Coleman and Norton interpreted the physical picture, that is, the propagating particle can be regarded as on mass shell with the time scale determined by $x_im_i$. \cite{Coleman:1965xm}. If every $x_i\ne0$, the equations give the leading order Landau singularity (LO-LS). For the leading order Landau singularity, the Cayley matrix is transformed as $Y_{ij}=[m_i^2+m_j^2-(p_i-p_j)^2]/2$. If some $x_i=0$, the equations give the lower order Landau singularity, in which the corresponding propagator $[(q+p_i)^2-m_i^2]$ can be contracted into a point. Hence, the lower order Landau singularity is just the LO-LS of the new loop integral with all the corresponding propagators of vanishing $x_i$ contracted. If only one of $x_i$ vanishes, there can be next leading order Landau singularity (NLO-LS). If two of $x_i$ vanish, there can be next next leading order Landau singularity (NNLO-LS).

Note that Landau equations only give us Landau singularities including the LO-LS and all lower order LS. However, there are also non-Landaunian singularities named as second type singularities sometimes. Furthermore, the discontinuity of amplitude can emerge when encountering the branch cut, which can be computed from the Cutkosky rules \cite{Cutkosky:1960sp}.

\subsection{Singularity of the $B_0$ function}
For the two-point scalar integral:
\begin{align}
&B_0(k^2,m_0^2,m_1^2)\sim\int d^Dq\frac{1}{(q^2-m_0^2)[(q+k)^2-m_1^2]},
\end{align}
the Landau equations are as follows:
\begin{align}\label{eqn:singularity:Landaueq:B0}
\left\{\begin{array}{ll}
&x_1(q^2-m_0^2)=0,\vspace{1ex}\\&x_2[(q+k)^2-m_1^2]=0,\vspace{1ex}\\&x_1q+x_2(q+k)=0.
\end{array}\right.
\end{align}

\begin{itemize}[itemindent=0em, leftmargin=10pt, listparindent=1em]
\item \textit{leading order Landau singularity}

For $x_1, x_2\ne0$, the Cayley matrix is
\begin{align}
Y=\left[\begin{array}{cc}
m_0^2&\ds\frac{m_0^2+m_1^2-k^2}{2}\vspace{1ex}\\\ds\frac{m_0^2+m_1^2-k^2}{2}&m_1^2
\end{array}\right].
\end{align}
Then, the equation $\mr{det}Y=0$ leads to $k^2=(m_0\pm m_1)^2$. The $k^2=(m_0+m_1)^2$ is called as normal threshold, while the $k^2=(m_0-m_1)^2$ is called as pseudo-threshold \cite{Eden:1966dnq, Zwicky:2016lka, Hannesdottir:2024hke}. Combining the following equations:
\begin{align}
\left\{\begin{array}{ll}
&\ds x_1Y_{11}+x_2Y_{12}=x_1m_0^2+x_2\frac{m_0^2+m_1^2-k^2}{2}=0,\vspace{1ex}\\&x_1+x_2=1,\vspace{1ex}\\&x_1q+x_2(q+k)=0,
\end{array}\right.
\end{align}
we have the singular points of
\begin{align}
x_1=\frac{k^2-m_0^2-m_1^2}{k^2+m_0^2-m_1^2},\quad x_2=\frac{2m_0^2}{k^2+m_0^2-m_1^2},\quad q=-\frac{2m_0^2}{k^2+m_0^2-m_1^2}k.
\end{align}
For $k^2=(m_0+m_1)^2$, we have 
\begin{align}
x_1=\frac{m_1}{m_0+m_1},\quad x_2=\frac{m_0}{m_0+m_1},\quad q=-\frac{m_0}{m_0+m_1}k.
\end{align}
In this case, both $x_1$ and $x_2$ lie between 0 and 1, which is the meaning of normal threshold.\vspace{1ex}\\
For $k^2=(m_0-m_1)^2$, we have 
\begin{align}
x_1=\frac{m_1}{m_1-m_0},\quad x_2=\frac{m_0}{m_0-m_1},\quad q=\frac{m_0}{m_1-m_0}k.
\end{align}
In this case, neither $x_1$ nor $x_2$ lie between 0 and 1. Then, this singular point is in the unphysical region, which is the meaning of pseudo-threshold.

\item \textit{next leading order Landau singularity}

For the NLO-LS, one of $x_1$ and $x_2$ vanishes. If $x_1=0$ and $x_2\ne0$, the Eq. \eqref{eqn:singularity:Landaueq:B0} leads to $(q+k)^2=m_1^2$ and $q+k=0$, which means that the singularity is $m_1=0$. If $x_2=0$ and $x_1\ne0$, the Eq. \eqref{eqn:singularity:Landaueq:B0} leads to $q^2=m_0^2$ and $q=0$, which means that the singularity is $m_0=0$.
\end{itemize}
\subsection{Singularity of the $C_0$ function}
For the three-point scalar integral:
\begin{align}
&C_0(k_1^2,k_{12}^2,k_2^2,m_0^2,m_1^2,m_2^2)\sim\int d^Dq\frac{1}{(q^2-m_0^2)[(q+k_1)^2-m_1^2][(q+k_2)^2-m_2^2]}~(k_{12}\equiv k_1-k_2),
\end{align}
the Landau equations are as follows:
\begin{align}\label{eqn:singularity:Landaueq:C0}
\left\{\begin{array}{ll}
&x_1(q^2-m_0^2)=0,\vspace{1ex}\\&x_2[(q+k_1)^2-m_1^2]=0,\vspace{1ex}\\&x_3[(q+k_2)^2-m_2^2]=0,\vspace{1ex}\\&x_1q+x_2(q+k_1)+x_3(q+k_2)=0.
\end{array}\right.
\end{align}

\begin{itemize}[itemindent=0em, leftmargin=10pt, listparindent=1em]
\item \textit{leading order Landau singularity}

For $x_1, x_2, x_3\ne0$, the Cayley matrix is
\begin{align}
Y=\left[\begin{array}{ccc}
m_0^2&\ds\frac{m_0^2+m_1^2-k_1^2}{2}&\ds\frac{m_0^2+m_2^2-k_2^2}{2}\vspace{1ex}\\
\ds\frac{m_0^2+m_1^2-k_1^2}{2}&m_1^2&\ds\frac{m_1^2+m_2^2-k_{12}^2}{2}\vspace{1ex}\\
\ds\frac{m_0^2+m_2^2-k_2^2}{2}&\ds\frac{m_1^2+m_2^2-k_{12}^2}{2}&m_2^2
\end{array}\right].
\end{align}
Then, $\mr{det}Y=0$ leads to the following equation:
\begin{align}\label{eqn:singularity:C0:Caydet}
&(k_1^2+k_2^2-k_{12}^2)(m_0^2k_{12}^2+m_1^2m_2^2)+(k_1^2-k_2^2+k_{12}^2)(m_1^2k_2^2+m_0^2m_2^2)\nonumber\\
&+(-k_1^2+k_2^2+k_{12}^2)(m_2^2k_1^2+m_0^2m_1^2)-m_0^4k_{12}^2-m_1^4k_2^2-m_2^4k_1^2-k_1^2k_2^2k_{12}^2=0.
\end{align}
This equation governs the anomalous threshold \footnote{The effects of triangle singularities have been discussed in hadronic reactions \cite{Liu:2015taa, Guo:2019twa}.} \cite{Eden:1966dnq, Zwicky:2016lka}. Combining the following equations:
\begin{align}\label{eqn:singularity:C0:x1x2x3q}
\left\{\begin{array}{lll}
&\ds x_1Y_{11}+x_2Y_{12}+x_3Y_{13}=x_1m_0^2+x_2\frac{m_0^2+m_1^2-k_1^2}{2}+x_3\frac{m_0^2+m_2^2-k_2^2}{2}=0,\vspace{1ex}\\
&\ds x_1Y_{21}+x_2Y_{22}+x_3Y_{23}=x_1\frac{m_0^2+m_1^2-k_1^2}{2}+x_2m_1^2+x_3\frac{m_1^2+m_2^2-k_{12}^2}{2}=0,\vspace{1ex}\\
&x_1+x_2+x_3=1,\vspace{1ex}\\
&x_1q+x_2(q+k_1)+x_3(q+k_2)=0,
\end{array}\right.
\end{align}
we have the singular points of
\begin{align}
&x_1=\frac{(m_1^2-m_0^2)(m_1^2-m_2^2)-m_0^2k_{12}^2-m_1^2(k_1^2-2k_2^2+k_{12}^2)-m_2^2k_1^2+k_1^2k_{12}^2}{m_0^2(k_1^2-k_2^2+k_{12}^2)+m_1^2(k_1^2+k_2^2-k_{12}^2)-2m_2^2k_1^2+k_1^2(-k_1^2+k_2^2+k_{12}^2)},\nonumber\\[1ex]
&x_2=\frac{(m_0^2-m_1^2)(m_0^2-m_2^2)-m_0^2(k_1^2+k_2^2-2k_{12}^2)-m_1^2k_2^2-m_2^2k_1^2+k_1^2k_2^2}{m_0^2(k_1^2-k_2^2+k_{12}^2)+m_1^2(k_1^2+k_2^2-k_{12}^2)-2m_2^2k_1^2+k_1^2(-k_1^2+k_2^2+k_{12}^2)},\nonumber\vspace{2ex}\\[1ex]
&x_3=\frac{-(m_0^2-m_1^2)^2+2(m_0^2+m_1^2)k_1^2-k_1^4}{m_0^2(k_1^2-k_2^2+k_{12}^2)+m_1^2(k_1^2+k_2^2-k_{12}^2)-2m_2^2k_1^2+k_1^2(-k_1^2+k_2^2+k_{12}^2)},\nonumber\\[1ex]
&q=-x_2k_1-x_3k_2.
\end{align}

\item \textit{next leading order Landau singularity}

For the NLO-LS, one of $x_1$, $x_2$, and $x_3$ vanishes. After contracting the corresponding propagator, the NLO-LS of $C_0$ is equivalent to LO-LS of $B_0$. \\
If $x_1=0$ and $x_2,x_3\ne0$, the Eq. \eqref{eqn:singularity:Landaueq:C0} leads to $k_{12}^2=(m_1\pm m_2)^2$. The $k_{12}^2=(m_1+m_2)^2$ is normal threshold, and the $k_{12}^2=(m_1-m_2)^2$ is pseudo-threshold.\\
If $x_2=0$ and $x_1,x_3\ne0$, the Eq. \eqref{eqn:singularity:Landaueq:C0} leads to $k_2^2=(m_0\pm m_2)^2$. The $k_2^2=(m_0+m_2)^2$ is normal threshold, and the $k_2^2=(m_0-m_2)^2$ is pseudo-threshold.\\
If $x_3=0$ and $x_1,x_2\ne0$, the Eq. \eqref{eqn:singularity:Landaueq:C0} leads to $k_1^2=(m_0\pm m_1)^2$. The $k_1^2=(m_0+m_1)^2$ is normal threshold, and the $k_1^2=(m_0-m_1)^2$ is pseudo-threshold.
\item \textit{next next leading order Landau singularity}

For the NNLO-LS, two of $x_1$, $x_2$, and $x_3$ vanish. If $x_1\ne0$ and $x_2=x_3=0$, the singularity is $m_0=0$. If $x_2\ne0$ and $x_1=x_3=0$, the singularity is $m_1=0$. If $x_3\ne0$ and $x_1=x_2=0$, the singularity is $m_2=0$.
\end{itemize}

\textbf{Application to $C_0$ in the muon MDM:} \\[-2ex]

In the light of the muon MDM contributions, we find that the relevant $C_0$ loop integral is in the form of $C_0(m_{\mu}^2,0,m_{\mu}^2,m_0^2,m_1^2,m_1^2)$. Now, let us apply the demonstration above to this special case. The variables in general $C_0(k_1^2,k_{12}^2,k_2^2,m_0^2,m_1^2,m_2^2)$ should be replaced as $(k_1^2\rightarrow m_{\mu}^2,k_{12}^2\rightarrow 0,k_2^2\rightarrow m_{\mu}^2,m_2\rightarrow m_1)$.\\
First of all, let us consider the LO-LS. We find that the Cayley determinant in Eq. \eqref{eqn:singularity:C0:Caydet} is zero automatically. The $x_{1,2,3}$ part of Eq. \eqref{eqn:singularity:C0:x1x2x3q} is corresponding to
\begin{align}
\left\{\begin{array}{lll}
&\ds x_1m_0^2+(x_2+x_3)\frac{m_0^2+m_1^2-m_{\mu}^2}{2}=0,\vspace{1ex}\\
&\ds x_1\frac{m_0^2+m_1^2-m_{\mu}^2}{2}+(x_2+x_3)m_1^2=0,\vspace{1ex}\\
&x_1+x_2+x_3=1.
\end{array}\right.
\end{align}
The non-zero solution exists only if $(m_0^2+m_1^2-m_{\mu}^2)^2-4m_0^2m_1^2=[m_{\mu}^2-(m_0+m_1)^2][m_{\mu}^2-(m_0-m_1)^2]=0$. Then, the singular points are
\begin{align}
&x_1=\frac{2m_1^2}{m_{\mu}^2+m_1^2-m_0^2},\qquad x_2+x_3=1.
\end{align}
As we can see, the $m_{\mu}^2=(m_0+m_1)^2$ is in the physical region because $0\le x_1\le1$. However, the $m_{\mu}^2=(m_0-m_1)^2$ is not in the physical region because $x_1$ does not lie between 0 and 1.\\
For the NLO-LS, the normal thresholds are $m_1=0$ and $m_{\mu}^2=(m_0+m_1)^2$, while the pseudo-threshold is $m_{\mu}^2=(m_0-m_1)^2$.\\
For the NNLO-LS, the singularities are $m_0=0$ and $m_1=0$. 

\bibliography{muongm2-arXiv_v3}
\end{document}